\documentclass{aa}  

\usepackage{graphicx}
\usepackage[varg]{txfonts}
\usepackage{natbib}
\bibpunct{(}{)}{;}{a}{}{,} 
\usepackage[colorlinks=true,allcolors=blue]{hyperref}
\usepackage{caption}
\usepackage{subcaption}
\usepackage{amsmath}
\usepackage{multirow}
\usepackage{algorithm}
\usepackage[noend]{algpseudocode}
\usepackage{algcompatible}
\usepackage{hyperref}
\usepackage{fontawesome} % Little github symbol
\usepackage{booktabs}
\usepackage{siunitx}
\usepackage{dcolumn}
\newcolumntype{d}[1]{D{.}{.}{#1}}

\DeclareMathOperator*{\argmin}{arg\,min}

\graphicspath{{imgs/}}

\newcommand{\nblink}[1]{\href{https://github.com/CosmoStat/mccd/notebooks/#1.ipynb}{\faFileCodeO}}
\newcommand{\github}{\href{https://github.com/CosmoStat/mccd}{\faGithub}}

\begin{document}

% autoref names
\renewcommand{\sectionautorefname}{Sect.}
\renewcommand{\subsectionautorefname}{Sect.}
\renewcommand{\subsubsectionautorefname}{Sect.}
\renewcommand{\figureautorefname}{Fig.}
\def\equationautorefname~#1\null{%
  Eq.~(#1)\null
}

 \title{Multi-CCD Point Spread Function Modelling}

   \titlerunning{Multi-CCD Point Spread Function Modelling}
   
   \authorrunning{T.~Liaudat et al.}

\author{T.~Liaudat$^{1}$, J.~Bonnin$^{1}$, J.-L.~Starck$^{1}$,  M.A.~Schmitz$^{2}$, A.~Guinot$^{1}$, M.~Kilbinger$^{1,3}$ and S.D.J.~Gwyn$^{4}$}

\institute{	$^{1}$ AIM, CEA, CNRS, Université Paris-Saclay, Université de Paris, F-91191 Gif-sur-Yvette, France\\
		$^{2}$ Department of Astrophysical Sciences, Princeton University, 4 Ivy Ln., Princeton, NJ08544, USA\\
		$^{3}$ Institut d'Astrophysique de Paris, UMR7095 CNRS, Universit\'{e} Pierre \& Marie Curie, 98 bis boulevard Arago, F-75014 Paris, France\\
		$^{4}$ NRC Herzberg Astronomy and Astrophysics, 5071 West Saanich Road, Victoria, BC V9E 2E7, Canada\\
        \email{tobias.liaudat@cea.fr}}

% \date{Date: To be defined ..}
\abstract{Galaxy imaging surveys observe a vast number of objects that are affected by the instrument's Point Spread Function (PSF). Weak lensing missions, in particular, aim at measuring the shape of galaxies, and PSF effects represent an important source of systematic errors which must be handled appropriately. This demands a high accuracy in the modelling as well as the estimation of the PSF at galaxy positions.} 
% Context
{Sometimes referred to as non-parametric PSF estimation, the goal of this paper is to estimate a  PSF at galaxy positions, starting from a set of noisy star image observations distributed over the focal plane. To accomplish this, we need our model to first of all, precisely capture the PSF field variations over the Field of View (FoV), and then to recover the PSF at the selected positions.} 
% Aims
{This paper proposes a new method, coined MCCD (Multi-CCD PSF modelling), that creates, simultaneously, a PSF field model over all of the instrument's focal plane. This allows to capture global as well as local PSF features through the use of two complementary models which enforce different spatial constraints. Most existing non-parametric models build one model per Charge-Coupled Device (CCD), which can lead to difficulties in capturing global ellipticity patterns.} 
% Methods
{We first test our method on a realistic simulated dataset comparing it with two state-of-the-art PSF modelling methods (PSFEx and RCA). We outperform both of them with our proposed method. Then we contrast our approach with PSFEx on real data from CFIS (Canada-France Imaging Survey) that uses the CFHT (Canada-France-Hawaii Telescope). We show that our PSF model is less noisy and achieves a $\sim22\%$ gain on pixel Root Mean Squared Error (RMSE) with respect to \texttt{PSFEx}.} % Results
{We present, and share the code of, a new PSF modelling algorithm that models the PSF field on all the focal plane that is mature enough to handle real data. \github} % Conclusions

\keywords{Astronomical instrumentation, methods and techniques -- Methods: data analysis -- Techniques: image processing -- Cosmology: observations -- Gravitational lensing: weak }
\maketitle

%-------------------------------------------------------------------

% Introduction
\section{Introduction}\label{sec:intro}
Current galaxy imaging surveys, such as DES \citep{jarvis2016}, KIDS \citep{kuijken2015}, CFIS \citep{cfis} or future surveys such as the Vera C. Rubin Observatory's LSST \citep{lsst}, the Euclid mission \citep{laureijs2011}, or the Roman Space Telescope require to estimate the Point Spread Function (PSF) of the instrument. For some scientific applications such as weak gravitational lensing \citep{kilbinger2015}, low-surface brightness studies \citep{lsb_psf}, or analysis of diffraction-limited images in crowded stellar fields \citep{sphere_psf}, the PSF must be reconstructed with high accuracy. 
A first approach is to derive a PSF model using known information about the instrument, where the model parameters are then chosen by fitting observed stars in the field to yield a PSF model. This has been widely used for the HST (Hubble Space Telescope) \citep[TinyTim software,][]{krist1995}, though it was later shown that a relatively simple PSF estimation from the data, which does not assumes a model for the instrument, provides better fits to stars for both photometry and astrometry measurements \citep{HST_PSF2017}. 
Furthermore, such a solution can not readily be applied to ground-based observations, where the atmosphere plays an important role and adds stochasticity to the PSF.
Other methods, based on imaging-data only, use unresolved stars in the field as direct measurements of the PSF, and reconstruct an accurate PSF from these observed stars. A very impressive range of methodologies have been proposed in the past to perform this task:  Moffat modelling \citep{moffat1988}, polynomial models \citep{Piotrowski2013,bertin2011}, principal component analysis \citep{pca_psf2007,schrabback2010,gentile2013}, sparsity \citep{ngole2015}, neural networks  \citep{herbel2018,neuralnet_psf2020,net_psf2020_2}, and 
optimal transport \citep{ngole2017,schmitz2018}.
The \texttt{PSFEx} software \citep{bertin2011} is the most widely used. 
The Resolved Components Analysis (RCA) method \citep{ngole2016,schmitz2020} was proposed in the framework of the Euclid space mission in order to deal with PSFs that are both undersampled  and spatially varying in the field.  
Cameras are often mosaics of several CCDs, but all mentioned methods can only build one PSF model per CCD, with the exception of the approach used by \cite{miller2013}, and the recently proposed approach by \cite{jarvis2020}. Since the models they build within each detector are independent from each other,
it is difficult to capture global patterns of variation in the PSF. For example, upon observing \texttt{PSFEx}'s shape residuals maps from the DES Year 1 results \citep[Fig. 8 in][]{zuntz2018}), we can see global patterns.

In this paper, we present a new method based on RCA that can capture large patterns spreading across several or all CCDs.
We compare the results with both RCA and \texttt{PSFEx} on simulations and real data.
Section~\ref{sec:RCA_psfex} reviews these two existing methods, while the proposed MCCD methods are described in section~\ref{sec:mccd}. Experiments on simulated images are shown in section~\ref{sec:num_exp}, 
tests on real data in section~\ref{sec:cfis}, and conclusions in section~\ref{sec:conclusion}.

\autoref{tb:variable} provides a glossary of variables used throughout this article.

\begin{table}
\begin{center}       
\begin{tabular}{cl} 
\hline \hline
\textbf{Variable} & \multicolumn{1}{c}{\textbf{Description}}   \\
\hline
\multicolumn{2}{c}{\textit{Observational model}} \\
\hline
$\mathcal{H}$           & PSF field  \\
$\mathcal{F}$           & degradation operator  \\
$u_{i}^{k}$             & 2-dimensional position of star $i$ in CCD $k$  \\
$n_{\text{star}}^{k}$   & number of observed stars in CCD $k$ \\
$N$                     & number of observed stars in all the CCDs \\
$n_{k, i}$              & noise realisation of star $i$ in CCD $k$  \\
$y_{k, i}$              & square star observation stamp $i$ on CCD $k$  \\
$n_y$                   & number of pixels on one dimension of $y_{k, i}$ \\
$D$                     & downsampling factor \\
$\mathbf{y}_{k, i}$     & 1-dimensional column representation of $y_{k, i}$ \\
$Y_{k}$                 & matrix stacking all the star observations $\mathbf{y}_{k, i}$ \\
\hline
\multicolumn{2}{c}{\textit{PSF model}} \\
\hline
$\hat{H}_{k}$                   & PSF model estimation of the observed stars $Y_{k}$  \\
$ r_k , \Tilde{r}_k$            & local and global number of eigenPSFs \\
$S_{k}$ , $\Tilde{S}$           & local and global \textit{eigenPSF} matrices \\
$A_{k}$ , $\Tilde{A}_k$         & local and global weight matrices \\
$\alpha_{k}$ , $\Tilde{\alpha}$ & local and global spatial constraint weights  \\
$V_k^\top$ , $\Pi_k$            & local and global spatial constraint dictionaries \\
$K_{\sigma}^{\text{Loc}}$ , $K_{\sigma}^{\text{Glob}}$  & local and global denoising parameters \\
$(e_{k,i} , a_{k,i})$           & RCA graph constraint parameters \\
$\mathbf{w}_{k,i}$ , $\mathbf{\Tilde{w}}_i$ & local and global weight vectors for \\
                                & the sparsity inducing term \\
$\Phi$                          & sparsity inducing transform \\
\hline
\multicolumn{2}{c}{\textit{PSF recovery}} \\
\hline
$\phi$                          & Radial Basis Function (RBF) kernel \\
$N_{\text{RBF}}$                & number of elements used to estimate the \\
                                & RBF interpolant \\
$(\lambda_i)_{i=1}^{N_{\text{RBF}}}$ & RBF interpolation weights \\
$A_{k}(N_{\text{RBF}})$         & weight matrix composed by the $N_{\text{RBF}}$ closest \\
                                & stars of a given target position \\
$A_{k,u}$ , $\Tilde{A}_{k,u}$   & local and global interpolated weight columns \\ 
                                & for a target position $u$ \\
$\hat{H}(u)$                    & recovered PSF at position $u$ \\
\hline 
\end{tabular}
\caption{Important variables used in this article.}
\label{tb:variable}
\end{center}
\end{table} 

%-------------------------------------------------------------------

% RCA and PSFEx
\section{PSFEx and RCA}
\label{sec:RCA_psfex}
\texttt{PSFEx} \citep{bertin2011} is a standard and widely-used software\footnote{{\url{https://github.com/astromatic/psfex}}}. RCA \citep{ngole2016} is a more recent method that was developed with the \textit{Euclid} Visible Imager's PSF in mind, to deal with the under-sampling of the observed star images. The software is also freely available\footnote{\url{https://github.com/CosmoStat/rca}}. It is important to remark that these two approaches rely solely on the  observed data: they are blind with respect to the optical system involved in the image acquisition process.

% --------- %
\subsection{The observation model}

Let us define $\mathcal{H}(u)$ as the PSF field involved in our problem. It is a continuous function of a 2-dimensional position $u=(x,y)$, which in principle could be image coordinates, based on the camera's CCD pixels, or could also be celestial coordinates such as right ascension and declination. Throughout this paper, we will consider that this PSF field accounts for the contribution of all effects from optical aberrations and diffraction to atmospheric distortions. 

Our observation model will consist of images $I_k$, the pixels in one CCD chip $k$, that will contain $n_{\rm star}^{k}$ noisy stars at positions $u_i^{k}$. 
We define a ``stamp'' as a square small image cutout centred on a single star.
Each star observation stamp $i$ on CCD $k$ can be written as:

\begin{equation}
  y_{k, i} = \mathcal{F}\left( \mathcal{H}(u_{i}^{k}) \right) + n_{k, i},
\end{equation}
where $n_{i , k}$ accounts for a noise image that we will consider to be white and Gaussian, and $\mathcal{F}$ is the degradation operator. Three main effects are taken into account in this operator: \textit{i)} the discrete sampling into a finite number of pixels, namely an image stamp of $n_y \times n_y$ pixels; \textit{ii)} a sub-pixel shift that depends on where the centroid of the image is placed with respect to the pixel grid; and \textit{iii)} a downsampling that affects the pixels in the stamp by a factor of $D$ leaving a $D \, n_y \times D \, n_y$ stamp.
For example, to handle the Euclid mission sampling rate \citep{cropper2013}, a factor $D = 1/2$ is required to achieve Nyquist sampling rate.
From now on, and throughout this article, we will use a unitary value for $D$.

We write each of these stamps into a 1-dimensional column vector and therefore $Y_k = [\mathbf{y}_{k, 1} \cdots \mathbf{y}_{k, n_{\rm star}^{k}}]$ is the matrix containing all the observed stamps in CCD $k$. It contains $n_{\rm star}^{k}$ columns and $D \, n_y \times D \, n_y$ rows. Finally, we  concatenate all CCD matrices and obtain $Y =  \left( \begin{array}{ccc} Y_1 & \cdots & Y_K \end{array} \right)$. 

% --------- %
\subsection{PSFEx}

For a given exposure, this method builds one independent model for each CCD. It was designed as a companion software for \texttt{SExtractor} \citep{bertin1996}, which builds catalogues of objects from astronomical images. Each object contains several measurements that \texttt{PSFEx} then uses to describe the variability of the PSF. Each selected attribute follows a polynomial law up to some user-defined maximum polynomial degree $d$.  The model for CCD $k$ can be written as:

\begin{equation}
  \hat{H}_{k}^{PSFEx} = S_{k} A_{k} ,
\end{equation}
where $A_{k}$ has $m$ rows corresponding to the number of polynomials used, and $n^{k}_{\rm star}$ columns corresponding to the number of observed stars used to train the PSF model. The matrix $S_k$ is learned during training and has $n_{y}^{2}$ rows (the number of pixels in each image), and $m$ columns.

For example, if $d$ is set to $2$ and the attributes chosen are the pixel coordinates $(x, y)$, each column $i$ of the $A_{k}$ matrix corresponding to the star $i$ at location $u^k_i = (x^k_i, y^k_i)$ is $ \mathbf{a}^{k}_{i} = [1, x^k_i, y^k_i, (x^k_i)^2, (y^k_i)^2, x^k_i y^k_i]^{T}$. The number of monomials $m$ corresponding to a maximum degree $d$ can be computed as $(d+1)(d+2)/2$.

The training of the model amounts to solving an optimisation problem of the form:

\begin{equation}
  \min_{ \Delta S_{k}} \left\{ \sum_{i = 1}^{n^{k}_{\rm star}} \left\lVert \frac{ \mathbf{y}_{k, i} - \mathcal{F}\left( (S_{0,k} + \Delta S_{k}) \mathbf{a}^{k}_{i} \right)  }{\hat{\sigma}_i}  \right\rVert_{2}^{2}  + \left\lVert T \Delta S_{k} \right\rVert_{F}^{2} \right\},
  \label{eq:psfex_def}
\end{equation}
where $\hat{\sigma}_i$ represents the estimated per-pixel variances, and $T$ is a scalar weighting. The matrix $S_{k}$ is decomposed as $S_{0,k} + \Delta S_{k}$, where the first term corresponds to a first guess of the PSF. The optimisation is carried out on the difference between this first guess and the observations. The second term in \autoref{eq:psfex_def} acts as a Tikhonov regularisation which, in this case, favours smoother PSF models. 

Finally, the PSF recovery at one galaxy position $u_j$ is straightforward and can be done by using the learned $S_k$ matrix and directly calculating a vector $\hat{a}_{k,j}$ corresponding to the monomials of the chosen attributes. The recovered PSF is then computed as

\begin{equation}
  \hat{\mathbf{h}}_{k,j}^{PSFEx} = S_k \hat{\mathbf{a}}_{k,j}.
\end{equation}

% --------- %
\subsection{Resolved Components Analysis}

The RCA method is based on a matrix factorisation scheme. It was first presented in \cite{ngole2016} and later evaluated on \textit{Euclid} image simulations in \cite{schmitz2020}. As with \texttt{PSFEx}, this method also builds independent models for each CCD within an exposure and is able to handle under-sampled images. Any observed star $i$ from CCD $k$ is modelled as a linear combination of PSF features, called eigenPSFs in the following, as

\begin{equation}
  \hat{\mathbf{h}}_{k,i}^{RCA} = S_{k} \mathbf{a}_{k,i} ,
\end{equation}
where $S_{k}$ is the matrix composed of the eigenPSFs, $\mathbf{a}_{k,i}$ a vector containing the set of linear weights and $\hat{\mathbf{h}}_{k,i}^{RCA}$ the reconstructed PSF. 

The modelling is recast into an optimisation problem were the $S_{k}$ and $A_{k}$ matrices are estimated simultaneously. The problem is ill-posed due to the under-sampling and the noise, meaning that many PSF fields can reproduce the observed stars. In order to break this degeneracy RCA uses a series of regularisers during the optimisation procedure to enforce certain mild assumptions on the PSF field: i) low-rankness of the solution, enforced by setting the number of eigenPSFs learned, $N$, to be small; ii) positivity of the reconstructed PSFs; iii) sparsity of the PSF representation on an appropriate basis; and iv) spatial constraints that account for imposing a certain structure within the $A_{k}$ matrix. This last constraint is imposed by a further factorisation of $A_{k}$ into $\alpha_{k}V_{k}^{T}$. The computation of the $V_{k}^{T}$ matrix will be addressed in section \ref{sec:local_model}. Finally, the PSF model reads:

\begin{equation}
  \hat{H}_{k}^{RCA} = S_{k} \alpha_{k} V_{k}^{T},
\end{equation}
and the optimisation problem that the RCA method solves is

\begin{align}
   \min_{S_{k}, \alpha_{k}} \Bigg\{ \frac{1}{2} \left\lVert Y_{k} - \mathcal{F}\left( S_{k} \alpha_{k} V_{k}^{T} \right)\right\rVert_{F}^{2} & \nonumber \\ 
   + \sum_{i = 1}^{N} \| w_{k, i} \odot \Phi \mathbf{s}_{k, i} \|_1 &+ \iota_+(S_{k} \alpha_{k} V_{k}^\top) + \iota_\Omega(\alpha_{k}) \Bigg\},
   \label{eq:optim_rca}
\end{align}
where $w_{k, i}$ are weights, $\Phi$ represents a transformation allowing the eigenPSFs to have a sparse representation, $\odot$ denotes the Hadamard product, $\iota_+$ is the indicator function of the positive orthant and $\iota_\Omega$ is the indicator function 
over a set $\Omega$ defined to enforce the spatial constraints.

The PSF recovery at a position $u_j$ is carried out by a Radial Basis Function (RBF) interpolation of the learned columns of the $A_{k}$ matrix, issuing a vector $\mathbf{a}_{k,j}$. In this way, the spatial constraints encoded in the $A_{k}$ matrix are preserved when estimating the PSF at galaxy positions. Finally, the reconstructed PSF is
\begin{equation}
  \hat{\mathbf{h}}_{k,j}^{RCA} = S_k \hat{\mathbf{a}}_{k,j}.
\end{equation}

%-------------------------------------------------------------------
% Multi-CCD method
\section{A new family of Multi-CCD methods}\label{sec:mccd}

The MCCD methods we propose here aim at exploiting all of the information available in a singe exposure, which
requires handling all CCDs simultaneously. 
The main advantage of this approach is the fact that we can build a more complex model since 
the number of stars available for training is much larger, compared to a model based on individual CCDs. 
We aim at a model able to capture PSF features following a global behaviour despite the fact that the PSF field is discontinuous at CCD boundaries. The main reason behind this discontinuity effect is the misalignments between different CCDs. Methods such as \texttt{PSFEx} or RCA, which process each CDD independently, avoid the discontinuity problem by construction, but have difficulties capturing global patterns of PSF variability that occur on scales larger than a single CCD.

The main idea behind our MCCD approach is to include both a global model which provides a baseline estimation of the PSF, and a local model that provides CCD-specific corrections. 

% --------- %
\subsection{The MCCD data model}\label{sec:data_model}

In a typical wide-field setting, the PSF field $\mathcal{H}$ exhibits a certain regularity that we translate into spatial correlations of the PSFs. The model we build for a specific CCD $k$ is the matrix $\hat{H}_{k} \in \mathbb{R}^{n_y^2 \times n_{\rm star}^{k}}$ composed by the concatenation of the estimations of the different stars encountered in that CCD. Each  postage stamp column of length $n_y^2$ corresponds to the model for a specific flattened star from the $n_{\rm star}^{k}$ stars present in CCD $k$. 

The PSF field at star positions is reconstructed as a linear combination of PSF features, called \textit{eigenPSFs}, learned from the observations. As previously stated we want to have both a global and a local component for the model, so we need different eigenPSFs for each component. Hence, the model is based on a matrix factorisation scheme as follows:

\begin{equation}
    \label{eq:local_decomposition}
    \hat{H}_k = \underbrace{S_k \; A_k}_{\text{Local: } \hat{H}_{k}^{\text{Loc}}} \; + \;  \underbrace{\Tilde{S} \; \Tilde{A}_k}_{\text{Global: } \hat{H}_{k}^{\text{Glob}}} ,
\end{equation}
where $S_k \in \mathbb{R}^{n_y^2 \times r_k}$ contains $r_k$ local eigenPSFs and $\Tilde{S} \in \mathbb{R}^{n_y^2 \times \Tilde{r}}$ contains $\Tilde{r}$ global eigenPSFs.  The matrices $A_k \in \mathbb{R}^{r_k \times n_{\rm star}^{k}}$ and $\Tilde{A}_{k} \in \mathbb{R}^{\Tilde{r} \times n_{\rm star}^{k}}$ correspond to the local and global weights of the linear combinations, respectively. We can see that for a given CCD $k$, the final model, $\hat{H}_{k}$, is made up of the sum of the contributions of the local, $\hat{H}_{k}^{\text{Loc}}$, and global, $\hat{H}_{k}^{\text{Glob}}$, models.

Now, let us build a single model for all the $K$ CCDs in the focal plane. We start by building a single matrix containing all the PSF models by concatenating the model $\hat{H}_{k}$ for each CCD as follows:

\begin{equation}
    \label{PSFs}
    \hat{H} = \left( \begin{array}{ccc} \hat{H}_1 & \cdots & \hat{H}_K \end{array} \right),
\end{equation}
where $\hat{H} \in \mathbb{R}^{n_y^2 \times N}$ and  $N = \sum_{k=0}^{K}  n_{\rm star}^{k}$ is the total number of stars in one camera exposure. Then, we can concatenate the different eigenPSF matrices $k$ into a single matrix:

\begin{equation}
    \label{components}
    S = \left( \begin{array}{cccc} S_1 & \cdots & S_K & \Tilde{S} \end{array} \right),
\end{equation}
where $S \in \mathbb{R}^{n_y^2 \times r}$ and we concatenated the global eigenPSF matrix, $\Tilde{S}$, at the end. This leaves a total of $r = \sum_{k = 1}^N r_k + \Tilde{r}$ columns for the $S$ matrix. We can follow a similar procedure to define $A$ as a block matrix:

\begin{equation}
    A = \left( \begin{array}{c c c c}
        A_1     & 0     & \cdots    & 0     \\ 
        0       & A_2   & \cdots    & 0     \\ 
        \vdots  &\vdots & \ddots    &\vdots \\ 
        0       & 0     & \cdots    & A_K   \\ 
        \Tilde{A}_1 & \Tilde{A}_2  & \cdots    & \Tilde{A}_K
    \end{array} \right),
\end{equation}
where $A \in \mathbb{R}^{r \times N}$ and $0$ is used for matrices made up of zeros. The last row of the $A$ block matrix is composed by the global model weights $\Tilde{A}_{k}$. Having already defined the Multi-CCD matrices, $\hat{H}$, $S$ and $A$, we can write the final model as:

\begin{equation}
  \hat{H} = S A,
  \label{eq:glob_decomp}
\end{equation}
where we include all the CCDs. Expanding it leads to a formula like \autoref{eq:local_decomposition} for each CCD.

% --------- %
\subsection{Inverse problem and regularisation}\label{sec:regularisations}

The estimation of our model, summarised in the matrices $S$ and $A$ of \autoref{eq:glob_decomp}, is posed as an inverse problem. Given the observation and MCCD data models presented above, this problem amounts to the minimisation of $\left\lVert Y - \mathcal{F}(S A) \right\rVert_{F}^{2}$, where $\lVert \cdot \rVert_{F}$ denotes the Frobenius matrix norm. This problem is ill-posed due to the noise in the observations and to the degradation operator $\mathcal{F}$, meaning that there are many PSF models that would match the star observations. In order to break this degeneracy we enforce several constraints, based on the basic knowledge we dispose of the PSF field, that regularise our inverse problem. Similarly to the ones exploited in the RCA method \citep{schmitz2020}, we use the following constraints: i) \textit{Low-rankness of the model}; ii) \textit{Positivity of the model}; iii) \textit{Sparsity in a given domain}; iv) \textit{Spatial variations}. We give a more detailed description of these in \autoref{appdx:rca_reg}. These constraints are used by both parts of our model, the global and the local components.

As mentioned above, the spatial constraint is enforced by further factorisation of the coefficient matrices $A$. However, since we want to enforce different properties for the global and the local contributions, the factorisation used will differ for each case.

% --------- %
\subsection{Global model}\label{sec:global_model}

We want the global component to provide a baseline estimation of the PSF and for that we propose that the coefficients follow a polynomial variation of the position. The global coefficient matrix $\Tilde{A}_k$ is factorised into $\Tilde{A}_k = \Tilde{\alpha} \Pi_k$ where $\Tilde{\alpha} \in \mathbb{R}^{\Tilde{r} \times \Tilde{r}}$ is a weight matrix and $\Pi_k \in \mathbb{R}^{\Tilde{r} \times n_{\rm star}^{k}}$ contains each considered monomials evaluated at global star positions. The dimension, $\Tilde{r}$, is determined by the maximum allowed degree in the polynomials: for all monomials of degree less than a given $d$, we have $\Tilde{r} = \binom{d + 2}{2} = \frac{(d + 1)(d + 2)}{2}$. For example, for $d = 2$ (i.e. $\Tilde{r} = 6$), we have:

\begin{equation}
    \Pi_k = \begin{pmatrix}
        1 & \cdots & 1 \\
        x_{k, 1} & \cdots & x_{k, n_{\rm star}^{k}} \\
        y_{k, 1} & \cdots & y_{k, n_{\rm star}^{k}} \\
        x_{k, 1}^2 & \cdots & x_{k, n_{\rm star}^{k}}^2 \\
        x_{k, 1} y_{k, 1} & \cdots & x_{k, n_{\rm star}^{k}} y_{k, n_{\rm star}^{k}} \\
        y_{k, 1}^2 & \cdots & y_{k, n_{\rm star}^{k}}^2
    \end{pmatrix} ,
    \label{eq:poly_matrix}
\end{equation}
where $(x_{k, i}, y_{k, i})_{1 \leq i \leq n_{\rm star}^{k}}$ are the global pixel coordinates of the observed stars distributed in the $k$\textsuperscript{th} CCD. The global component of the model for a specific CCD $k$ are as follows:

\begin{equation}
     \hat{H}_{k}^{\text{Glob}} = \Tilde{S} \; \Tilde{\alpha} \; \Pi_k .
     \label{eq:global_contribution}
\end{equation}
It is important to mention that despite our choice, throughout this article, to use position polynomials for building the global space constraint, the model is not necessarily restricted to that choice. The $\Pi_k$ matrix could be constructed using other parameters of the observations in order to facilitate the capture of other dependencies and could also follow other types of functions.

% --------- %
\subsection{Local model}\label{sec:local_model}

It is possible to define different types of local models. In this article we discuss three options that depend on how we enforce the local spatial constraint. More specifically, they depend on how we factorise the local $A_k$ matrix in the relation: 

\begin{equation}
    \hat{H}_{k}^{\text{Loc}} = S_k \; A_k.
\end{equation}
Nevertheless, the MCCD framework does not restrict us to these three options, and it is possible to define other local models.

It is worth remarking that all the framework and optimisation procedures are maintained throughout the different flavours of the MCCD algorithms. The main difference is the way the spatial constraints are enforced in the local and global models.

\subsubsection{MCCD-RCA}

One motivation for the local model is to provide CCD-specific corrections, and to do so our first choice is RCA's spatial constraint strategy which leads to the MCCD-RCA algorithm. The motivation for this choice is the capability of the RCA spatial constraint to handle different types of PSF variations. 
On the one hand it can capture smooth variations over the CCD and on the other hand it can account for localised changes that affect a reduced number of PSFs. If the PSFs were sampled on a regular grid this would mean to capture variations occurring at different spatial frequencies. Unfortunately, the PSF locations do not coincide with a regular grid but on what could be seen as a fully connected undirected weighted graph where the weights can be defined as a function of the distance between the different nodes (PSF locations)\footnote{A graph $G$ can be defined as a pair $(V,E)$, where $V$ is the set of vertices and $E$ the set of edges that connects the different vertices. In our case, each star position constitutes a vertex and there is one edge for each pair of vertices. The edges have no preferred direction and its value depends on the distance between the two vertices it connects.}. 
However, the RCA spatial constraint exploits the graph harmonics in order to capture the different PSF variations. These harmonics are represented by the eigenvectors of the graph's Laplacian matrix \citep{chung1997}, which will depend on how we define the graph's weights. A parametric function of the PSF distances can serve as the graph's weights as in \cite{schmitz2020} and the selection of the function's parameters can be done following \cite{ngole2016}. 
For each local model (i.e. each CCD in the mosaic), we define $r_k$ graphs, each corresponding to one of the $r_k$ local parameters. For each graph, we can extract the $m_k$ most useful eigenvectors of its Laplacian matrix and gather all of them as columns of a matrix $V_k^{\text{RCA}} \in \mathbb{R}^{n_{\rm star}^{k} \times r_k m_k}$. This way, we can write:

\begin{equation}
    A_{k}^{\text{RCA}} = \alpha_{k}^{\text{RCA}} V_{k}^{\top \; \text{RCA}} ,
\end{equation}
where $\alpha_{k}^{\text{RCA}} \in \mathbb{R}^{r_k \times r_k m_k}$ is a weight matrix that is used to enforce the spatial constraints. In other words, the sparsity of $A_{k}^{\text{RCA}}$'s rows in the dictionary $V_{k}^{\top \; \text{RCA}}$. Full details are available in \cite{ngole2016} and \cite{schmitz2020}.

\subsubsection{MCCD-POL}

The second local model, referred to as MCCD-POL, follows a polynomial spatial constraint. Similar to \texttt{PSFEx}, we factorise the the local weights into two matrices as follows:

\begin{equation}
    A_{k}^{\text{POL}} = \alpha_{k}^{\text{POL}} \Pi_{k}^{\text{POL}},
    \label{eq:pol_extension}
\end{equation}
where $\Pi_{k}^{\text{POL}}$ has the same form as the matrix in \autoref{eq:poly_matrix}, with the difference that in this case the positions are represented in local coordinates of its corresponding CCD $k$. As with $d$ in the global model, a parameter is chosen to define the maximum order of the polynomial used.

\subsubsection{MCCD-HYB}

The third option consists in using the two local models we presented above, RCA and polynomial, to work together in an hybrid algorithm we will refer to as MCCD-HYB. The idea behind it is that the addition of the polynomial space constraint could help the original graph constraint to capture the different features found. In this case we will factorise the local weights with block matrices as

\begin{equation}
    A_{k}^{\text{HYB}} = \alpha_{k}^{\text{HYB}} V_{k}^{\top \; \text{HYB}} = \left( \begin{array}{cc}
        \alpha_{k}^{\text{RCA}}     & 0     \\ 
        0       & \alpha_{k}^{\text{POL}}   \\ 
    \end{array} \right) \begin{pmatrix}
        V_k^\top        \\
        \Pi_{k}^{\text{POL}}  \\
    \end{pmatrix},
    \label{eq:hybrid_extension}
\end{equation}
where $\alpha_{k}^{\text{POL}}$ and $\Pi_{k}^{\text{POL}}$ are the matrices defined in the polynomial version, and $\alpha_{k}^{\text{RCA}}$ and $V_k^\top$ are the matrices defined in the original MCCD-RCA algorithm.

Finally, generically including the spatial constraints in \autoref{eq:local_decomposition} we get the following description of our model for a specific CCD:
\begin{equation}
    \hat{H}_k = S_k \alpha_k V_k^\top + \Tilde{S} \Tilde{\alpha} \Pi_k ,
\end{equation}
that we can also write in a global form $\hat{H} = S \alpha V^\top$, where $\hat{H}$ and $S$ have already been defined in \autoref{PSFs} and \autoref{components}, and where $\alpha$ and $V^\top$ are the following matrices:

\begin{equation}
    \alpha = \left( \begin{array}{cccc} 
    \alpha_1& 0     & \cdots    & 0         \\ 
    0       &\ddots &           & \vdots    \\ 
    \vdots  &       & \alpha_N  & 0         \\ 
    0       &\cdots & 0         & \Tilde{\alpha} \end{array} \right)
    \quad , \quad
    V^\top = \left( \begin{array}{cccc} 
    V_1^\top    &  0        & \cdots    & 0         \\
    0           &V_2^{\top} &           & \vdots    \\ 
    \vdots      &           & \ddots    & 0         \\ 
    0           & \cdots    & 0         & V_N^\top  \\ 
    \Pi_1       & \Pi_2     & \cdots    & \Pi_N \end{array} \right).
\end{equation}

% --------- %
\subsection{Optimisation problem}

Combining the regularisations enumerated in \autoref{sec:regularisations} and the data model described in \autoref{sec:data_model} we can construct the optimisation problem in an elegant way by reformulating the \eqref{eq:optim_rca}.
However, we can split the optimisation problem into a more convenient way

\begin{align}
    \label{eq:min_problem}
    \min_{\substack{S_1, \ldots, S_N, \Tilde{S} \\ \alpha_1, \ldots, \alpha_N, \Tilde{\alpha}}} \Bigg\{ & \sum_{k=1}^N \Bigg( \frac{1}{2} \| Y_k - \mathcal{F}_k(S_k \alpha_k V_k^\top + \Tilde{S} \Tilde{\alpha} \Pi_k) \|_F^2 + \nonumber \\
    &  \sum_{i = 1}^{r_k} \| \mathbf{w}_{k,i} \odot \Phi \mathbf{s}_{k,i} \|_1 + \iota_+(S_k \alpha_k V_k^\top + \Tilde{S} \Tilde{\alpha} \Pi_k) + \iota_{\Omega_k}(\alpha_k) \Bigg)  \nonumber \\
    & + \sum_{i = 1}^{\Tilde{r}} \| \Tilde{\mathbf{w}}_i \odot \Phi \Tilde{\mathbf{s}}_i \|_1 + \iota_{\Tilde{\Omega}}(\Tilde{\alpha}) \Bigg\}. 
\end{align}
In the previous equation, the columns of $Y_k \in \mathbb{R}^{D^2 n_y^2 \times n_{\rm star}^{k}}$ are the stars distributed in the $k$\textsuperscript{th} CCD sensor, $\mathcal{F}_k$ is the degradation operator, $\mathbf{w}_{k,i}$ and $\mathbf{\Tilde{w}}_i$ are weight vectors, $\Phi$ is a transform that allows a sparse representation of our eigenPSFs, and $\Omega_k$ and $\Tilde{\Omega}$ are sets to enforce sparsity and normalisation of the rows of $\alpha_k$ and $\Tilde{\alpha}$, respectively. The indicator function of a set $\mathcal{C}$ is written as $\iota_{\mathcal{C}}(\cdot)$, that is equal to 0 if the argument belongs to $\mathcal{C}$ and $+ \infty$ otherwise. For example, $\iota_{+}$ is the indicator function over the positive orthant. 
More explicitly, the sets $\Omega_k$ and $\Tilde{\Omega}$ are defined the following way:

\begin{align}
    &\Omega_k =  \{ \alpha_k \ | \ \forall i \in \{1, \ldots, r_k\}, \| (\alpha_k^\top)_i \|_0 \leq \eta_{k,i} \land \| (\alpha_k V_k^\top )_i \|_2 = 1 \} ,  \\
    &\Tilde{\Omega} =  \{ \Tilde{\alpha} \ | \ \forall i \in \{1, \ldots, \Tilde{r}\}, \| (\Tilde{\alpha}^\top)_i \|_0 \leq \Tilde{\eta}_i \land \| (\Tilde{\alpha}\Pi_k)_i \|_2 = 1 \},
\end{align}
where $(\eta_{k,i})_{1 \leq i \leq r_k}$ and $(\Tilde{\eta}_i)_{1 \leq i \leq \Tilde{r}}$ are appropriately chosen integers, and $\lVert \cdot \rVert_0$ is the pseudo-norm $\ell_0$ that returns the number of non-zero elements of a vector. So we are enforcing, in the global case, the row $i \in \{1, \ldots, \Tilde{r} \}$ of $\Tilde{\alpha}$ to have at most $\Tilde{\eta_{i}}$ non-zero elements. An interpretation could be that we are forcing each eigenPSF to follow  a small number of positional polynomials as $\Tilde{A}$'s rows will be sparsely represented over the $\Pi_k$ matrices.

The $\Phi$ transform used throughout this paper is the starlet transform \citep{starck2011}. We enforce the sparsity on the different decomposition levels excluding the coarse scale.
The $\ell_1$ term promotes the sparsity of the eigenPSFs with respect to $\Phi$ while the weights $\mathbf{w}_{k,i}$ and $\Tilde{\mathbf{w}}_i$ regulate the sparsity penalisation against the other constraints and should adapt throughout the optimisation algorithm depending on the noise level.
 
The second term in each of the $\Omega$ sets (e.g. $\|(\alpha_k V_k^\top )_i \|_2 = 1$) was not mentioned in the regularisation section \ref{sec:regularisations}, but they are needed to avoid a degenerated solution, for example $ \lVert S_k \rVert_F \rightarrow \infty$ and $\lVert A_k \rVert_F \rightarrow 0$, due to the usual scale indeterminacy when doing a matrix factorisation. To avoid this, we normalise $A_k$ and $\Tilde{A}$ columns.
This translates to forcing the normalisation of the eigenPSF weights contributing to model each observed star. This does not mean that the eigenPSF weights will be the same for each star, but that the norms of the weight vectors are equal.

% --------- %
\subsection{Algorithm}

The optimisation in \autoref{eq:min_problem} is non-convex as we are facing a matrix factorisation problem. To overcome this situation we use an alternating minimisation scheme where we optimise one variable at a time, iterating over the variables as studied in \cite{xu2013} or \cite{bolte2014}. In consequence, we can at most expect to converge towards a local minima. The main iteration is performed over the different variables occurring in \autoref{eq:min_problem}, first over the global $\Tilde{S},  \Tilde{\alpha}$  and then over the local $S_1, \alpha_1 , \ldots, S_K, \alpha_K$.

The method is shown in Algorithm \ref{alg:mccd}, which contains the four main optimisation problems derived from the alternating scheme. There exists a wide literature on minimisation schemes involving non-smooth terms, specifically proximal methods \citep{parikh2014}, that we can exploit in order to handle the four cases. Notably, we use the algorithm proposed by \cite{condat2013} for the problems (II), (III) and (IV). For the problem (I) we use the method proposed by \cite{liang2018} which is an extension of the well-known FISTA algorithm \citep{beck2009}. Even though the $\ell_0$ pseudo-norm is non-convex and therefore not adapted to the general scenario of the aforementioned algorithms, we can alleviate this fact by combining the use of its proximal operator and a given heuristic.

Concerning the algorithm's initialisation, we start by a preprocessing where we reject stars that are strong outliers in terms of shape or size. We run the shape measurement algorithm mentioned in \autoref{sec:quality_criteria} on the train stars and discard the ones that are several sigmas away from nearby stars. At this moment we can assign a specific weight for each train star. There are three available options: i) use a unitary weight for each train star; ii) use a weight provided by the user; iii) compute a weight $\omega_i$ as a function of the star's Signal-to-Noise-Ratio (SNR) based on $\omega_i \propto \text{SNR}_i / (\text{SNR}_i + \text{median}(\text{SNR}))$ and bounded to a specific interval to avoid bright stars from dominating the optimisation.
% that can be provided by the user or can follow a suggestion based on the star Signal-to-Noise-Ratio (SNR).

Next, we continue with all the local eigenPSFs set to zero, as seen in line 4 of Algorithm \ref{alg:mccd}; and the $\Tilde{\alpha}$ matrix set to the identity, favouring the specialisation of each global eigenPSF to one specific monomial. By following this procedure, we are training a global polynomial model that fits the stars as best as it can. Later on, the local models will work with the residuals between the observed stars and the global model, trying to capture variations missed in the previous step.

There are four iteration loops in algorithm \ref{alg:mccd}. On line 8, the main iteration, and on line 15 the iteration over the CCDs for the training of the local model. The other two iterations on lines 9 and 14 correspond to a refinement of the estimation. Our objective is to correctly estimate the global and the local contributions for the model, and to do this we alternate the minimisation between the global and the local contributions, which we call outer minimisation. On top of that, each of these two contributions include an inner alternating minimisation scheme as we are performing a matrix factorisation for the local and for the global models. For example, we are simultaneously minimising over $S_k, \alpha_k$ for the local model and over $\Tilde{S}, \Tilde{\alpha}$ for the global model.
We want to refine this inner minimisation, meaning that the optimisation of the two variables separately approaches the joint optimisation of both variables. To accomplish this, we need to go through a small number of iterations, which are described by the $n$ superscript variables, before continuing the iteration of the next alternating scheme. The optimisation strategy can be seen as a compound alternating minimisation scheme considering the outer and the inner alternations.

More information about the optimisation strategy can be found in \autoref{appdx:optim}.

\begin{algorithm*}
    \caption{Multi-CCD Resolved Components Analysis}
    \begin{algorithmic}[1]
        \Statex
        \Statex \textbf{Initialisation:}
        
        \State Preprocessing()
        
        \FOR{$k = 1$ to $K$}
            \State Harmonic constraint parameters $(e_{k,i}, a_{k,i})_{1 \leq i \leq r_k}$ $\to$ $V_k^\top$, $\alpha_k^{(0,0)}$
            \State $0_{n_y^2 \times r_k }$ $\to$ $S_{k}^{(0,0)}$
        \ENDFOR
        
        \State Global coordinates $\to$ $\Pi_k$, $\Tilde{\alpha}^{(0,0)}$ \hfill ($\Tilde{\alpha}^{(0,0)} = I$)
        \State $0_{n_y^2 \times \Tilde{r} }$ $\to$  $\Tilde{S}^{(0,0)}$
        \Statex
        
        \Statex \textbf{Alternate minimisation:}
        \FOR{$l = 0$ to $l_{max}$} \hfill Algorithm's main iterations
        \Statex
            \FOR{$n = 0$ to $n_{G}$} \hfill Global alternating iterations
                \State Noise level, $\Tilde{\alpha}^{(l,n)}$ $\to$ update $\Tilde{W}^{(l,n)}$
                
                \State $\tilde{S}^{(l+1, n+1)} = \argmin_{\Tilde{S}} \{ \sum_{k=1}^K \frac{1}{2} \| Y_k - \mathcal{F}_k(S_k^{(l,0)} \alpha_k^{(l,0)} V_k^\top + \Tilde{S} \Tilde{\alpha}^{(l,n)} \Pi_k) \|_F^2 + \sum_i \| \Tilde{\mathbf{w}}_{i}^{(l,n)} \odot \Phi \Tilde{\mathbf{s}}_i \|_1 \}$ \hfill (I)
                    
                \State $\Tilde{\alpha}^{(l+1, n+1)} = \argmin_{\Tilde{\alpha}} \{ \sum_{k=1}^K \frac{1}{2} \| Y_k - \mathcal{F}_k(S_k^{(l,0)} \alpha_k^{(l,0)} V_k^\top + \Tilde{S}^{(l+1, n+1)} \Tilde{\alpha} \Pi_k) \|_F^2 + \iota_{\Tilde{\Omega}}(\Tilde{\alpha}) \}$ \hfill (II)
                \ENDFOR
                
            \Statex
            
            \FOR{$n = 0$ to $n_{L}$} \hfill Local alternating iterations
                \FOR{$k = 1$ to $K$} \hfill CCD iterations
                   \State Noise level, $\alpha_k^{(l,n)}$ $\to$ update $W_{k}^{(l,n)}$
            
                    \State $S_k^{(l+1,n+1)} = \argmin_{S_k} \{ \frac{1}{2} \| Y_k - \mathcal{F}_k(S_k \alpha_k^{(l,n)} V_k^\top + \Tilde{S}^{(l+1,n_G)} \Tilde{\alpha}^{(l+1,n_G)} \Pi_k) \|_F^2 $
                    \Statex $\qquad \qquad \qquad \qquad \qquad \qquad \qquad + \sum_i \| \mathbf{\mathbf{w}}_{k,i}^{(l,n)} \odot \Phi \mathbf{s}_{k,i} \|_1 + \iota_+(S_k \alpha_k^{(l,n)} V_k^\top + \Tilde{S}^{(l+1,n_G)} \Tilde{\alpha}^{(l+1,n_G)} \Pi_k) \}$ \hfill  (III)
                
                    \State $\alpha_k^{(l+1,n+1)} = \argmin_{\alpha_k} \{ \frac{1}{2} \| Y_k - \mathcal{F}_k(S_k^{(l+1,n+1)} \alpha_k V_k^\top + \Tilde{S}^{(l+1,n_G)} \Tilde{\alpha}^{(l+1,n_G)} \Pi_k) \|_F^2 + \iota_{\Omega_k}(\alpha_k) \}$ \hfill (IV)
                
                \ENDFOR
            \ENDFOR
    
     \Statex
             
        \ENDFOR
        
    \end{algorithmic}
    \label{alg:mccd}
\end{algorithm*}

% --------- %
\subsection{PSF recovery}

Once the training of the model on the observed stars is done, we can continue with the problem of estimating the PSF field at galaxy positions. We call this problem PSF recovery. \cite{gentile2013} conduct a study on PSF interpolation techniques and \cite{ngole2017} propose a sophisticated approach based on optimal transport theory \citep{peyre2018}. We will follow a RBF (Radial Basis Function) interpolation scheme with a thin plate kernel\footnote{Where the kernel is defined as $\phi (r) = r^2 \ln (r)$.}, as in \cite{schmitz2020}, due to its simplicity and good performance. This choice comes with the assumption that the influence of each observation does not depend on the direction but only on the distance to the target which is well described by the RBF kernel.

The RBF interpolation of a function $f$ on a position $u$ works by building a weighted linear combination of RBF kernels ($\phi(\cdot)$) centred in each of the available training star positions $u_i$.
The interpolation function reads

\begin{equation}
  \hat{f}(u) = \sum_{i=1}^{N_{\text{RBF}}} \lambda_i \phi \left( \lVert u - u_i \rVert \right),
  \label{eq:RBF_interpolant}
\end{equation}
where $(\lambda_i)_{i=1}^{N_{\text{RBF}}}$ are the linear weights that need to be learnt and $N_{\text{RBF}}$ is the number of elements used to estimate the interpolant. In order to learn the weights, we force the exact reconstruction of the interpolant on the known positions, that is $\hat{f}(u_i) = f(u_i) \, \forall i \in \{1, \ldots , N_{\text{RBF}} \}$. By fixing the aforementioned constraint we have a system of $N_{\text{RBF}}$ equations with $N_{\text{RBF}}$ unknown that are the $\lambda_i$ weights. Once the system is solved, it is just a matter of evaluating the interpolant on the desired position $u$ following \autoref{eq:RBF_interpolant}.

At this point, we need to choose over which function $f$ we will interpolate. A straightforward choice would be to use the reconstructed PSFs at the training positions as the $f(u_i)$. Nevertheless, this would not take into account the specificities and structure of our model. Following the discussion in Sect. 4.2 of \cite{schmitz2020} we will use the learnt $A_k$ and $\Tilde{A}_k$ matrices. They encompass all the spatial distribution properties of the learned features, our eigenPSFs, and it is natural for our framework to use these values as the function to interpolate.

\begin{figure}
    \centering
    \includegraphics[width=.85\linewidth]{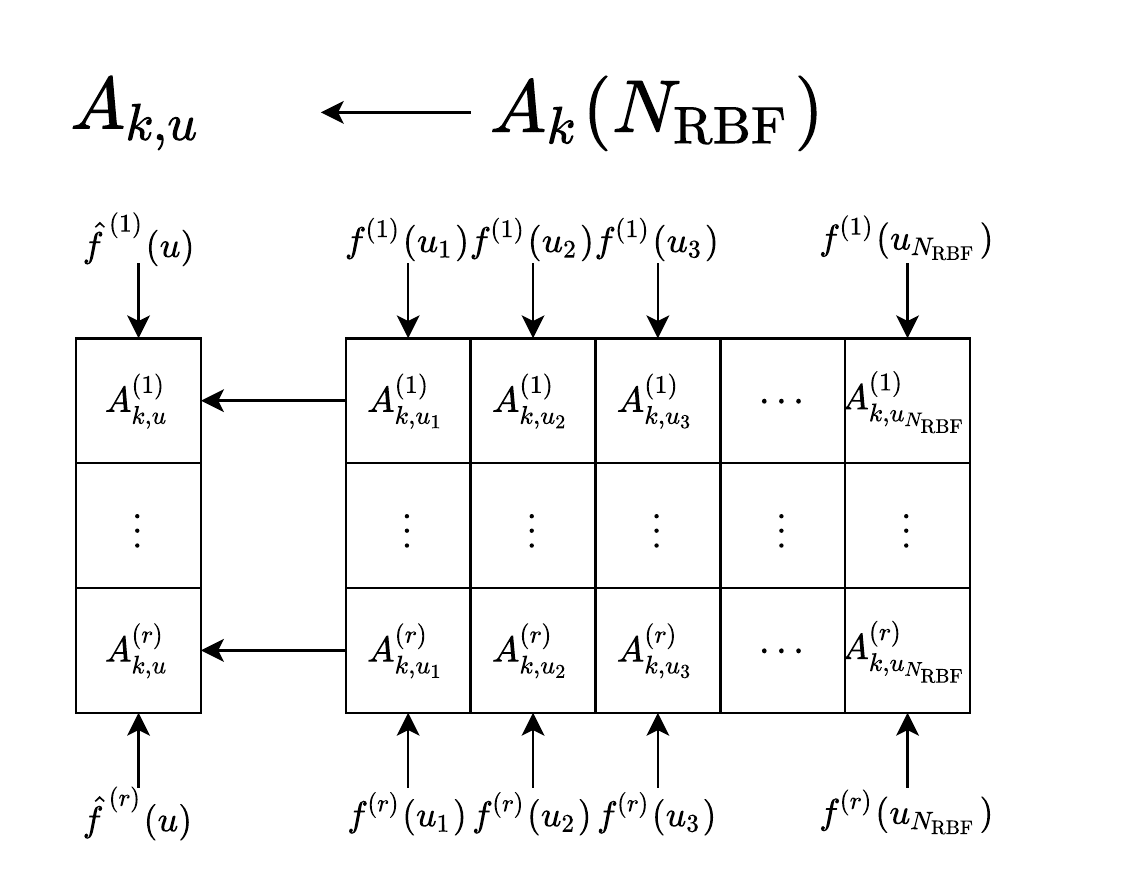}
    \caption{Example of the interpolation procedure involved in the PSF recovery.}    
    \label{fig:mccd_psf_recovery}
\end{figure}

We continue with a brief explanation of the interpolation procedure.
For one given target position $u$ in CCD $k$, we consider the $N_{\text{RBF}}$ closest observed stars to that position that also belong to the CCD $k$. We call $A_k(N_{\text{RBF}})$ to the $A_k$ matrix composed only with the columns of the aforementioned $N_{\text{RBF}}$ stars. We want to estimate the interpolated column vector $A_{k,u}$. For this, we use a RBF interpolation scheme for each row of the $A_k(N_{\text{RBF}})$ matrix. The elements of the row $t$ represent the $(f^{(t)}(u_i))_{i=1}^{N_{\text{RBF}}}$ evaluations and the element $A^{(t)}_{k,u}$ represents the interpolated value $\hat{f}^{(t)}(u)$. The same procedure is repeated for each row of the $A_k(N_{\text{RBF}})$ matrix so as to obtain the column vector $A_{k,u}$. This is illustrated in \autoref{fig:mccd_psf_recovery}.
We repeat the procedure with the global component matrix, $\Tilde{A}_k$, in order to obtain $\Tilde{A}_{k,u}$, another column vector with the interpolated values. At this point we note that we handle the global and the local contributions independently. Once we have calculated the two interpolated vectors, the reconstructed PSF is obtained following the MCCD data model as can be seen in the next equation

\begin{equation}
  \hat{H}(u) = \Tilde{S} \Tilde{A}_{k,u} + S_k A_{k,u} \, .
\end{equation}
We found that restricting the $N_{\text{RBF}}$ neighbours to a single CCD for the global components gave better results.
This might be due to the fact that the global components are able to capture some of the discontinuities from one CCD to another and therefore the interpolation is degraded when using stars from different CCDs. The number of neighbours $N_{\text{RBF}}$ should be chosen as a function of the available number of stars per CCD in the training set and as the RBF kernel chosen. From now on, and given the training set we handle in this article, $N_{\text{RBF}}$ is set to $20$.

% ------------------------------ %
% Experiences
\section{Numerical experiments}\label{sec:num_exp}

% --------- %
\subsection{Data}\label{sec:exp_data_set}
\begin{figure}
    \centering
    \includegraphics[width=.85\linewidth]{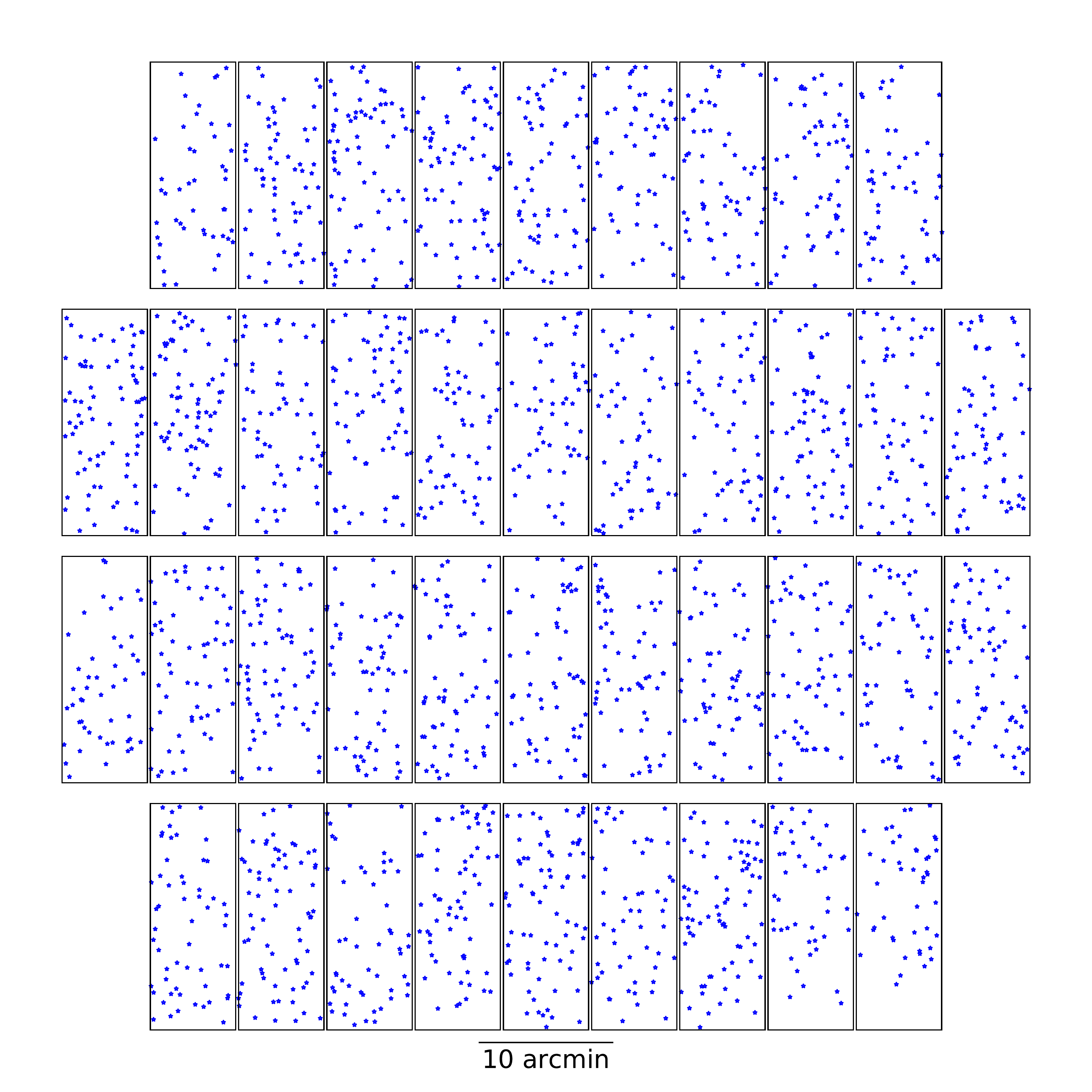}
    \caption{Star positions in CFHT's MegaCam used for the simulated dataset. The positions were taken from a real CFIS exposure.}    
    \label{fig:sim_star_positions}
\end{figure}

The simulated data set we create to evaluate MCCD set is based on a CFIS\footnote{\url{http://www.cfht.hawaii.edu/Science/CFIS/}} (Canada-France Imaging Survey) MegaCam\footnote{\url{http://www.cfht.hawaii.edu/Instruments/Imaging/MegaPrime/}} exposure from the CFHT (Canada-France-Hawaii Telescope). 
It contains $2401$ stars distributed along $40$ CCDs over a field of view of $\sim 1\;\text{deg}^{2}$ as shown in \autoref{fig:sim_star_positions}. Each CCD consists of a matrix of $2048$ by $4612$ pixels with some given gaps between the different CCDs. The horizontal gap length consist of $\sim70$ pixels while vertical gaps of $\sim425$ pixels.

\subsection{Training set}

Our simulation pipeline considers a Moffat PSF profile with normalised flux drawn using the \texttt{Galsim} software\footnote{\url{https://github.com/GalSim-developers/GalSim}} \citep{rowe2015} for each position in the exposure. 
To simulate the PSF shape variation, we used two radial analytic functions which define our ground truth 
shape ellipticities distortions. Shearing stars leads naturally to a size variation. \autoref{fig:analytic_moments} shows the resulting $e_1$, $e_2$ and size maps.
Our pipeline performs the following steps:
\begin{enumerate}
\item Simulate Moffat stars with a size fixed to the mean size measured in the real exposure. 
\item Shear the simulated stars as a function of their position using the two analytical functions.
\item Apply a random sub-pixel shift following a uniform distribution centred in zero.
\item Apply a binning to get a $51 \times 51$ pixels image, with a pixel size equivalent to CFIS MegaCam's maps, i.e.  $0.187$ arcsec.
\item Add a constant white Gaussian noise to the images, with standard deviation $\sigma$, derived from the desired SNR level

\begin{equation}
  \text{SNR} = \frac{\lVert y \rVert_{2}^{2}}{\sigma^{2} p^2},
  \label{eq:SNR_def}
\end{equation}
where $y$ is the image postage stamp consisting of $p^2$ pixels. Each experience will consist of a constant SNR value, as we will later see, that will be drawn from the set $\{10,\; 30,\; 50,\; 70  \}$.

\end{enumerate}

As \texttt{PSFEx} was designed as a companion software of \texttt{SExtractor} we need to follow a different procedure to generate the simulated data. We first need to process our simulations with \texttt{SExtractor}, so that the catalogue produced can be used as inputs for \texttt{PSFEx}.
To accomplish this we mimic a complete CCD so that \texttt{SExtractor} is able to process it.
We create star images as we already described for the MCCD method but without noise as it will be added later. 
Then we distribute them on a mock image of $2048 \times 4612$ pixels. The corresponding positions will be the pixel coordinates that are presented in \autoref{fig:sim_star_positions}. Once the mock image is created,
we add the noise value according to the desired SNR to the whole image. When the mock image is created, we run \texttt{SExtractor} in order to have a star catalogue that \texttt{PSFEx} can use as input. 

% --------- %
\subsection{Testing data set}

\begin{figure*}
        \centering
    \begin{subfigure}{0.33\linewidth}
        \centering
        \includegraphics[width=1.\textwidth]{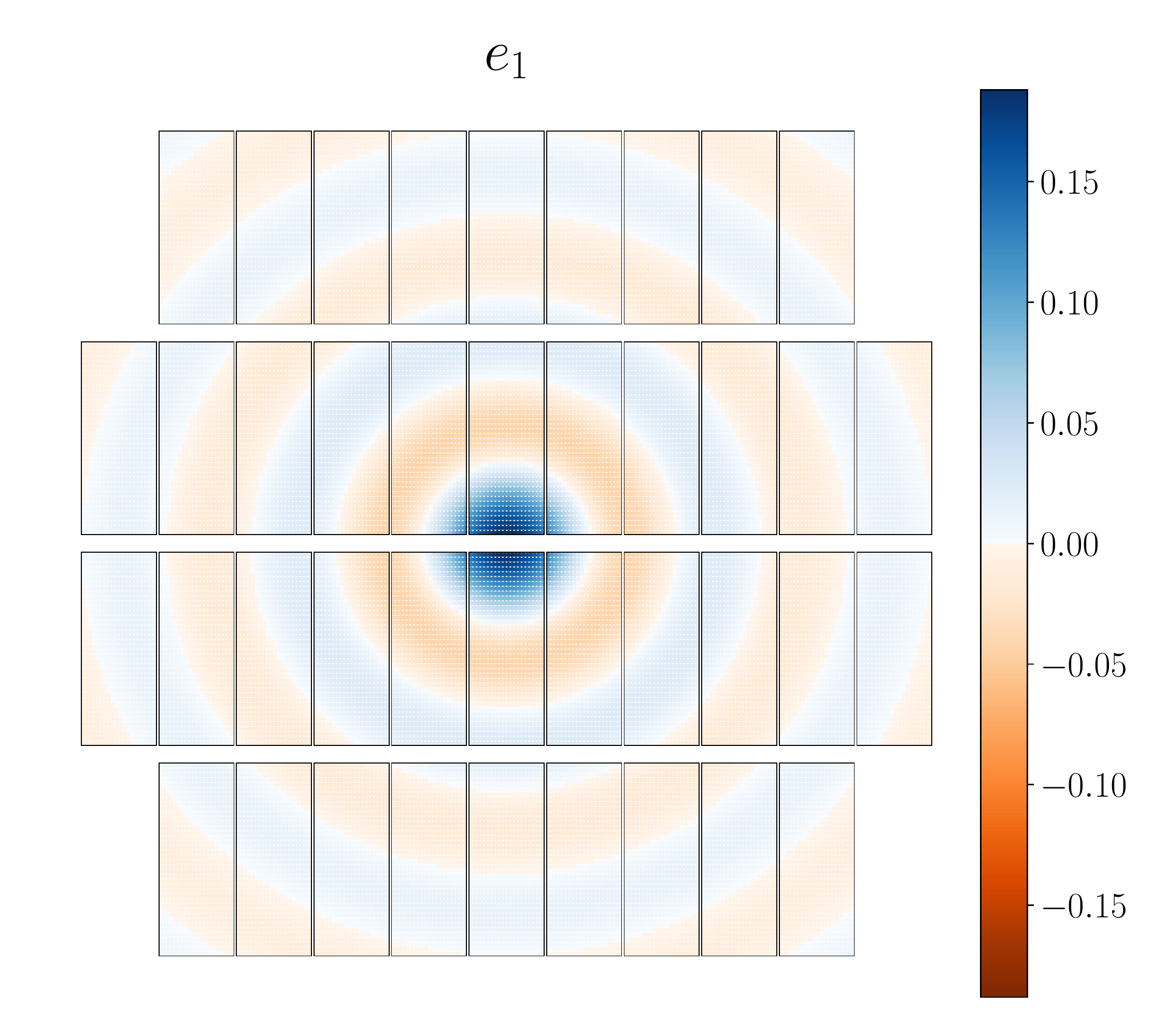}
    \end{subfigure}
    \begin{subfigure}{0.33\linewidth}
        \centering
        \includegraphics[width=1.\textwidth]{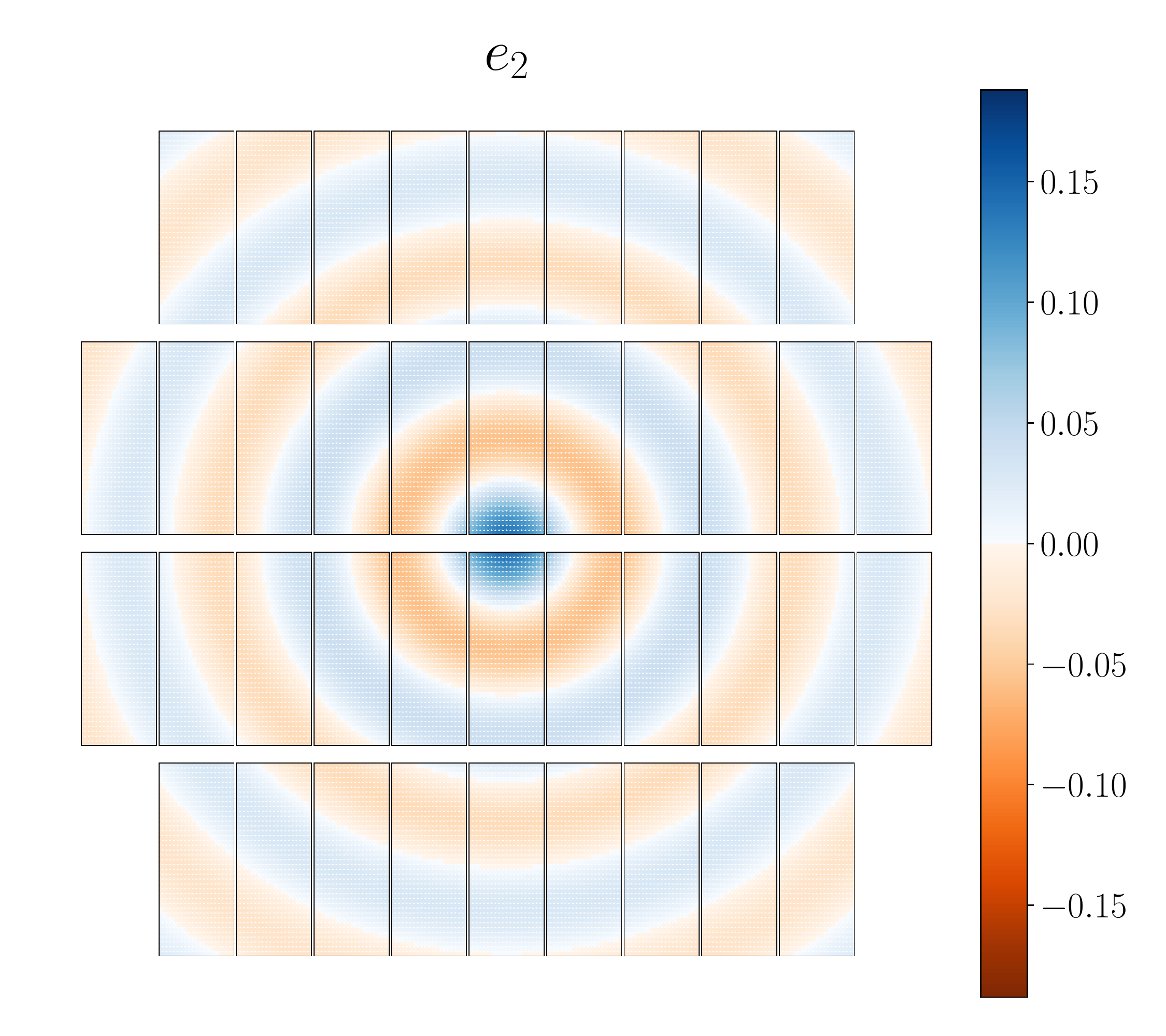}
    \end{subfigure}
    \begin{subfigure}{0.33\linewidth}
        \centering
        \includegraphics[width=1.\textwidth]{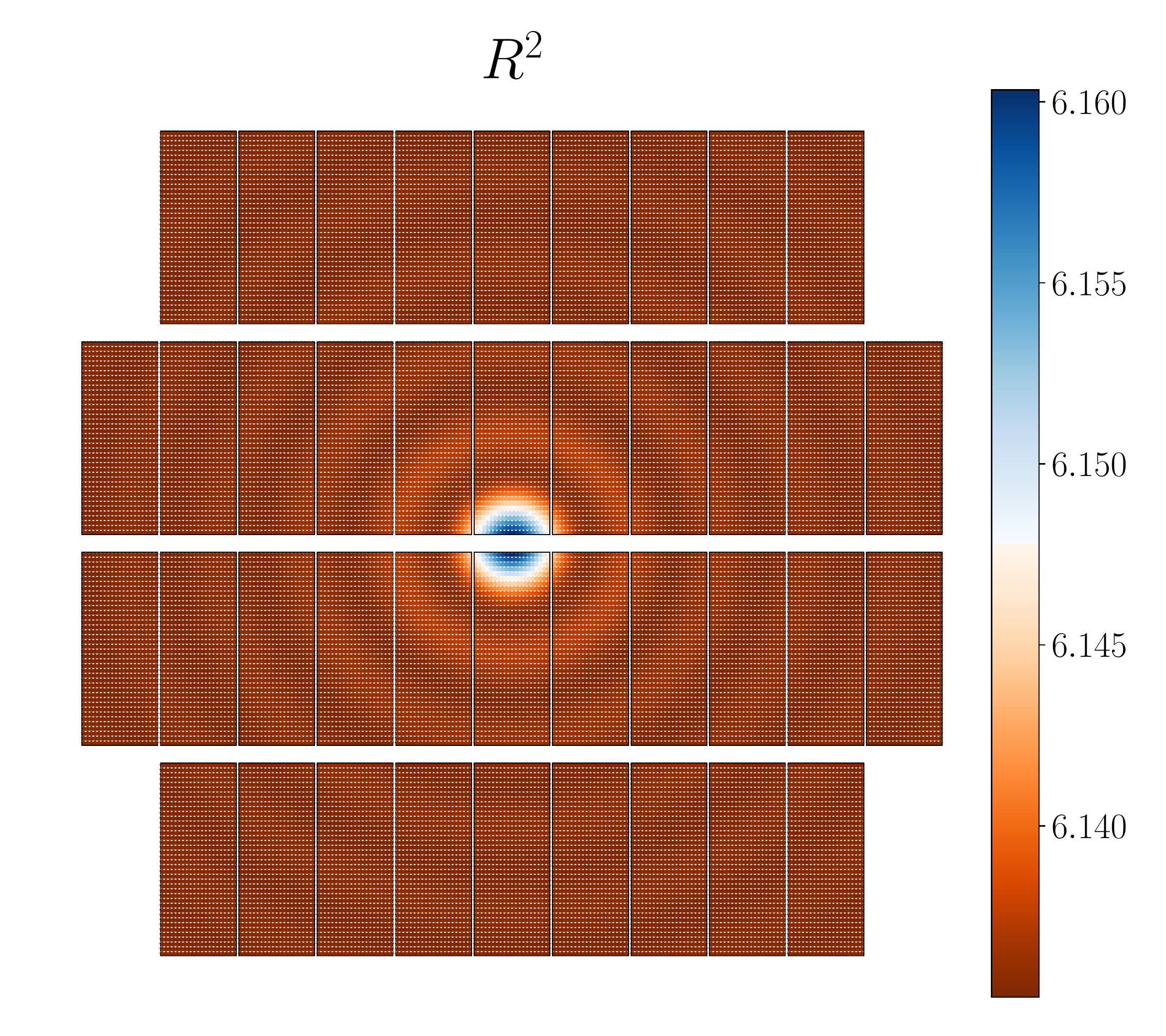}
    \end{subfigure}
     \caption{Shape measurement results of the simulated test star catalogue following the analytical ellipticities.}
    \label{fig:analytic_moments}
\end{figure*}

For the testing, we want to observe how well the different models capture the ellipticity maps when trained on real star positions.
Therefore, the positions in each CCD are taken from a regular grid of $20 \times 40$ and considering that the total amount of CCDs is $40$, we finally have a total of $32000$ stars to test our model. 
These stars are simulated following the same ellipticity maps (see  \autoref{fig:analytic_moments}), without any sub-pixel shift and without any noise. The goal is now to use the training data (i.e. simulated observed stars) to learn the model, and then to predict the PSFs at positions of test stars. As we have the ground truth at these positions, without noise and sub-pixel shift, it is easy to get a 
robust evaluation of model predictions.   

\subsection{Quality criteria}\label{sec:quality_criteria}

In order to correctly assess the performance of our PSF modelling algorithm, 
we consider several criteria:
\begin{itemize}

    \item Pixel Root Mean Square (RMS) error : calculated between the pixel images of the recovered PSFs and the noiseless test stars. The expression of the pixel RMS error is the following:
\begin{equation}
  Y_{pix}^{RMS} = \sqrt{ \langle  (Y^{*} - \hat{Y} )^{2} \rangle }= \left( \frac{1}{N n_{y}^{2}} \sum_{i=1}^{N} \sum_{j=1}^{n_{y}^{2}} (Y_{i,j} - \hat{Y}_{i,j} )^{2}  \right)^{\frac{1}{2}},
\end{equation}
where $Y_{i,j}$ is the pixel $j$ of test star $i$ that has a total of $n_{y}^{2}$ pixels, $N$ is the total number of test stars, $\hat{Y}_{i,j}$ is the estimation of the test star's pixel and $\langle \cdot \rangle$ denotes the mean over all the elements in the array.

    \item Shape (ellipticity) error: 
    We estimate the ellipticities of reconstructed stars using 
    the adaptive moments' ellipticity estimator 
    from \texttt{Galsim}'s HSM module \citep{hirata2003, mandelbaum2005}. The shape and size definitions can be found in \autoref{appdx:shape_defs}. For each of the ellipticity components, the RMS error is calculated as:
\begin{equation}
  e^{RMS} = \sqrt{ \langle (e^{*} - \hat{e})^{2} \rangle } = \left( \frac{1}{N} \sum_{i=1}^{N} (e_{i} - \hat{e}_{i} )^{2}  \right)^{\frac{1}{2}}.
\end{equation}   
    
    \item Size error: We use the measurements from HSM and the definition in \autoref{appdx:shape_defs} to compute the following RMS error:
\begin{equation}
  R^{2,RMS} = \sqrt{ \langle (R^{2,*} - \hat{R}^{2})^{2} \rangle } = \left( \frac{1}{N} \sum_{i=1}^{N} (R^{2}_{i} - \hat{R}^{2}_{i} )^{2}  \right)^{\frac{1}{2}}.
\end{equation} 

    \item Moment residual maps: To visualise the shape and size errors we plot these quantities as a function of their position on the focal plane.

\end{itemize}

When comparing two methods we define the relative gain concerning metric $m$ of method $1$ with respect to the method $2$ as:
\begin{equation}
    G_{1/2}(m) = \frac{m_2 - m_1}{m_2} \times 100\% .
\end{equation}

% --------- %
\subsection{Model parameters}\label{sec:sims_psf_modelling}

Based on experiments with simulated and real data, we have chosen the following parameters
\begin{itemize}
    \item{\texttt{PSFEx}:} we use the following configuration:
\begin{verbatim}
PSF_SAMPLING    1.0               
PSF_SIZE        51,51           
PSFVAR_KEYS     XWIN_IMAGE,YWIN_IMAGE 
PSFVAR_GROUPS   1,1            
PSFVAR_DEGREES  2              
\end{verbatim}
    \texttt{PSFVAR\_DEGREES} refers to the maximum polynomial degree, and, \texttt{XWIN\_IMAGE} and \texttt{YWIN\_IMAGE}, to the windowed centroid positions in pixel coordinates.
    The \texttt{PSFEx} software\footnote{\url{https://www.astromatic.net/software/psfex}} does not include publicly an interpolation method, so we use an available \texttt{PSFEx} interpolation module\footnote{\url{https://github.com/esheldon/psfex}}.
    
    \item{RCA:} we set $r$ equal to $8$ local components,  the denoising parameters $K_{\sigma}^{RCA}$ to $1$, and the other parameters to their default value from its official repository\footnote{\url{https://github.com/CosmoStat/rca}}.
    
    \item{MCCD:} we use the same parameters as RCA for the local component, and a maximum polynomial degree of $8$ for the global components. The denoising parameters $K_{\sigma}^{\rm Loc}$ and $K_{\sigma}^{\rm Glob}$ are set to $1$ for the local and the global contributions.
\end{itemize}

The MCCD parameters that most affect its behaviour were mentioned above. Their choice greatly relies on the training data set used. Depending on the number of stars available and the complexity of the instrument's PSF one may tend to prefer a more complex model by augmenting the number of local components, $r$, and the maximum polynomial degree. However, if the stars are not enough to constraint the model one may end with a model that is overfitting the training stars.
A proper selection of the denoising parameters can control the bias-variance tradeoff in the estimation. A high value of the denoising parameter, i.e. $3$, will output an extremely denoised model. It will contain a high estimation bias that can be related with a model that cannot capture some spatial variations and fine details of the PSF. On the contrary, if the denoising parameter is close to zero, the only denoising performed by the MCCD is due to the low-rank constraint and therefore the estimations can be rather noisy.

 \begin{figure*}
    \centering
    \includegraphics[width=.49\linewidth]{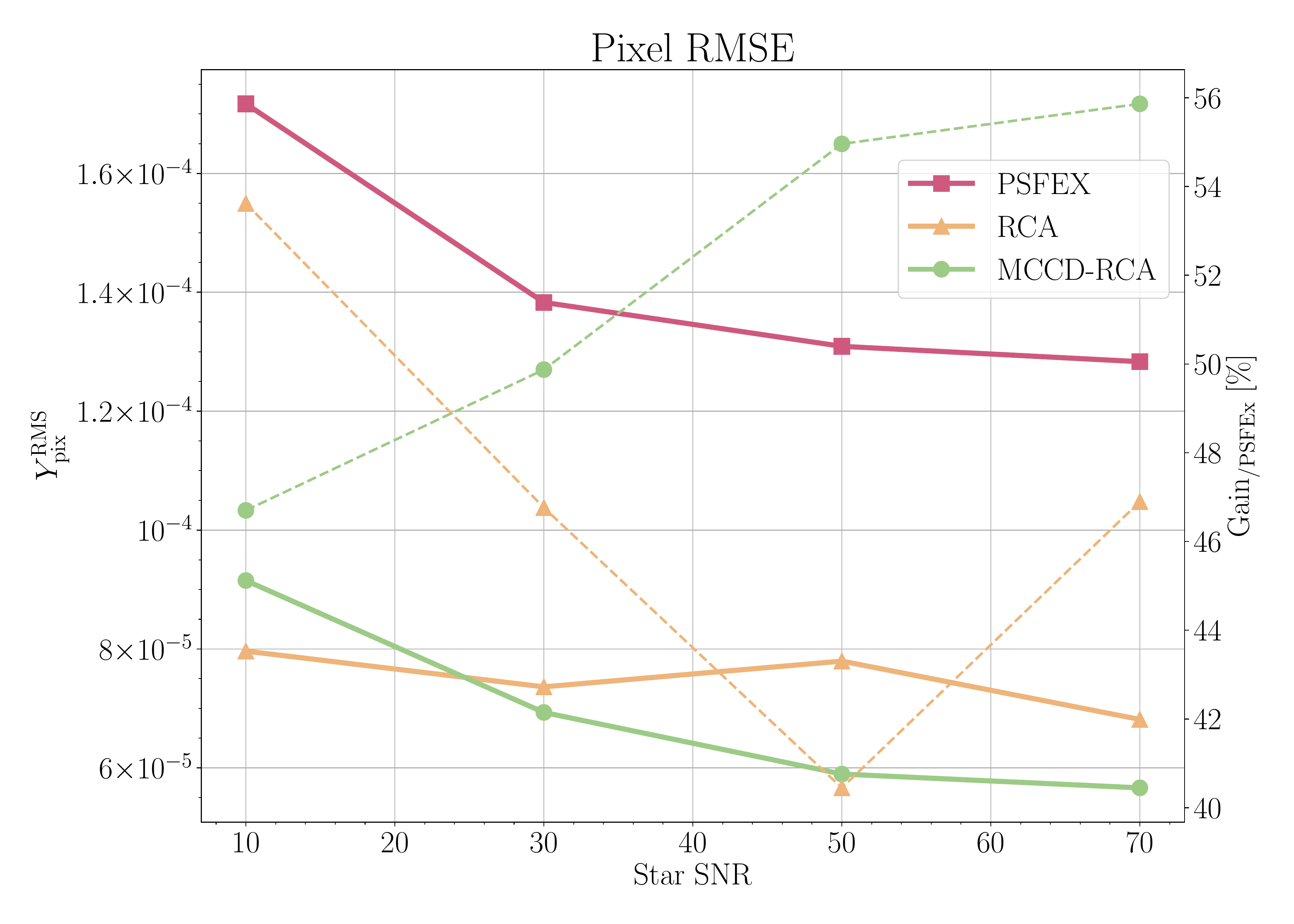}
    \includegraphics[width=.49\linewidth]{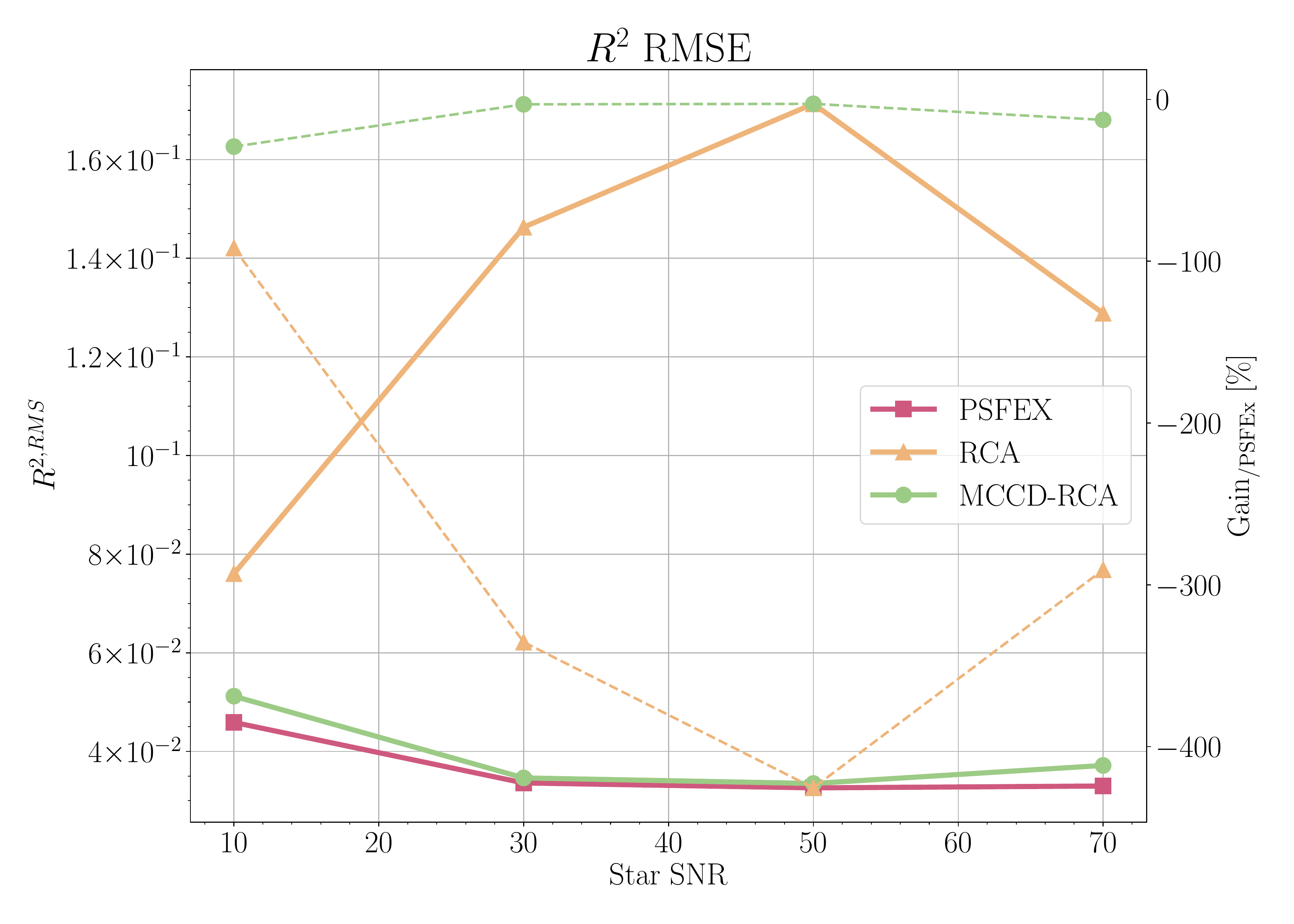}
    \includegraphics[width=.49\linewidth]{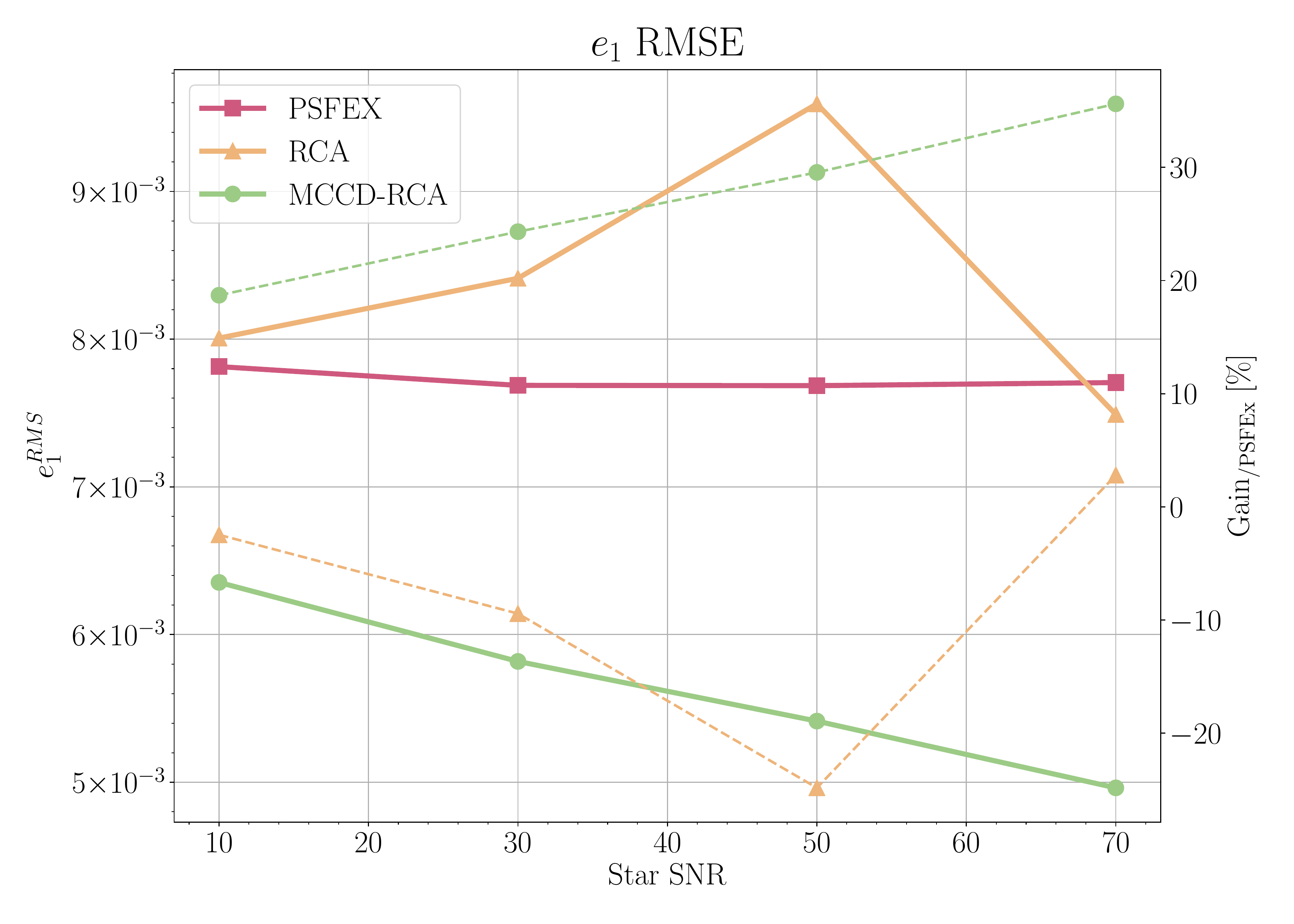}
    \includegraphics[width=.49\linewidth]{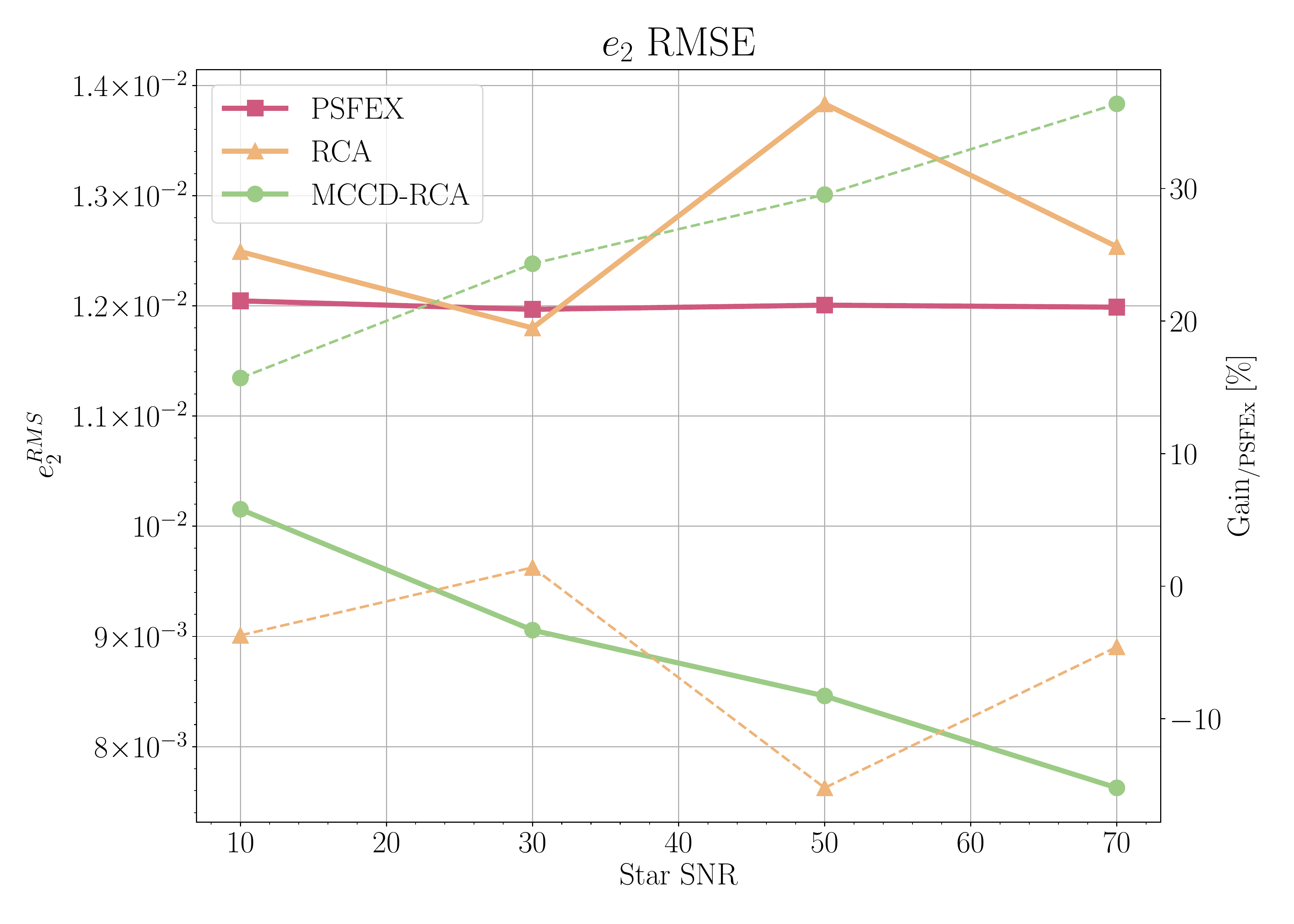}
    \caption{RMS errors on pixels, shape and size metrics as a function of stars SNR for the three main methods. The RMS errors are plotted with solid lines and the gain of the methods with respect to \texttt{PSFEx} are plotted with dashed lines.}
    \label{fig:sim_benchmark_plots}
\end{figure*}

 \begin{figure*}
    \centering
    \includegraphics[width=.49\linewidth]{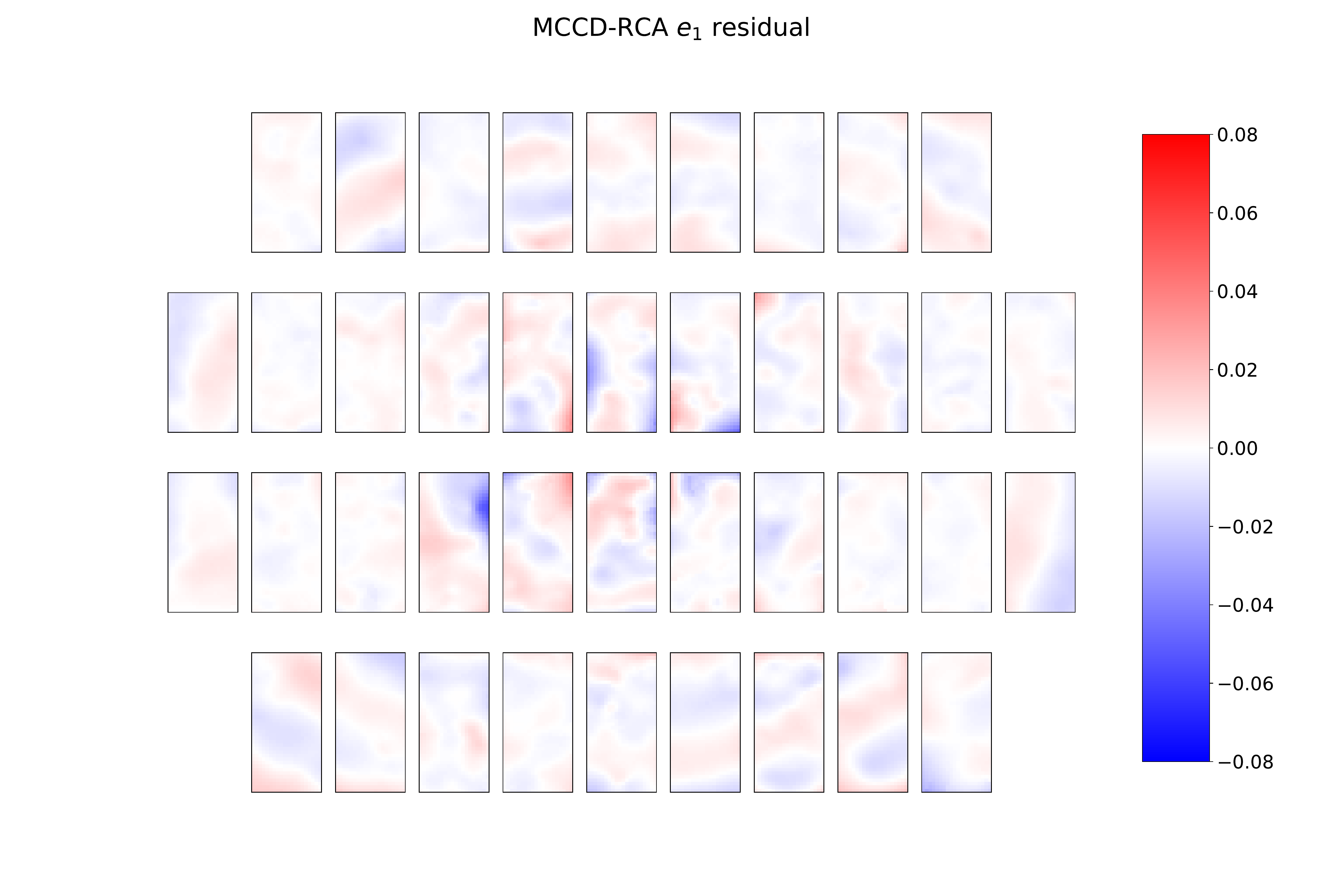}
    \includegraphics[width=.49\linewidth]{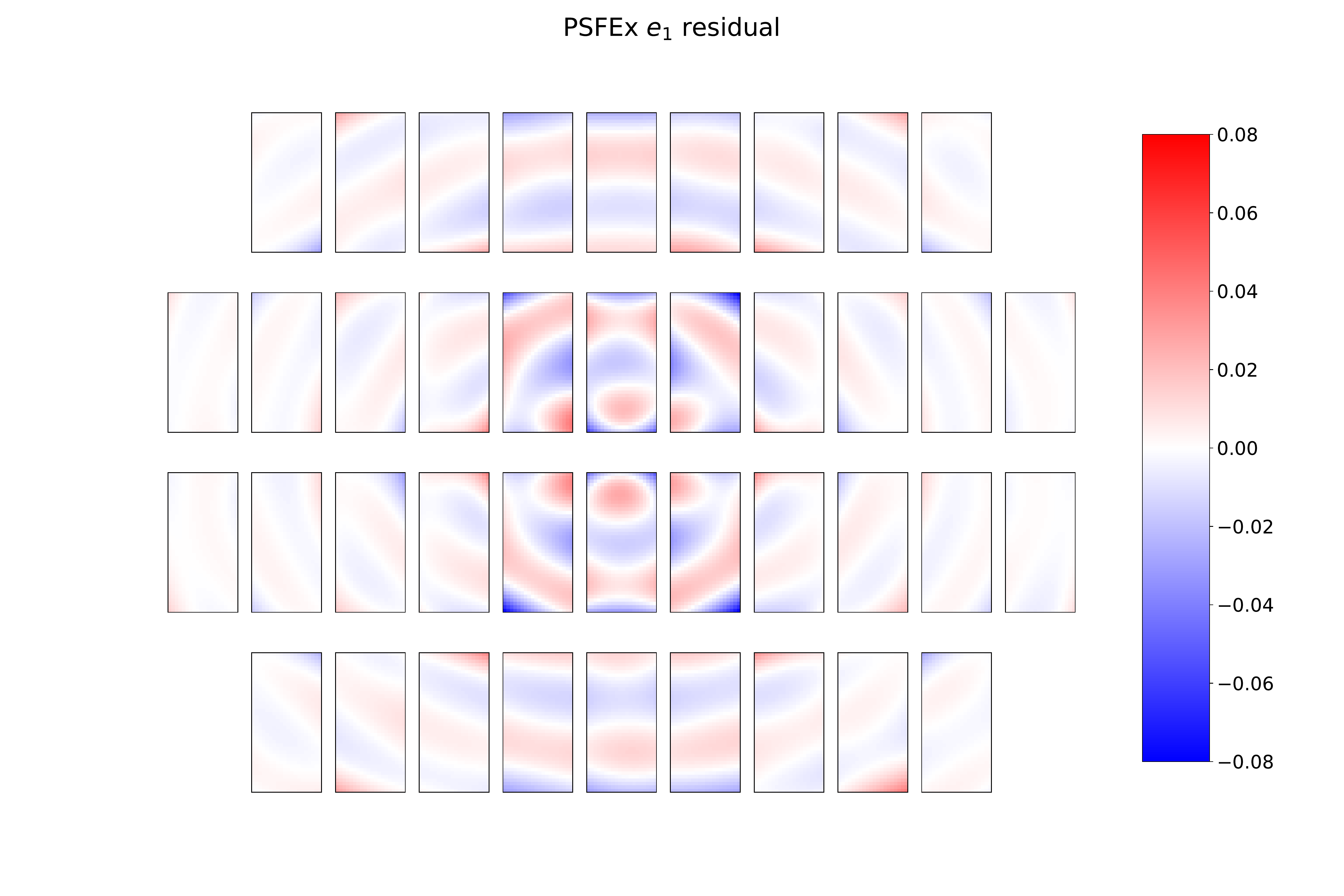}
    \includegraphics[width=.49\linewidth]{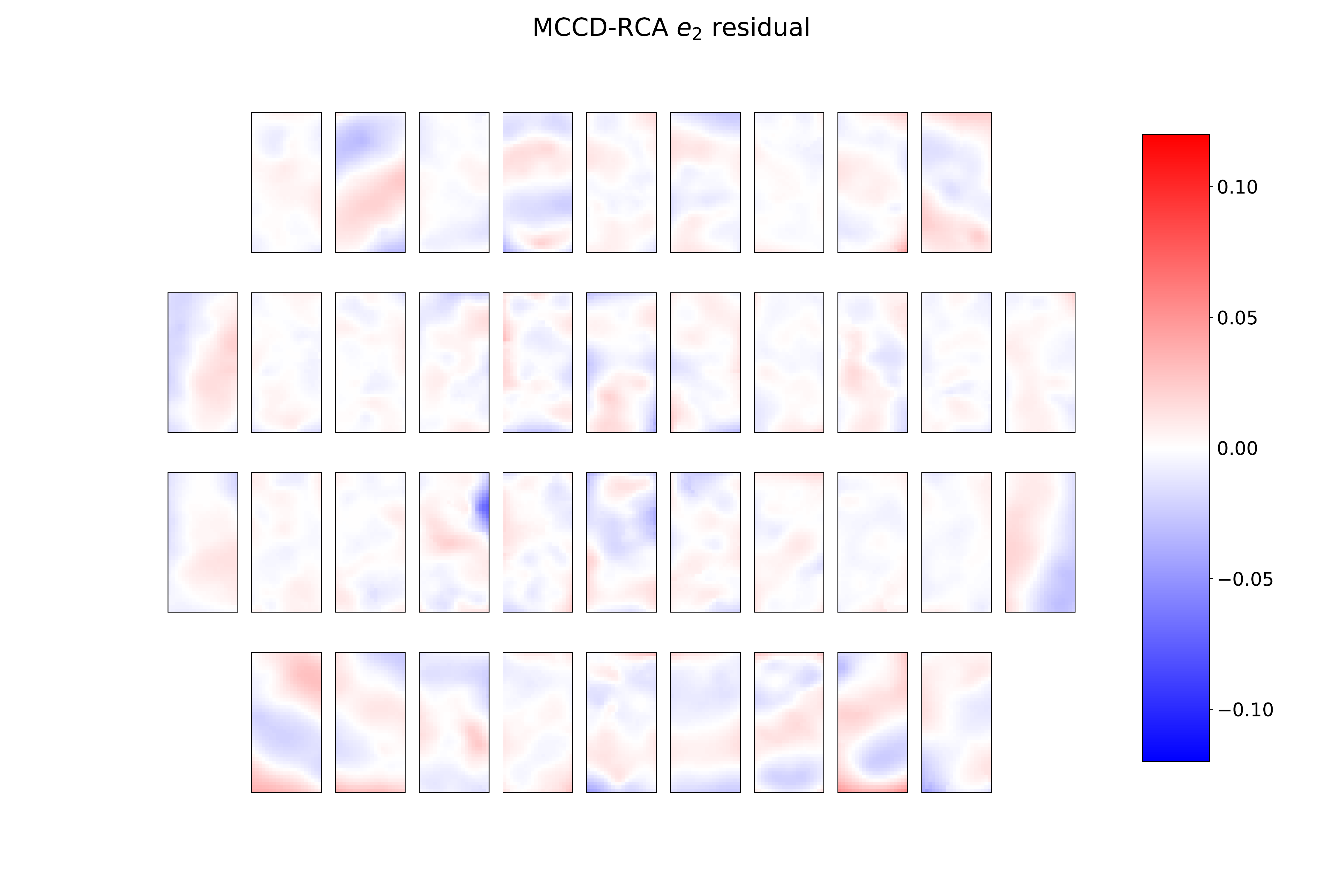}
    \includegraphics[width=.49\linewidth]{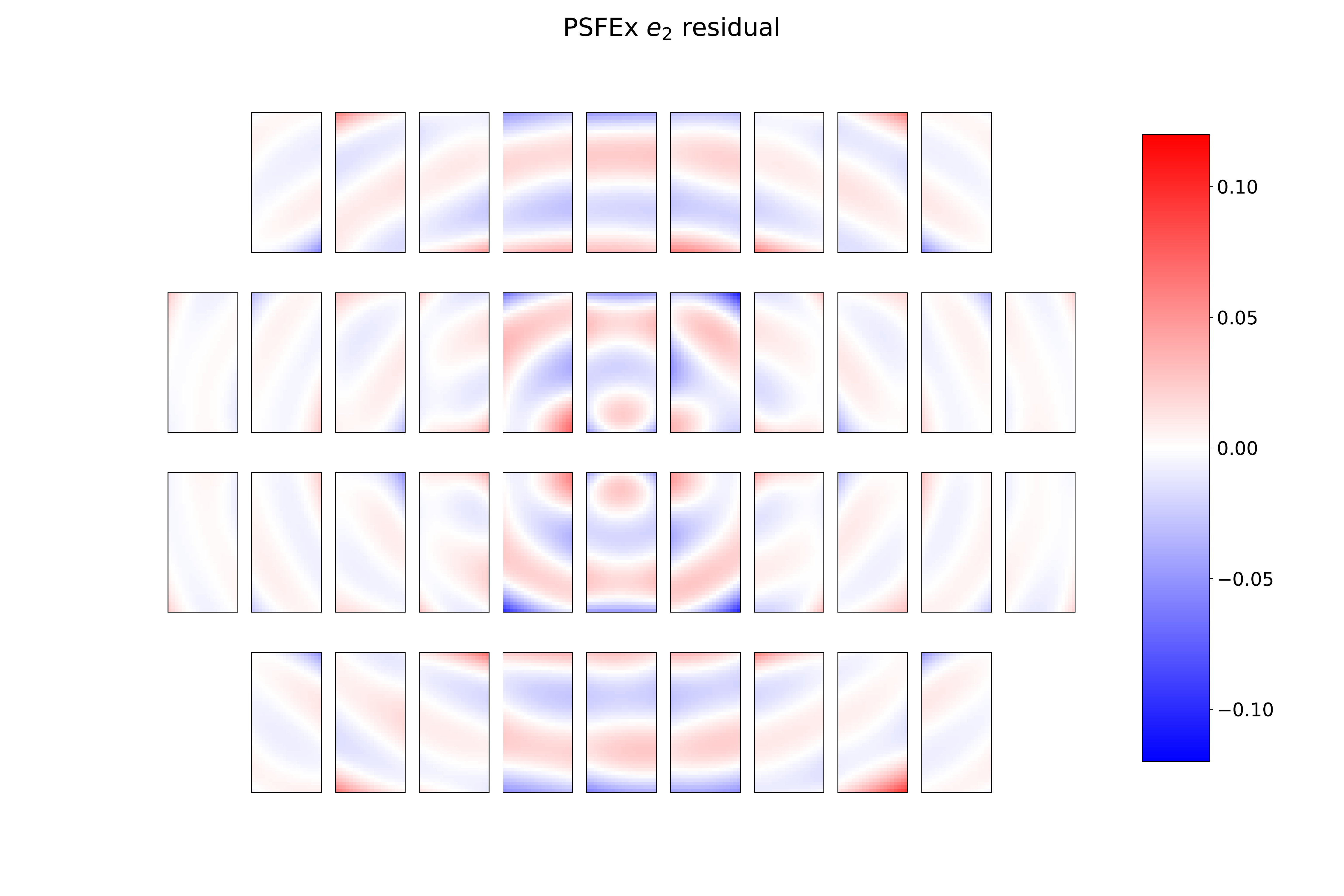}
    \includegraphics[width=.49\linewidth]{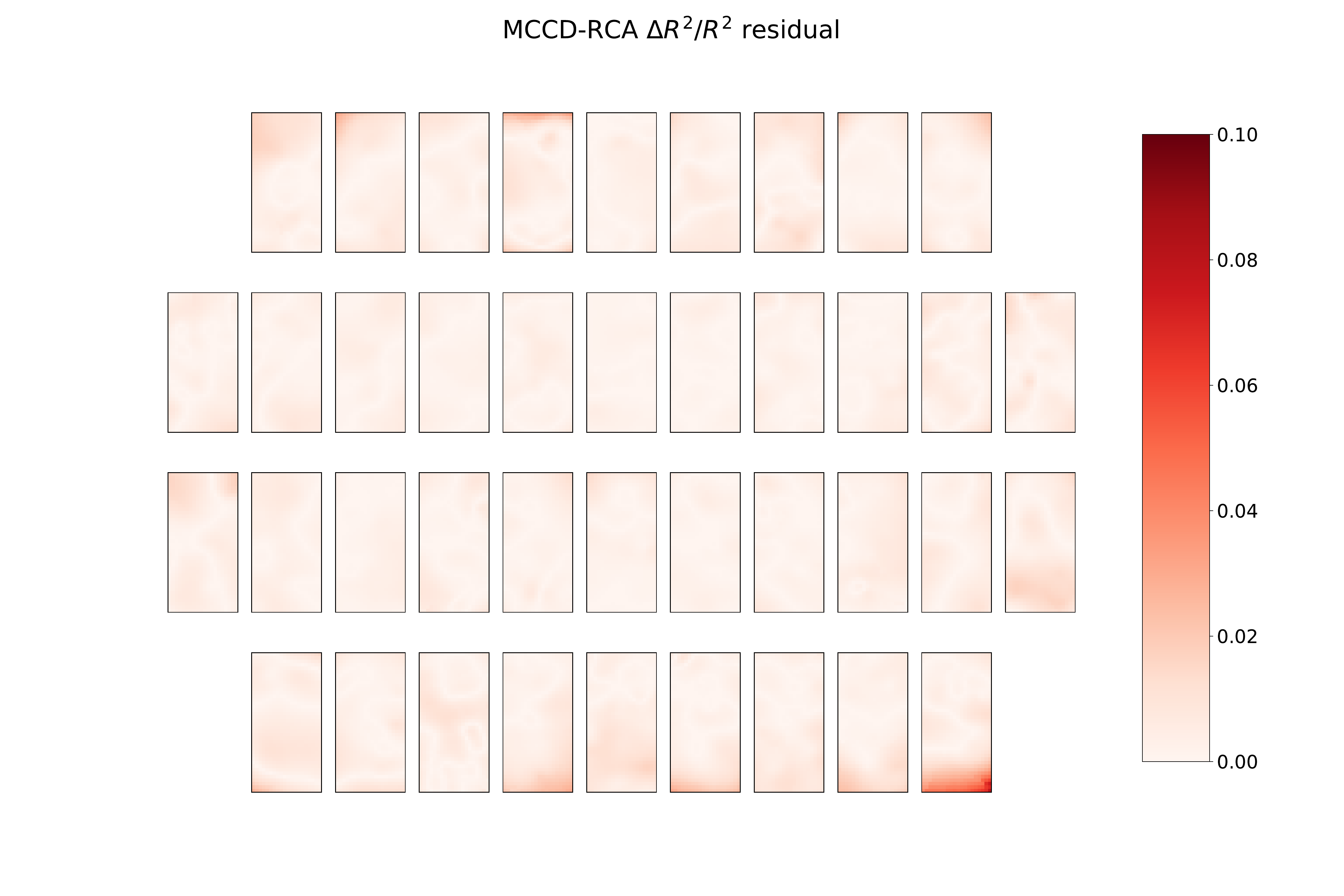}
    \includegraphics[width=.49\linewidth]{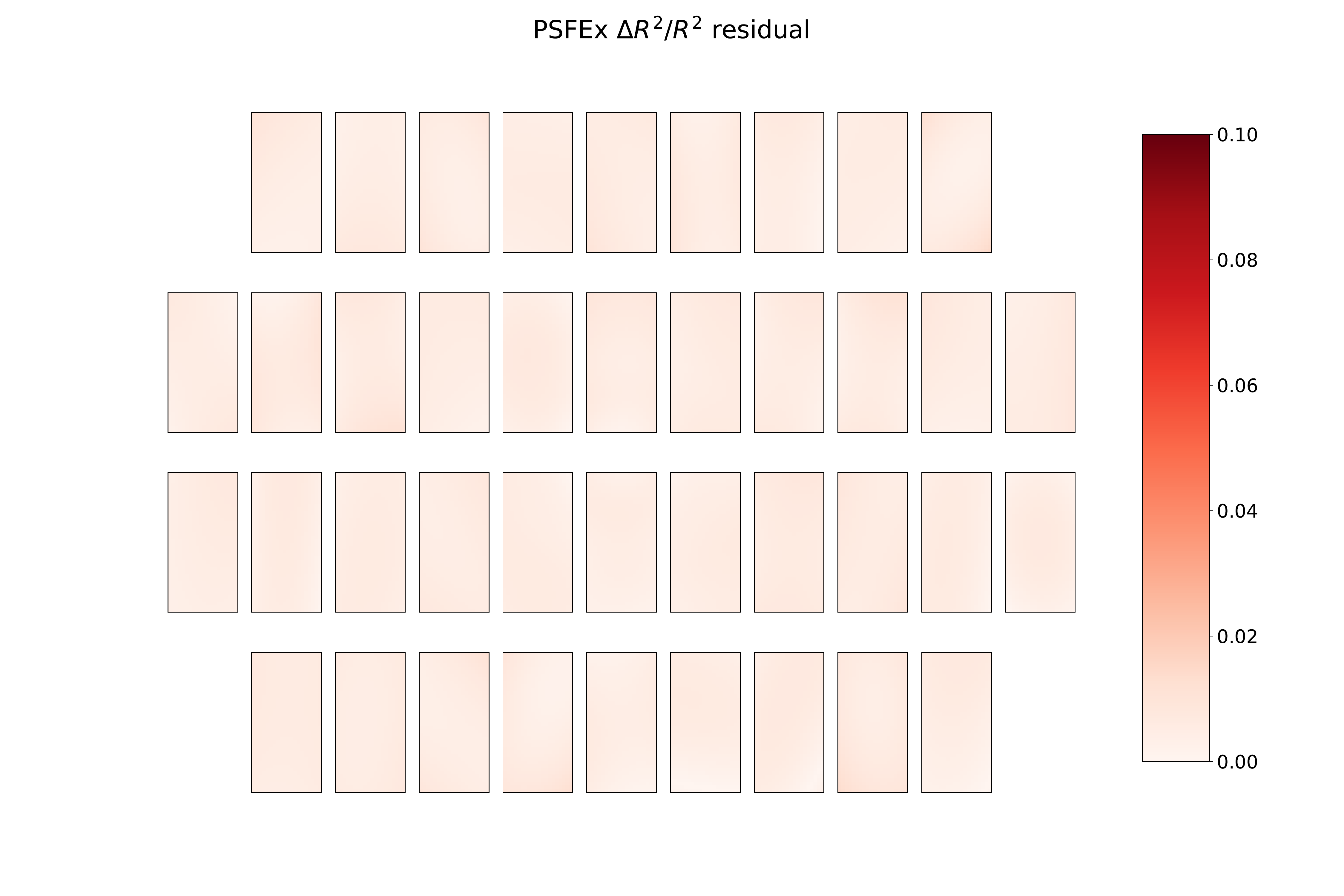}
    \caption{Moment residual maps comparing the MCCD-RCA algorithm on the left and the \texttt{PSFEx} algorithm on the right. They are obtained by subtracting the model's and the test star's measured shape and size metrics and plotting them on their corresponding position over the focal plane. The SNR value of the star dataset is $50$.}
    \label{fig:exp_residual_meanshapes}
\end{figure*}

% --------- %
\subsection{Results}

\subsubsection{Comparison between PSFEX, RCA and MCCD-RCA}

The first results can be seen in \autoref{fig:sim_benchmark_plots} 
and \autoref{fig:exp_residual_meanshapes} where we compare the \texttt{PSFEx}, RCA and MCCD-RCA algorithms. 
We observe that MCCD-RCA outperforms the other methods, with an average pixel RMS improvement over \texttt{PSFEx} of $51\%$  and ellipticity RMS improvement ranging from $15\%$ for stars with SNR 10 to $36\%$ for a SNR of 70. 
RCA is almost as good as MCCD-RCA for the pixel error, but does not provide good results 
for the other metrics. This behaviour can be explained by the fact that the model strongly deteriorates for some CCDs, giving extreme ellipticities and sizes values. These deteriorations of the model are not strong enough to produce a large pixel error but causes much more significant errors on the moments. 
We include in \autoref{appdx:plots} RCA's $R^2$ residual map that shows the catastrophic failure in the modelling of some CCDs.

One can see on the right column of the residual maps in \autoref{fig:exp_residual_meanshapes} that \texttt{PSFEx}'s ellipticity residuals follow the global pattern from the dataset. This means that is not captured in the model, showing some of the difficulties found when modelling a global ellipticity pattern using independent models for each CCD. The MCCD-RCA algorithm, which builds up a model for the whole focal plane, does a better job in capturing the global ellipticity pattern.  MCCD-RCA's residuals are smaller and less correlated with the dataset's pattern. 
Concerning the third row of \autoref{fig:exp_residual_meanshapes}, where the size of the simulated PSFs is practically constant, we observe that MCCD-RCA has slightly larger errors when the training star density is low, as in the bottom-right corner (see \autoref{fig:sim_star_positions}).

\subsubsection{Comparison between MCCD-POL, MCCD-RCA and MCCD-HYB}
 \begin{figure*}
    \centering
    \includegraphics[width=.49\linewidth]{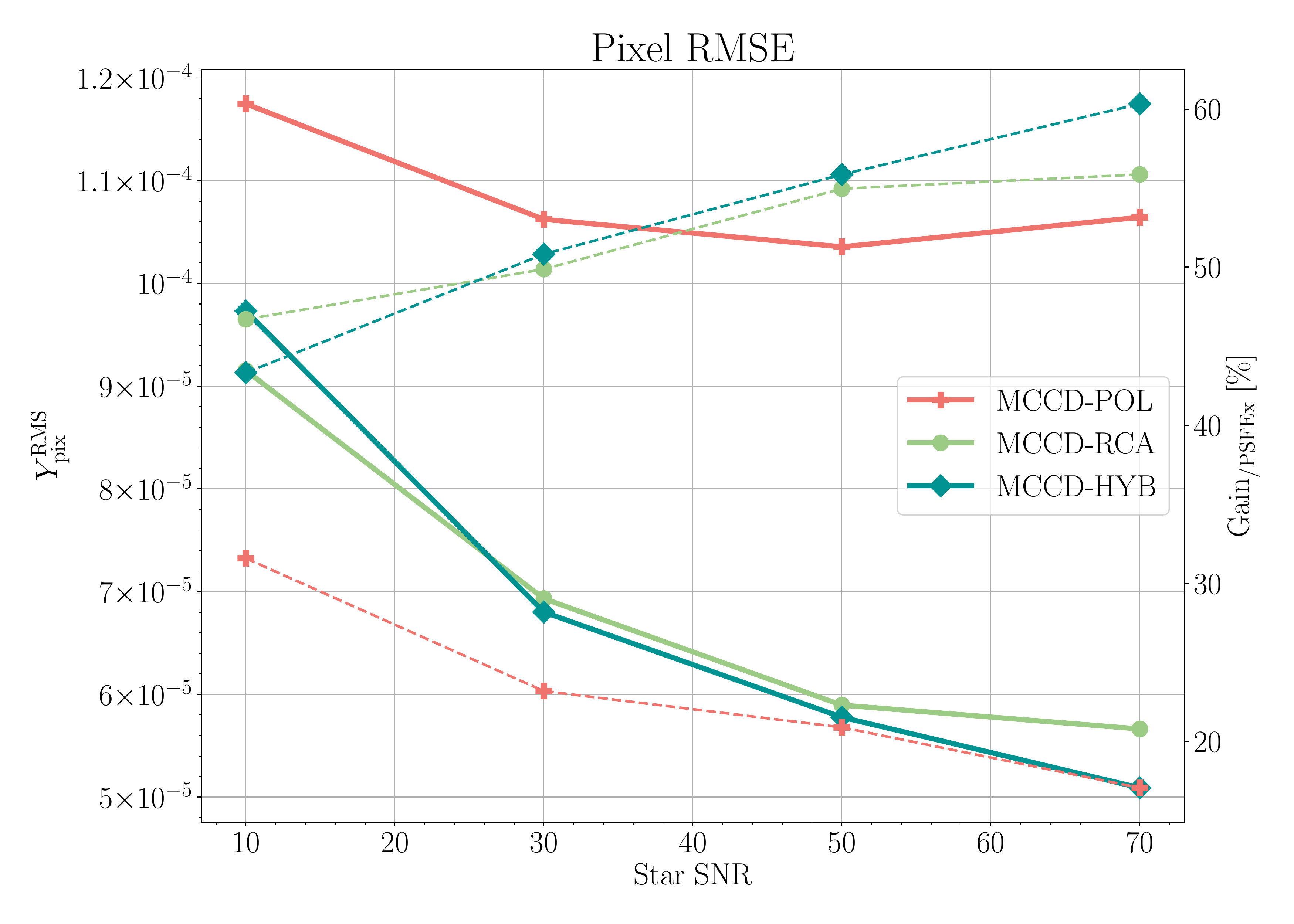}
    \includegraphics[width=.49\linewidth]{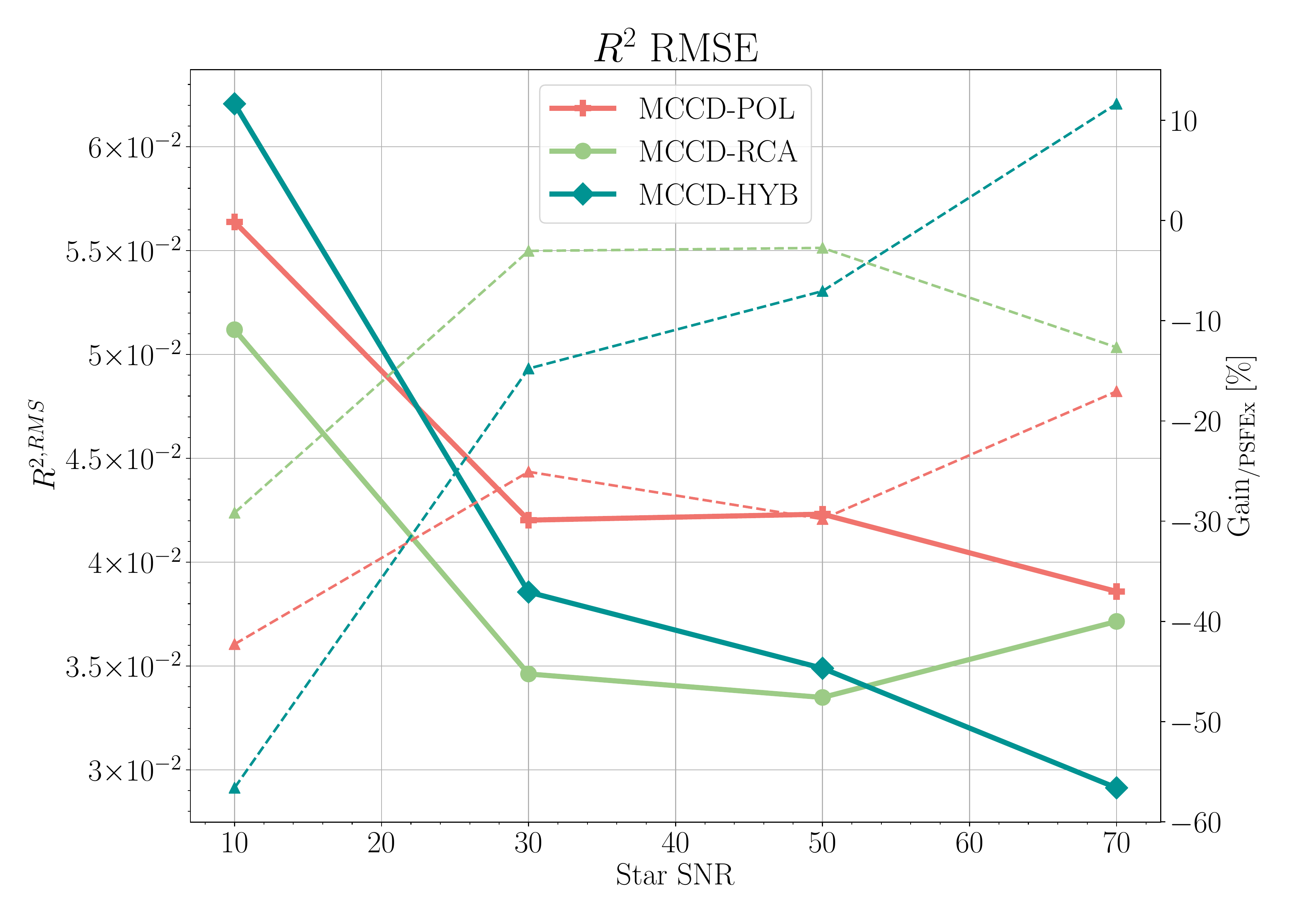}
    \includegraphics[width=.49\linewidth]{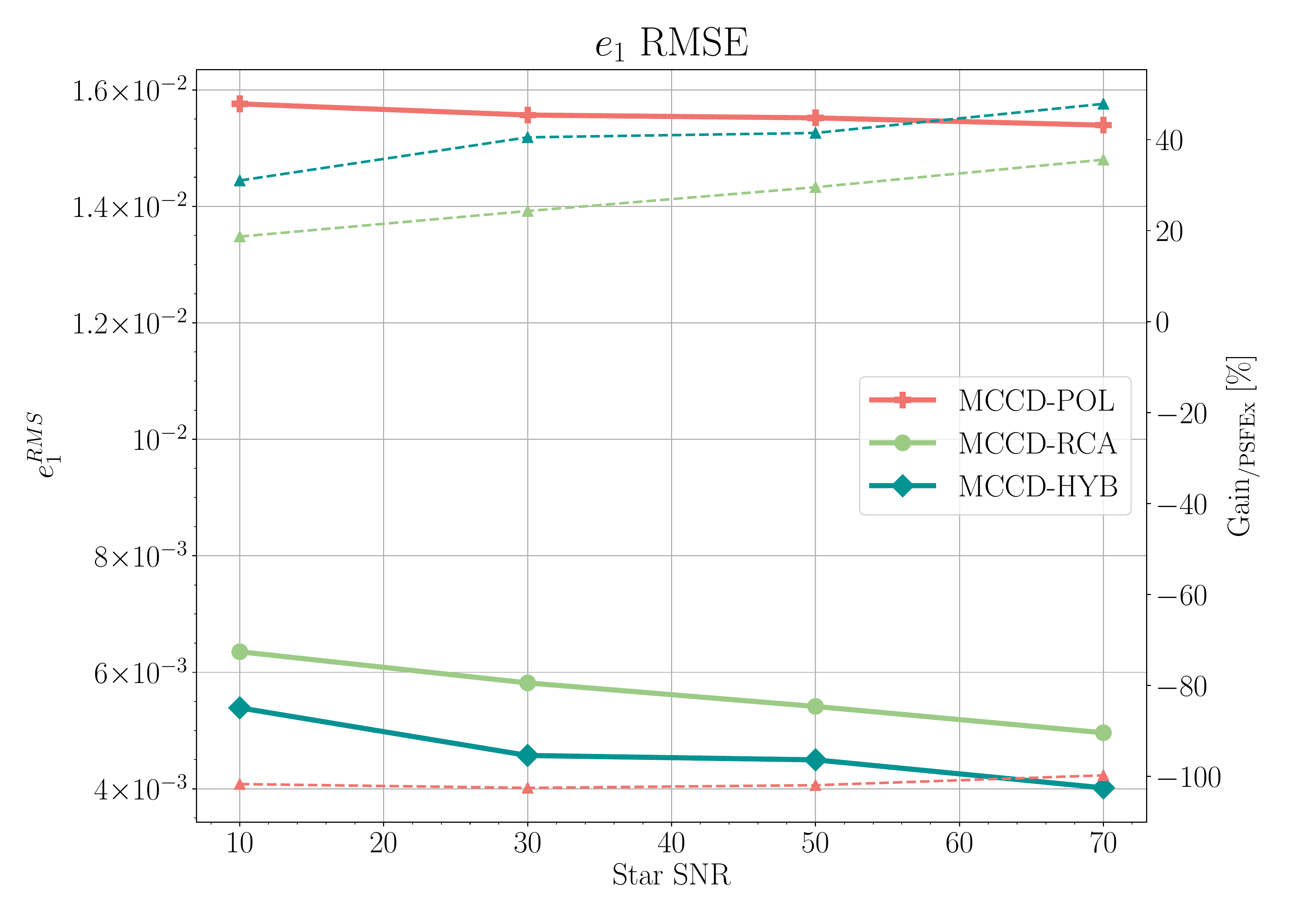}
    \includegraphics[width=.49\linewidth]{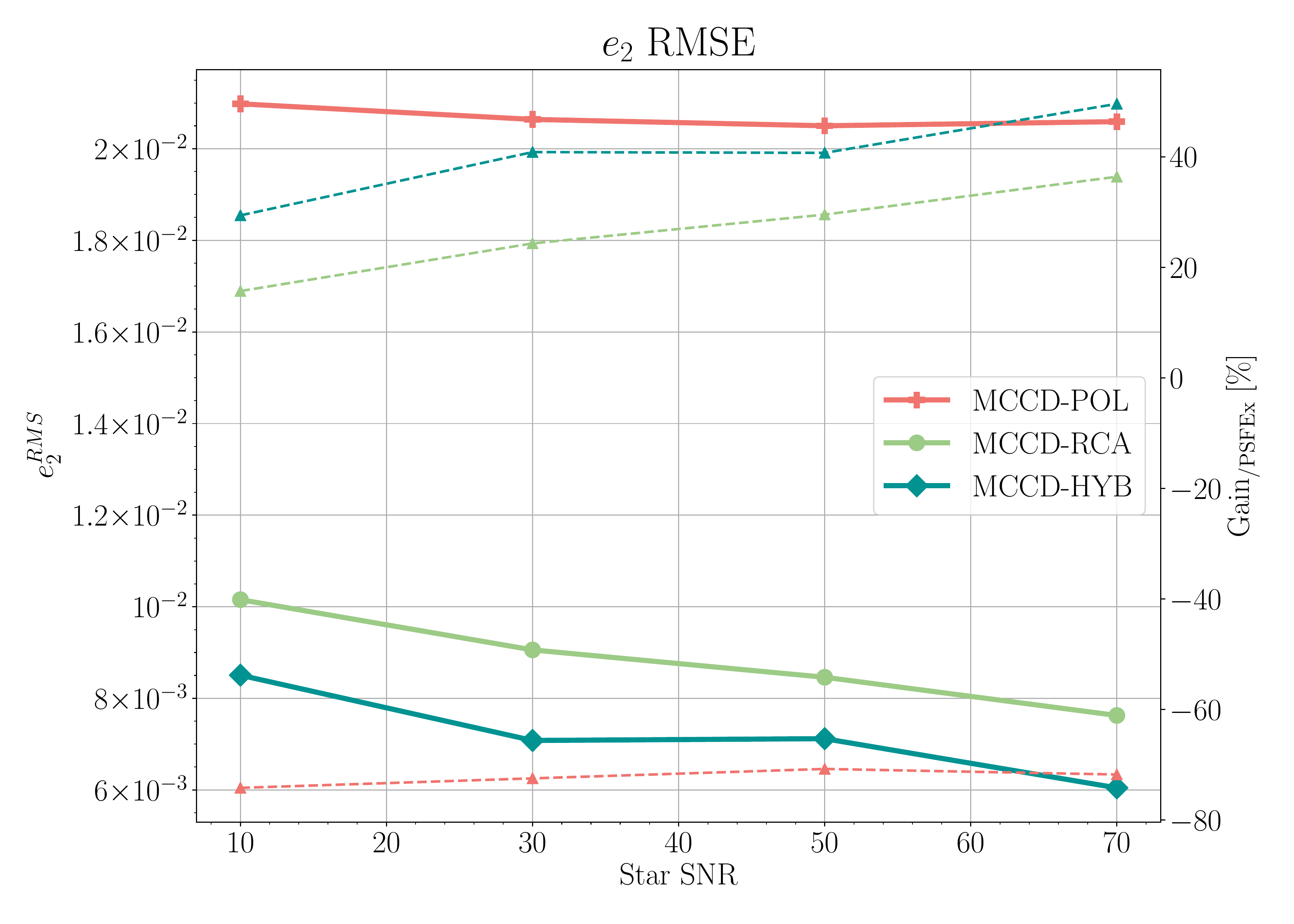}
    \caption{Comparison of the performance of MCCD-POL, MCCD-RCA and MCCD-HYB methods in terms of the RMS errors on the pixels, the shape and the size metric as a function of star SNR. The RMS errors are plotted on a solid line and the gain of the method with respect to \texttt{PSFEx} is plotted on dashed lines.}
    \label{fig:sim_comparison_extensions}
\end{figure*}

The comparison between MCCD-POL, MCCD-RCA and MCCD-HYB methods is shown in \autoref{fig:sim_comparison_extensions}. 
First, we notice that MCCD-POL presents poor performance in most of the metrics. This indicates that the local polynomial model is not able to capture the PSF variations that are left from the difference of the global model and the observed stars. Hence, even if MCCD-POL has a lower pixel error than \texttt{PSFEx} (see \autoref{fig:sim_benchmark_plots}), it has larger ellipticity errors. Capturing these PSF variations properly is essential to obtain good ellipticity performances.
MCCD-RCA and MCCD-HYB have similar behaviours, but MCCD-HYB uses a mixed approach of a polynomial and graph-based local model outperforms the original MCCD-RCA method in terms of ellipticity components. The average gain in both components of MCCD-HYB with respect to MCCD-RCA is around $18\%$, proving the utility of using the hybrid approach. 
This suggests that there are some features related to the PSF shape that can be captured by a simple polynomial model and not by the graph-based model alone. 

Examples of global and local eigenPSF from the MCCD-HYB model can be seen in \autoref{appdx:plots}.

% --------- %
\subsection{Comparison of computing resources}
The MCCD methods take $\sim2.9\times$ more CPU-time than \texttt{PSFEx} when compared on the same machine.
We evaluate it on the fitting and validation procedures, that is, the estimation of the PSF model and the recovery of PSF at test positions.
It is relevant to mention that the \texttt{PSFEx} package is coded in the C programming language, while the MCCD methods are completely coded in Python. 

%-------------------------------------------------------------------
% CFIS experiments
\section{UNIONS/CFIS experiments}\label{sec:cfis}

In this section we compare the MCCD-HYB method with \texttt{PSFEx} using real data from the Ultra-violet Near-Infrared Optical Northern Sky (UNIONS) survey, which is a collaboration between the Panoramic Survey Telescope and Rapid Response System (Pan-STARRS) and CFIS. We use the \textit{r}-band data from the latter.

% --------- %
\subsection{Dataset}

\begin{figure}
    \centering
    \includegraphics[width=.99\linewidth]{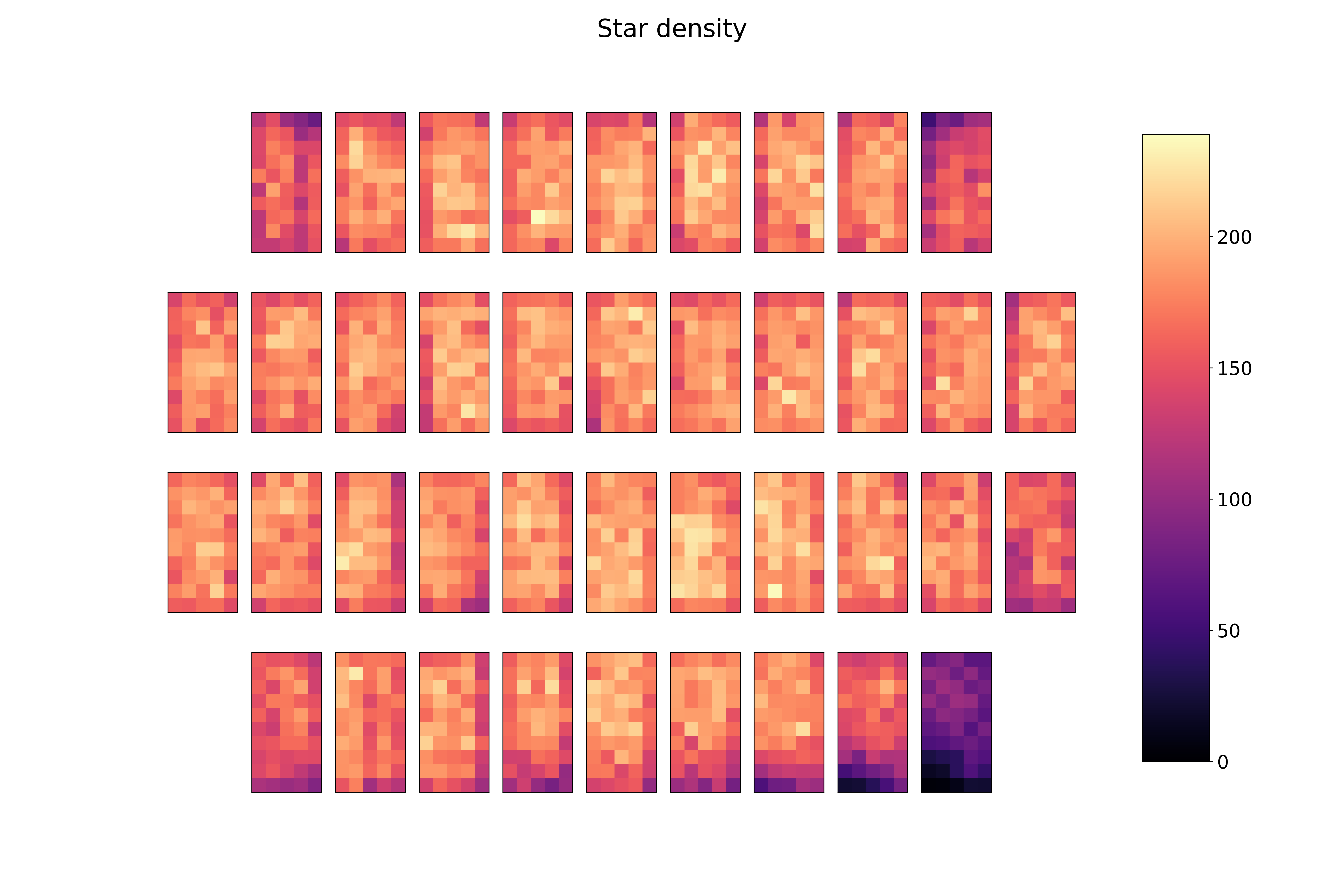}
    \caption{The star density of all the training dataset with respect to their position on the MegaCam's focal plane. We have on average $1560$ training stars per exposure.}    
    \label{fig:CFIS_star_density}
\end{figure}

We analyse a subset of around $50 \, \text{deg}^{2}$ from the whole CFIS survey area, that, in total, will span $5000 \, \text{deg}^{2}$. It corresponds to the subset named W3 described in \cite{erben2013}, and includes $217$ exposures.
Each CCD from each exposure has been processed independently with \texttt{SExtractor}. The stars were selected
in a size-magnitude diagram, in the magnitude range between $18$ and $22$, and a Full-Width Half Maximum (FWHM) range between $0.3$ and $1.5$ arcsec. In order to validate the PSF models, we randomly split the stars into a testing and a training dataset, trying to estimate the first set of stars while constructing our model only with the second. The training dataset is composed of $80\%$ of the detected stars and the test dataset of the remaining $20\%$.
We consider a fixed threshold on the number of training stars per CCD, meaning that if the number of training stars in a given CCD is less than $30$, we discard the CCD. The star density of the training dataset is presented in \autoref{fig:CFIS_star_density}.

\begin{figure}
    \centering
    \includegraphics[width=.99\linewidth]{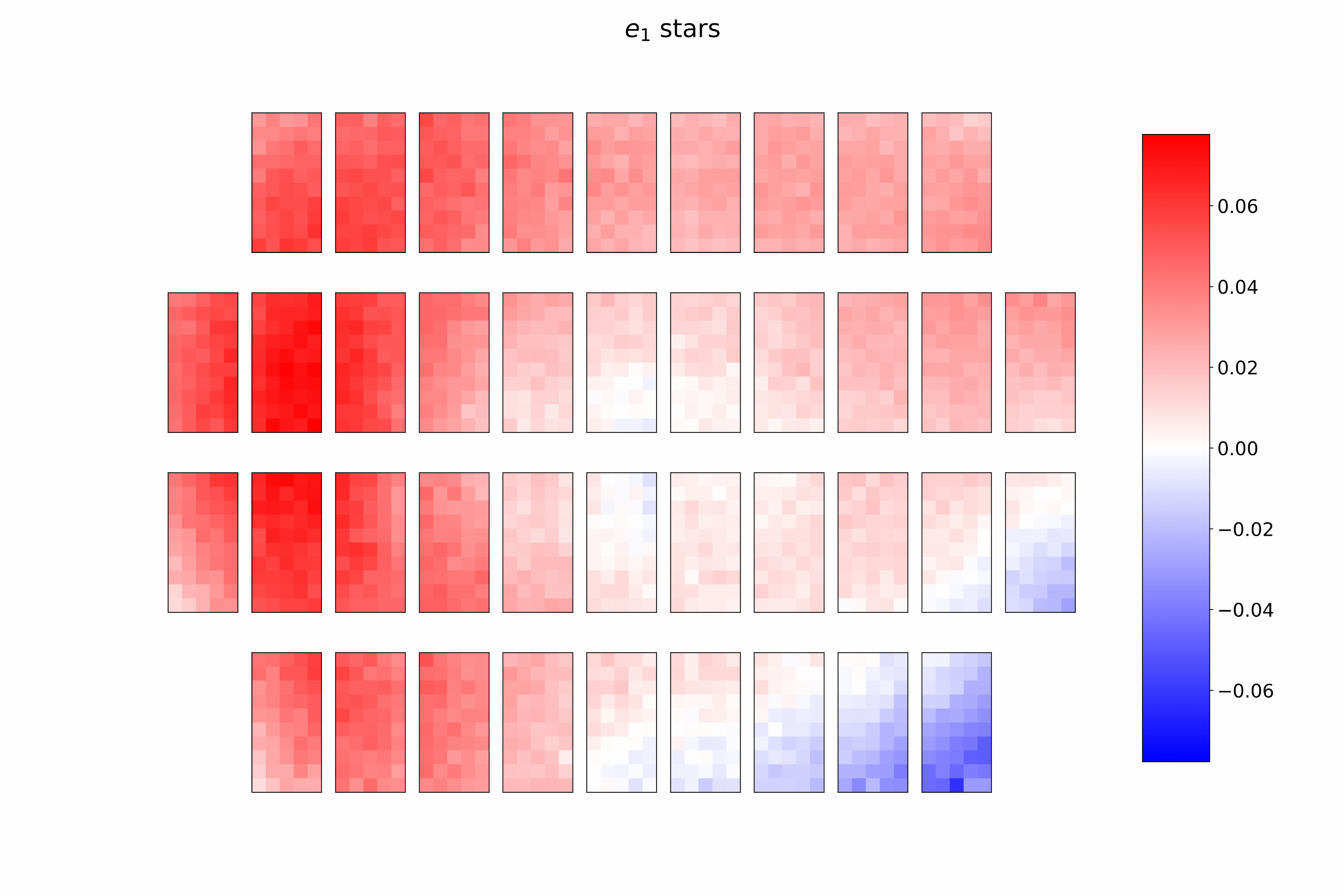}\\
    \includegraphics[width=.99\linewidth]{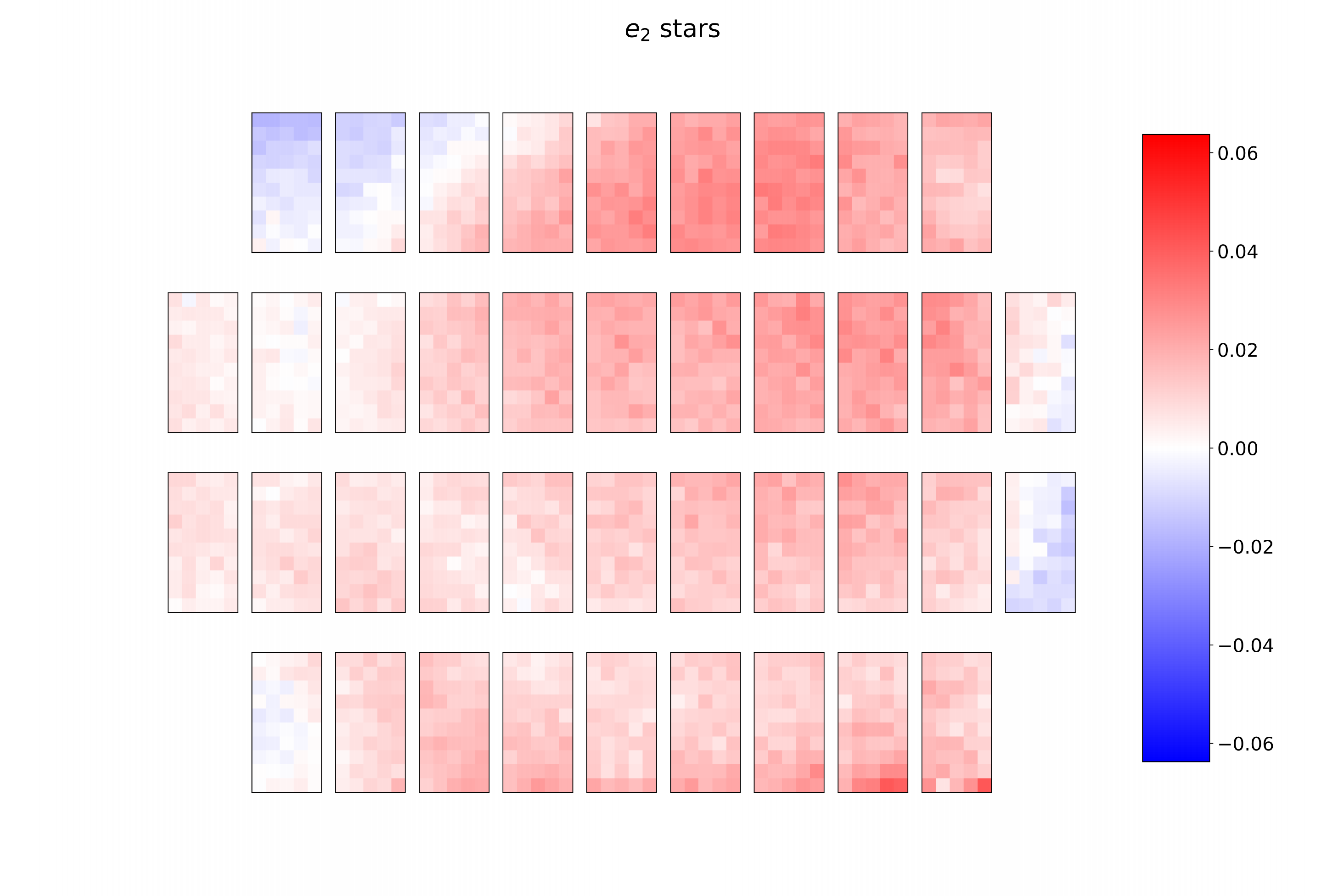}\\
    \includegraphics[width=.99\linewidth]{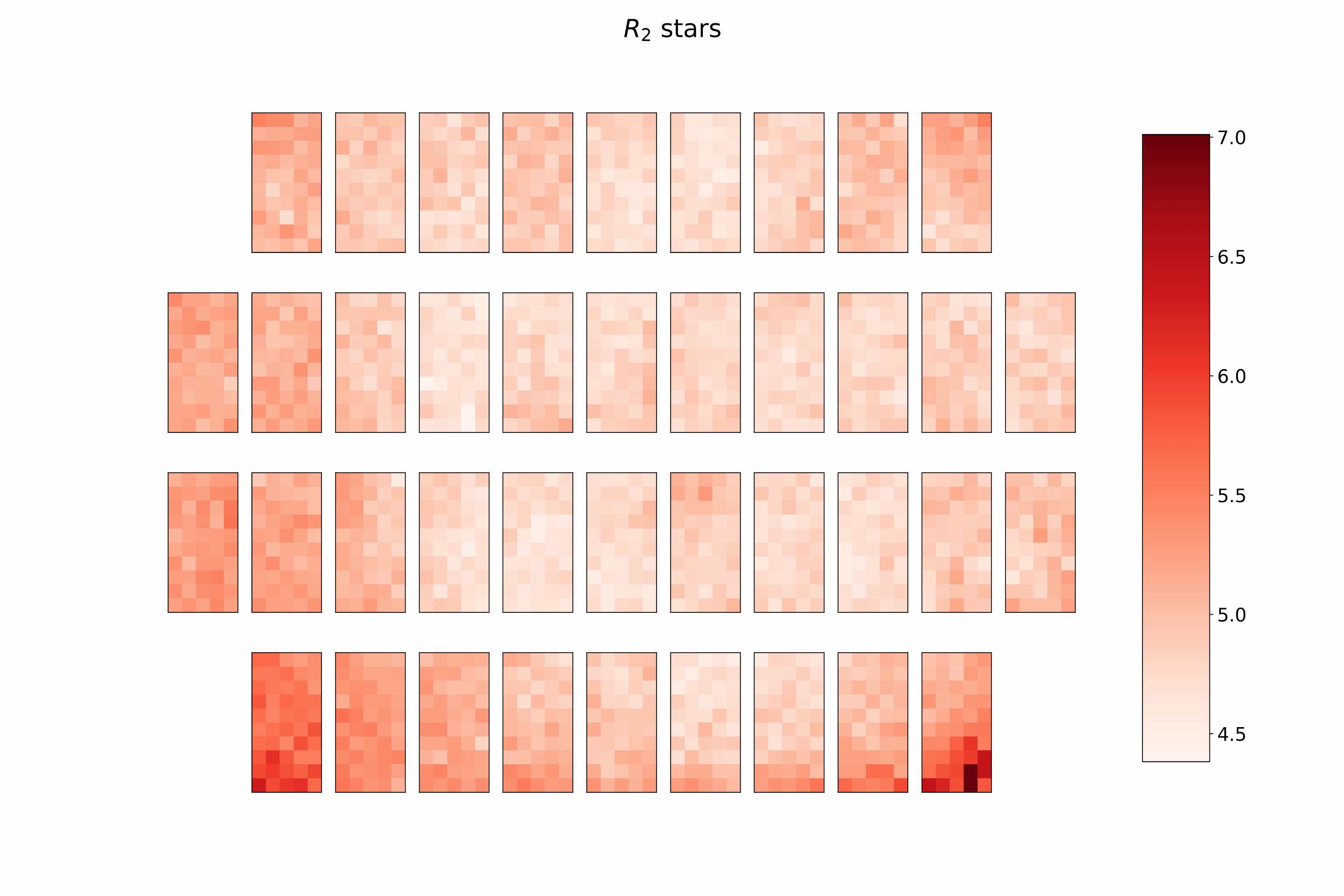}
    \caption{Ellipticities and size maps measured on the training stars of our CFIS dataset.}
    \label{fig:CFIS_meanshapes}
\end{figure}

The ellipticity and the size of the training stars can be seen in \autoref{fig:CFIS_meanshapes}. Each bin represents the mean shape measurement over all the stars with a centroid located within the bin.

\subsection{Model parameters}

The setup of \texttt{PSFEx} for this experiment is similar to the one used for the simulated images that can be found in \autoref{sec:sims_psf_modelling}. The MCCD-HYB method uses a maximum global polynomial degree of $3$, $16$ local components and the denoising parameters $K_{\sigma}$ set up to $0.1$.
In order to compare the star images with the different methods (\texttt{PSFEx} and MCCD-HYB), 
the models need to match the flux as well as the centre of the star. 
Hence, after estimating a PSF model at a given star location, the PSF is normalised and shifted to match the star. 
For this purpose, we use the same intra-pixel shift and flux estimation methods for both PSF models: 
i) we estimate the star and the PSF centroids,  
ii) we calculate the shift needed by the PSF to match the star and construct a shifting kernel, and 
iii) the PSFs are convolved by their corresponding shifting-kernel. 
To match the flux, we calculate an $\alpha$ parameter for each test star and PSF that corresponds to the argument that minimises the function $f(\alpha) = \| I_{1} - \alpha I_{2}  \|_{2}$, where $I_1$ and $I_2$ are the star and the PSF, respectively.

% --------- %
\subsection{Metric on real data: the $Q_p$ criteria}

Performing a comparison between two PSF models with real data is an arduous task since 
we do not know the shapes and pixel values of the observed stars. However, subtracting our estimated model from an observed star (i.e. pixel residual) should lead to a residual map containing only noise if the model is perfect. The probability of having our model correlated with the noise is extremely small.
Therefore, from this point of view, the method with the smallest pixel RMS residual error can be considered as the best.
Using all the test stars $y_s$ and our estimates $\hat{y}_s$, we calculate the pixel RMS residual error: 
$\text{Err} = \sqrt{ \frac{1}{N_i N_s} \sum_{s} \sum_{i} (y_{s,i} - \hat{y}_{s,i})^2}$, 
where $N_s$ is the number of stars and $N_i$ is the number of pixels we consider in a given image when we use a $10$ pixel radius circle from the centre of the residual images. The noise standard deviation $\sigma_{\text{noise}}$ is calculated from the stars only using the pixels outside the aforementioned circle. For a perfect modelling, we would have $\text{Err} \approx \sigma_{\text{noise}}$, and we define the $Q_{p_1}$ metric as:

\begin{equation}
    Q_{p_1} =  \left( \text{Err}^{2} - \sigma^{2}_{\text{noise}} \right)^{1/2}.
\end{equation}  
We next introduce two metrics to quantify how noisy the models are. 
The variance of the PSF model for the test stars $s$ reads $\sigma_{s}^{2} = [ \text{Var}(y_{s} - \hat{y}_{s}) - \sigma_{\text{noise}}^{2}(y_{s}) ]_{+}$, 
where $\text{Var}(\cdot)$ is a usual variance estimator, 
the operator $[\cdot]_+$ sets to zero negative values and $\sigma_{\text{noise}}^{2}(y_{s})$ is the noise variance estimation for a single star. We present the $Q_{p_2}$ and $Q_{p_3}$ metrics in the following equations:

\begin{equation}
    Q_{p_2} =  \left( \frac{1}{N_{s}} \sum_{s} \sigma_{s}^{2} \right)^{1/2}, \qquad Q_{p_3} =  \left( \frac{1}{N_{s}} \sum_{s} (\sigma_{s}^{2} - Q_{2}^{2})^{2}  \right)^{1/4}. 
\end{equation} 

The $Q_{p_2}$ metric represents the modelling error expectation for a given star, and the $Q_{p_3}$ metric 
indicates the fluctuation of the modelling error. 
A perfect PSF model would give values close to zero for the three metrics.

\subsection{Results}

\begin{table}[ht]
  \centering
  \begin{tabular}{lccc}
    \noalign{\smallskip} \toprule  \noalign{\smallskip}
    Method                  & $Q_{p_1}$             & $Q_{p_2}$             & $Q_{p_3}$             \\ \hline   \noalign{\smallskip}
    PSFEx                   & $15.56$           & $8.13$            & $14.31$           \\          \noalign{\smallskip}
    \textbf{MCCD-HYB}       & $\mathbf{12.14}$  & $\mathbf{6.68}$   & $\mathbf{10.86}$  \\ \hline   \noalign{\smallskip}
    $\text{Gain}_{PSFEx}$   & $22\%$            & $18\%$            & $24\%$            \\
    \noalign{\smallskip} \hline \hline                                                  \\
    Noise Std. Dev. ($\sigma_{\text{noise}}$)   & \multicolumn{3}{c}{$15.83$}           \\
    \noalign{\smallskip} \bottomrule \noalign{\smallskip}                               \\
  \end{tabular}
  \caption{$Q_p$ criterion using all test stars of the W3 dataset from CFIS. The gain of the MCCD-HYB with respect to \texttt{PSFEx} and the noise standard deviation are also presented.}
  \label{tab:CFIS_main_results}
\end{table}

The main results of the experiment are synthesised in \autoref{tab:CFIS_main_results} where the $Q_{p}$ criteria are given. 
In the $Q_{p_1}$ column, we can observe a $22\%$ gain of the MCCD-HYB method with respect to \texttt{PSFEx}. 
From $Q_{p_2}$ and $Q_{p_3}$ metrics, we also conclude that the MCCD-HYB model is considerably less noisy than the one from \texttt{PSFEx}.

\begin{figure}[ht]
    \centering
    \includegraphics[width=.49\linewidth]{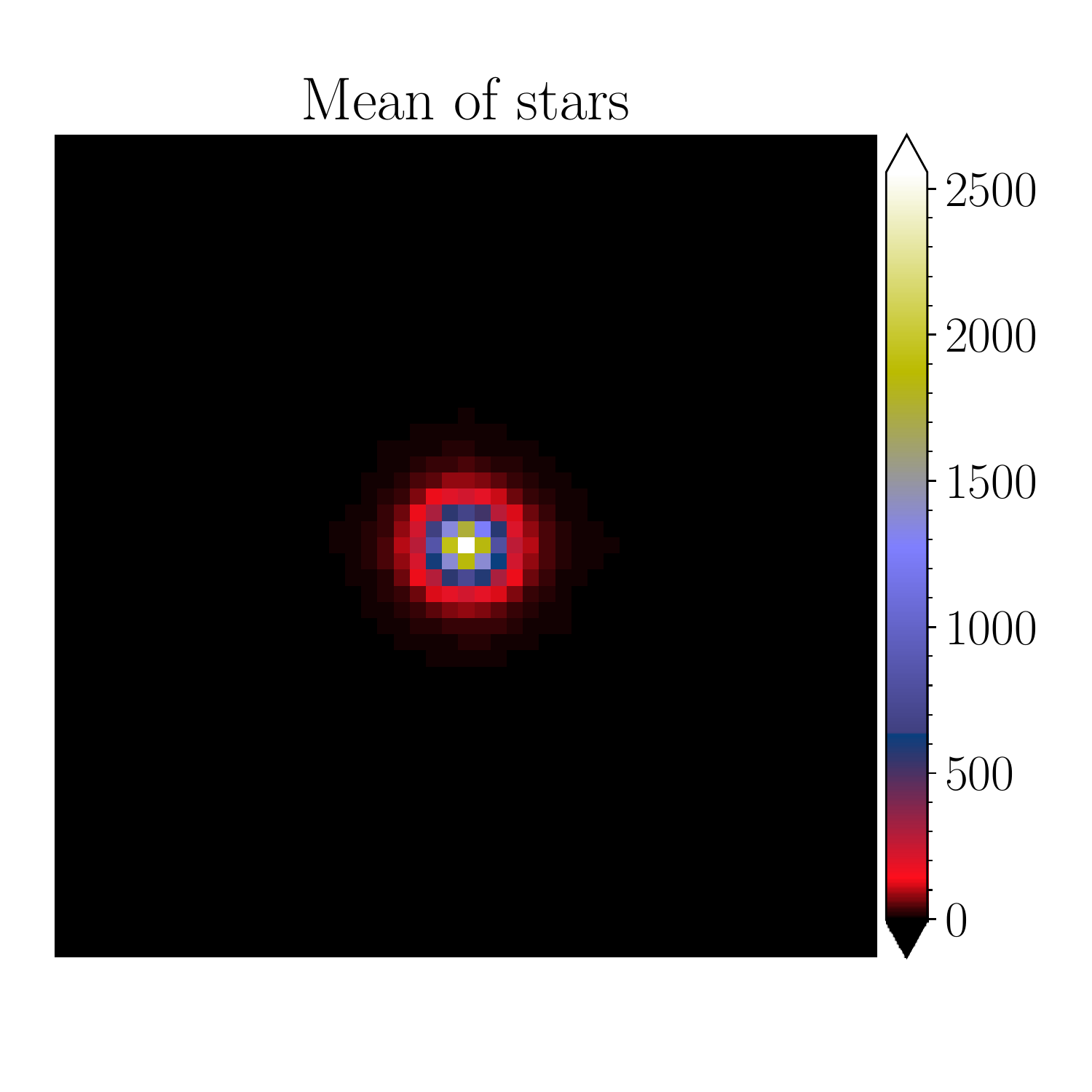} \\
    \includegraphics[width=.49\linewidth]{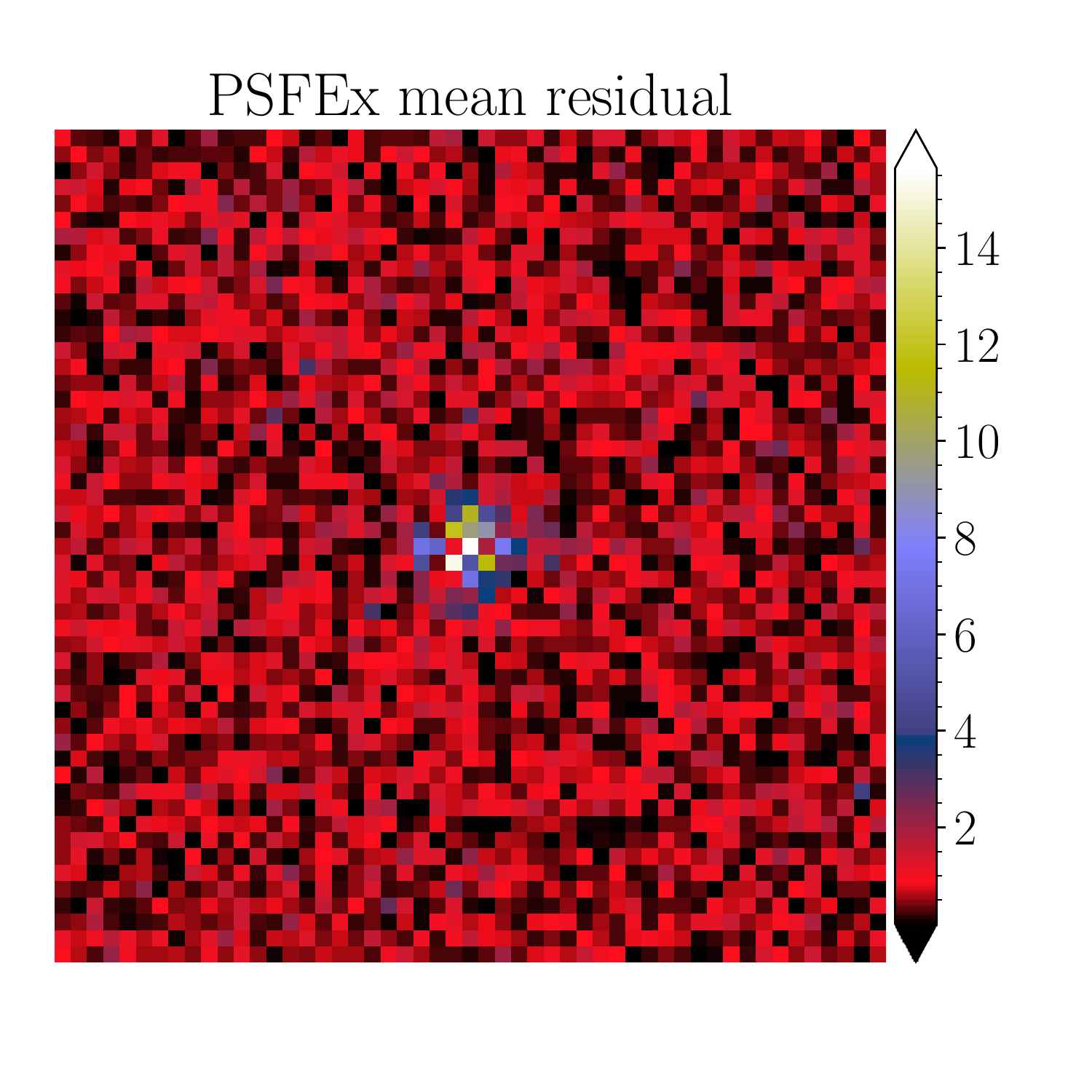}
    \includegraphics[width=.49\linewidth]{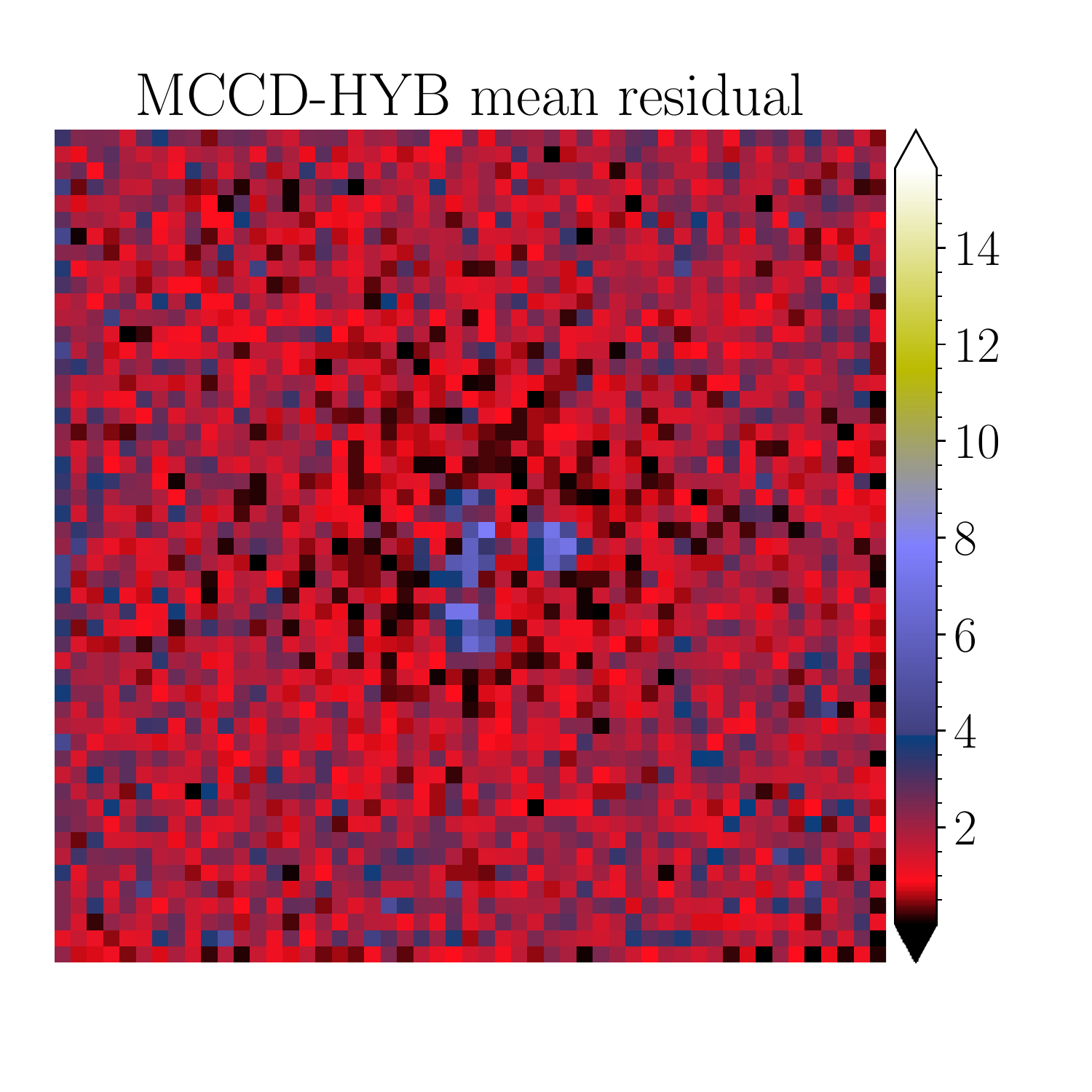}\\
    \includegraphics[width=.49\linewidth]{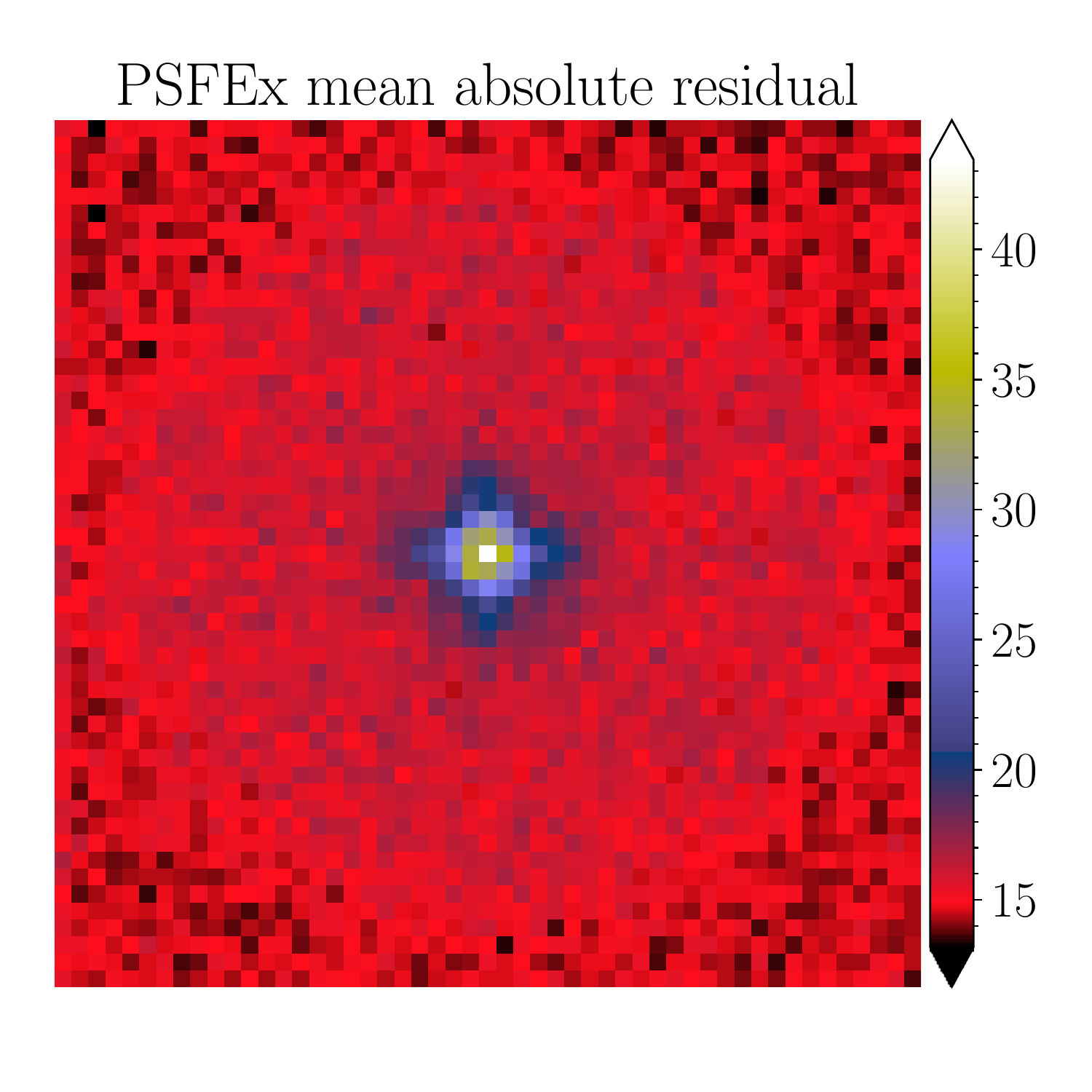}
    \includegraphics[width=.49\linewidth]{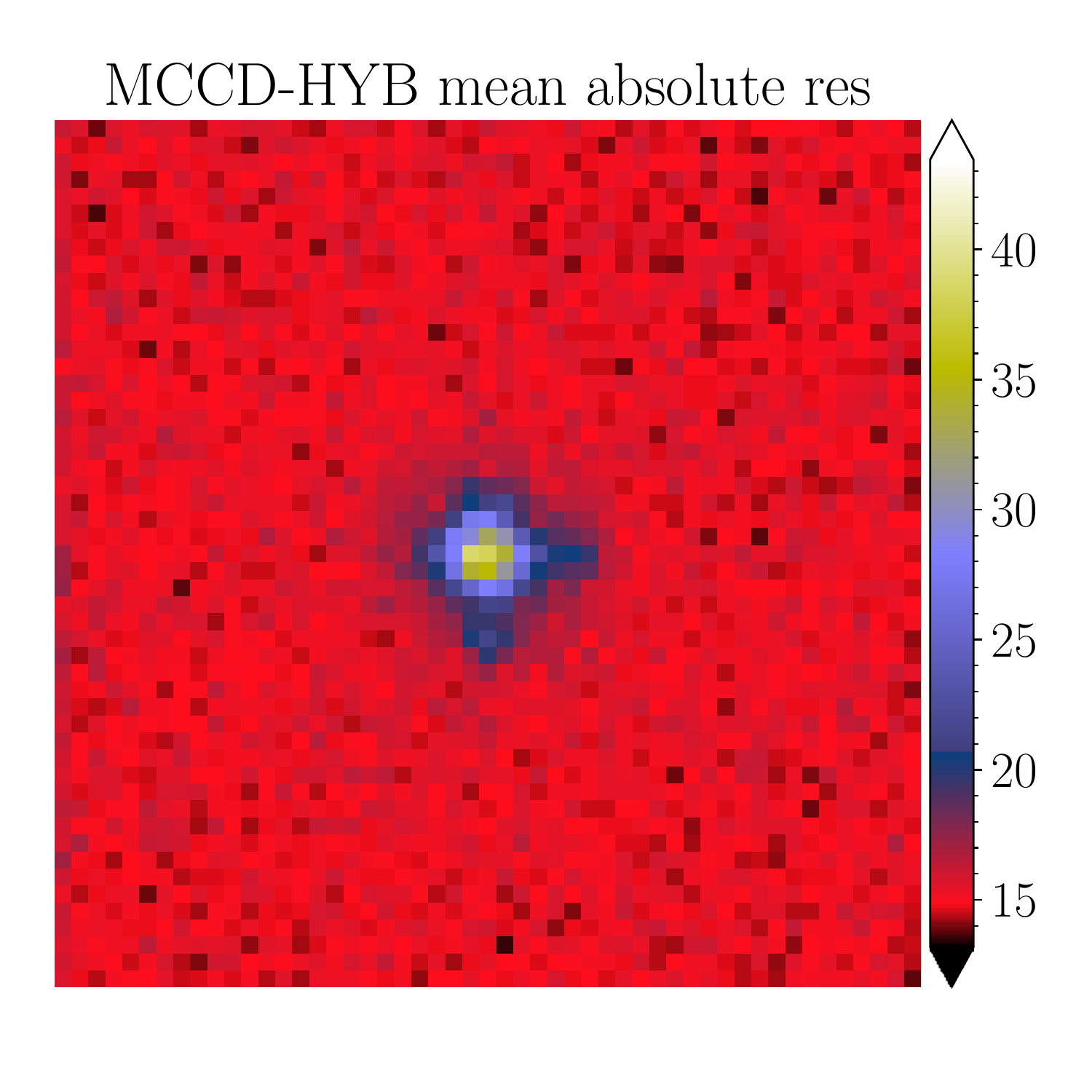}
    \caption{Stacked star profile from all 534 test stars in a random CFIS exposure (id. 2099948) (top), corresponding stacked residuals after subtraction by the PSFEx (middle left) or MCCD-HYB (middle right) PSF models. The bottom row includes the stacked absolute value of the residuals.}
    \label{fig:CFIS_mean_residuals}
\end{figure}

In order to explore potential remaining structure in the residuals, we stack together the residuals for all $534$ test stars from a random exposure. These are shown, along with the stacking of the test stars themselves, in \autoref{fig:CFIS_mean_residuals}.
We can see that \texttt{PSFEx} has a sharper stacked error compared to MCCD-HYB. This could indicate that our algorithm is better at capturing the size of the PSF, as the peak of the residual is directly related to it. Considering that there is no trace of shifting errors and that we are calculating the flux optimally, a greater mismatch in the size of the PSF equals to a greater peak pixel error on the residual. The third row presents the mean of the stacked absolute value of the residuals for both of the PSF models so that the residuals can not cancel themselves. We observe the same behaviour described above with the \texttt{PSFEx} pixel error distribution being sharper but more centred. It is also possible to notice the higher noise \texttt{PSFEx} has when compared to the MCCD-HYB model.

\begin{figure*}
    \centering
    \includegraphics[width=.49\linewidth]{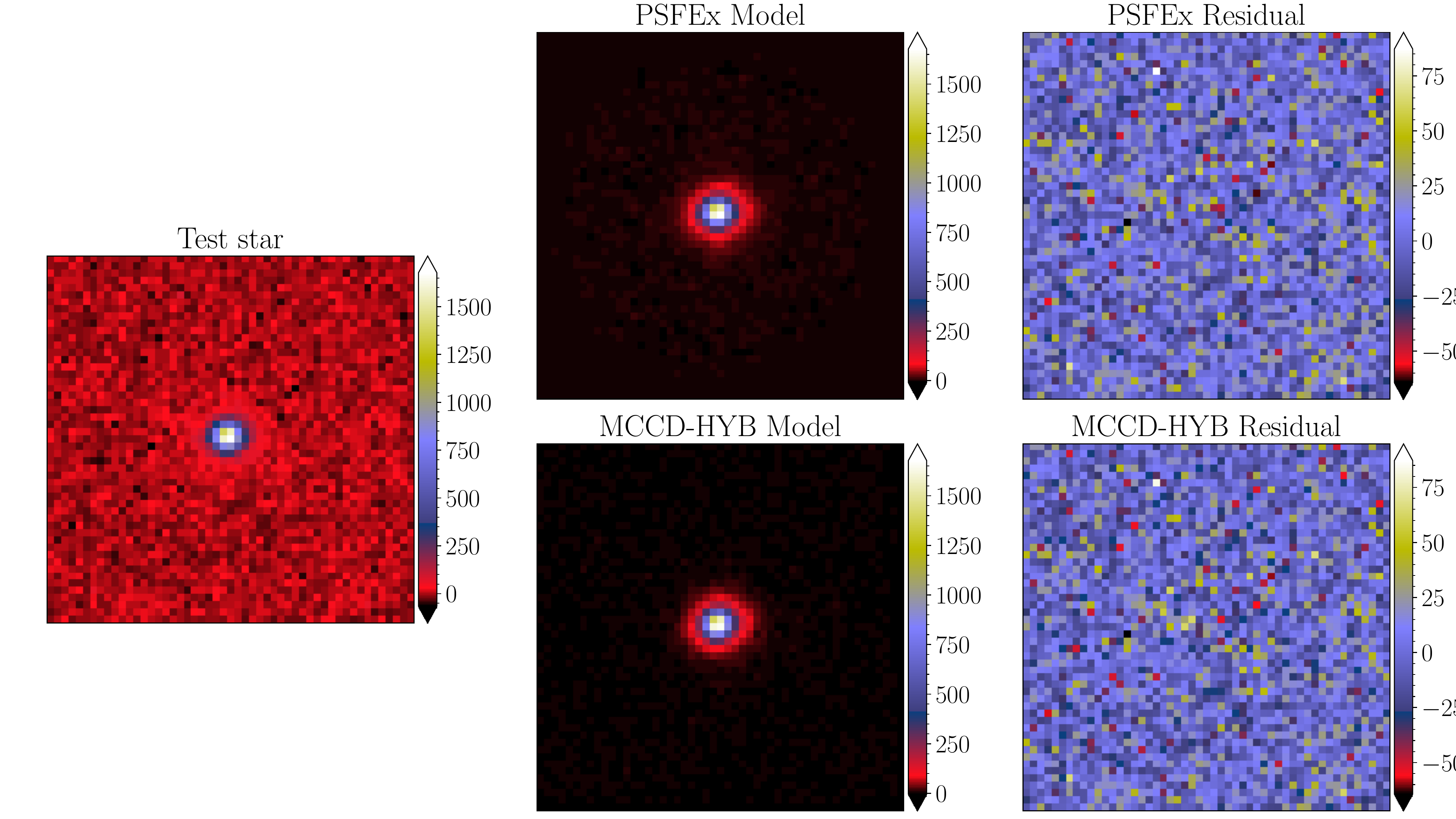}
    \includegraphics[width=.49\linewidth]{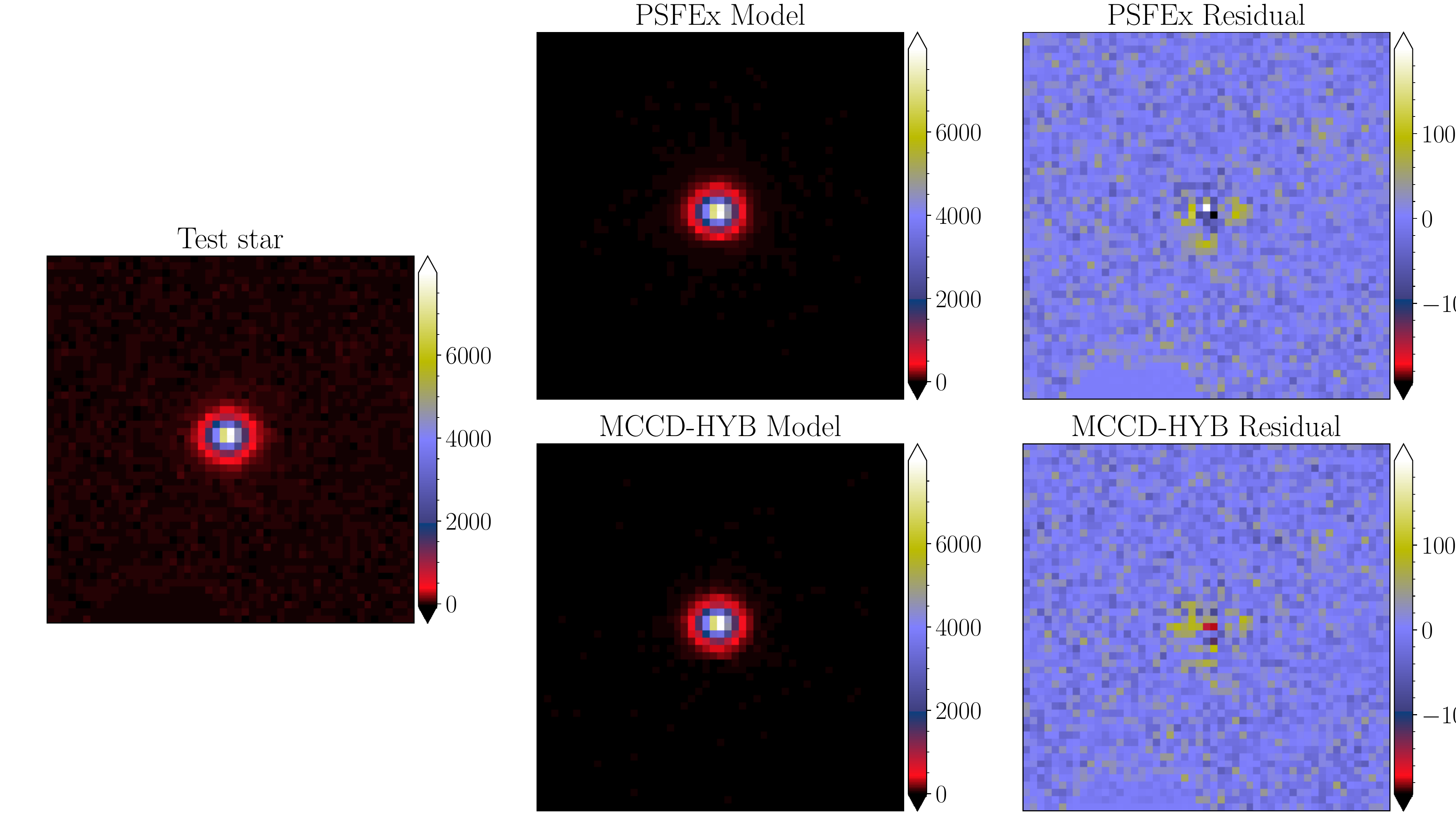}\\
    \includegraphics[width=.49\linewidth]{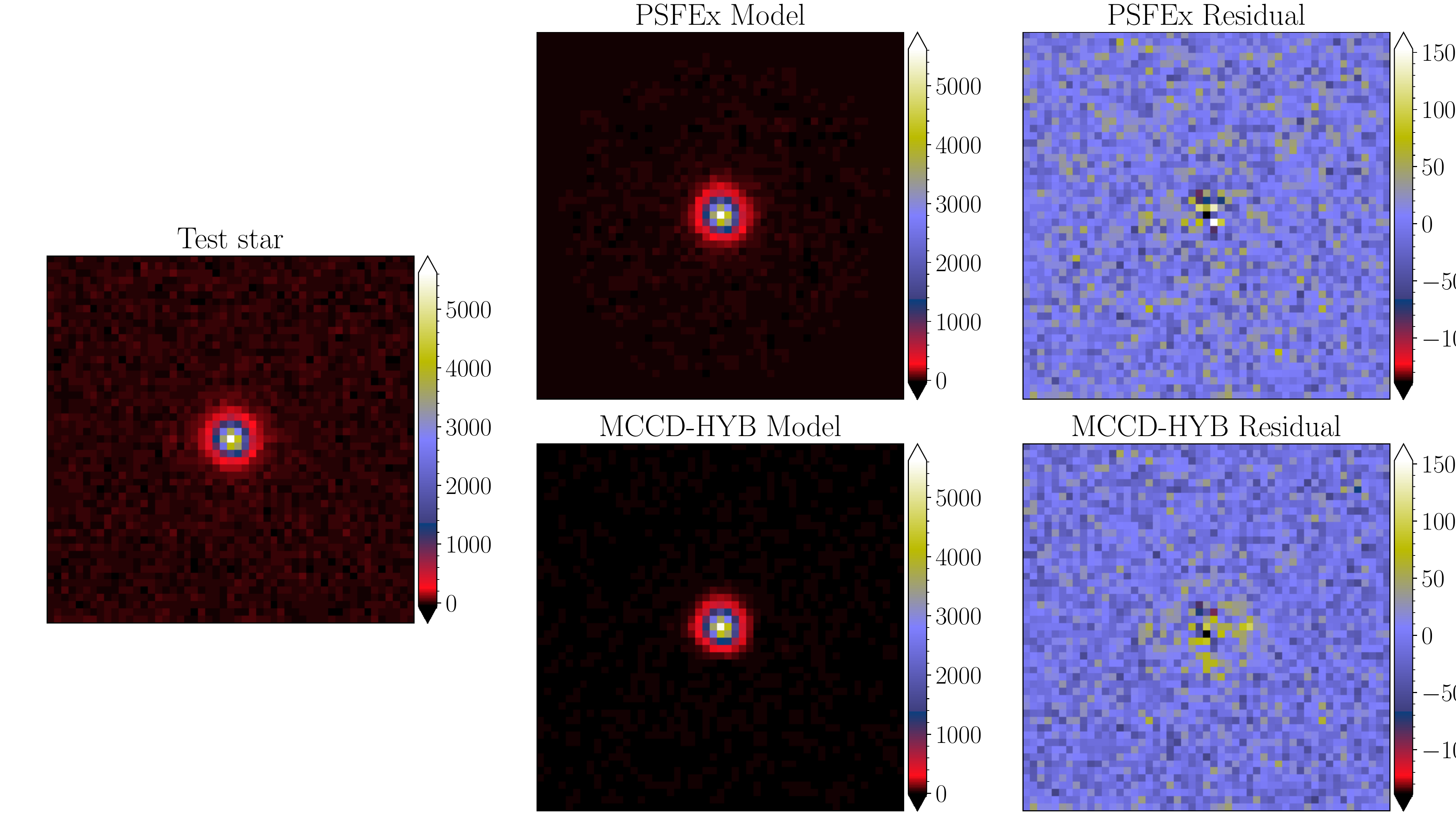}
    \includegraphics[width=.49\linewidth]{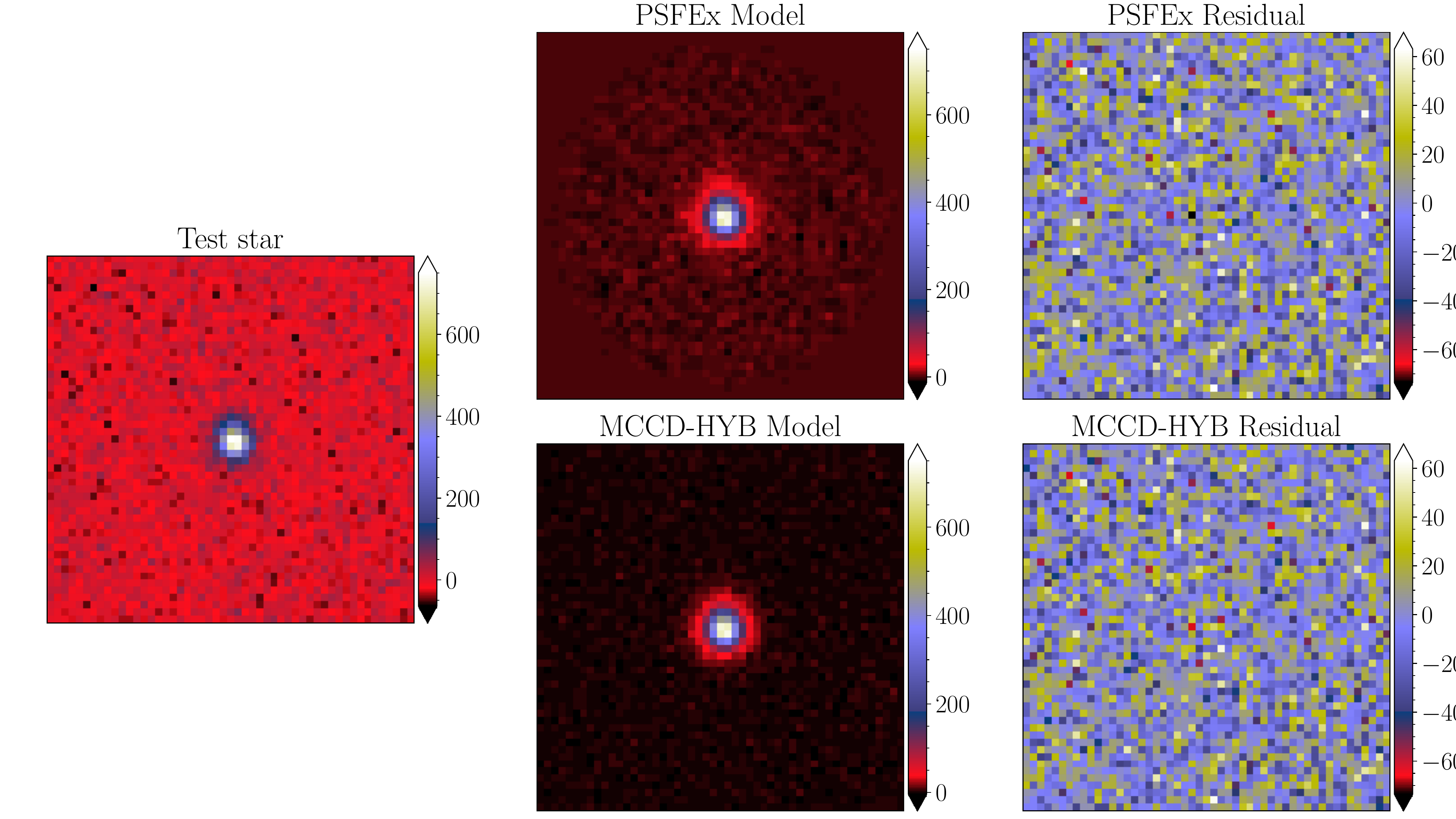}\\
    \caption{Examples of real CFIS test stars, the estimations of both methods and the pixel residuals. We used the same exposure as in \autoref{fig:CFIS_mean_residuals}. 
    We present four test stars with the estimated PSF models and the corresponding residuals. The top-left star corresponds to a star extracted from the top-left corner of the focal plane with a SNR of $19$. The top-right star corresponds to a star leading to a high error for both methods. The bottom-left star corresponds to a star located in the centre of the focal, with a relatively high SNR of $160$. The bottom-right star corresponds to a star located in the bottom-right corner of the focal plane, with a low SNR of $4$.}
    \label{fig:CFIS_reconstruction residuals}
\end{figure*}

\autoref{fig:CFIS_reconstruction residuals} presents examples of star image reconstructions by the two different PSF models, \texttt{PSFEx} and MCCD-HYB, and their corresponding residuals. The proposed method yields a near noiseless model when compared to \texttt{PSFEx}, as can clearly be seen on the top-left and bottom-right stars of \autoref{fig:CFIS_reconstruction residuals}, where the stars have low SNRs of $19.3$ and $4.2$ respectively.
Both models share a good estimation of the bottom-right star, which comes a low-stellar-density region of the focal plane (the bottom-right corner, as can be seen in \autoref{fig:CFIS_star_density}). On the bottom-left star of \autoref{fig:CFIS_reconstruction residuals}, we observe a similar type of error as that appearing in \autoref{fig:CFIS_mean_residuals}.

It is difficult to derive conclusions of different PSF model performances based on the shape measurement of noisy stars due to its high stochasticity. Nevertheless, driven by the comments from DES Y1 \citep{zuntz2018} on the residual mean size offset from the \texttt{PSFEx} model, we conducted a study with our data. We measured the size from the training stars and from both calculated PSF models, \texttt{PSFEx} and MCCD-HYB, and then computed the residual. The RMS residual size of the $\Delta R^2 /R^2$ value gave $4.82\times 10^{-2}$ for \texttt{PSFEx} and $4.02\times 10^{-2}$ for MCCD-HYB. This represents a $16\%$ gain of our proposed algorithm.

\begin{figure*}
    \centering
    \includegraphics[width=.49\linewidth]{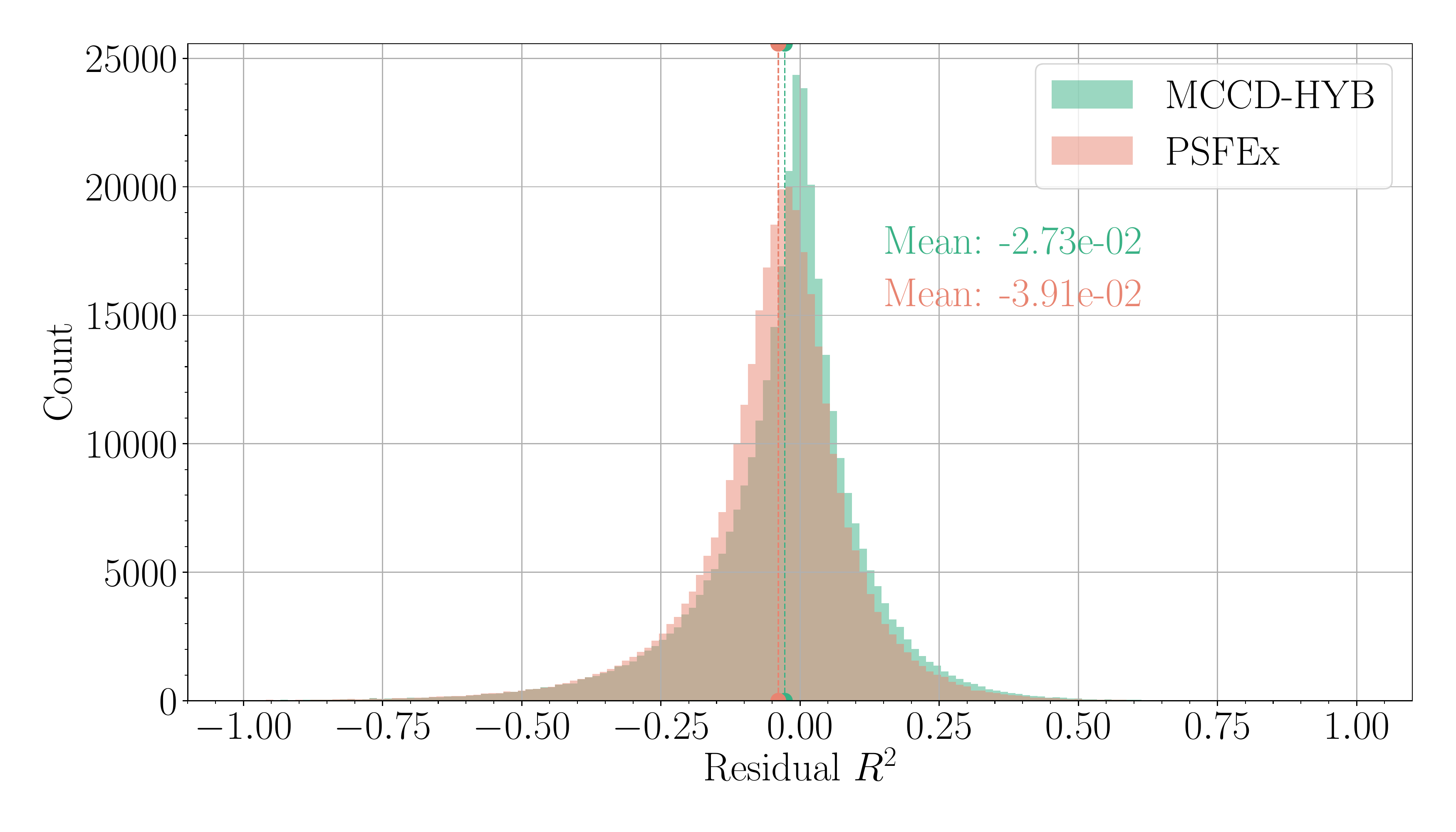}
    \includegraphics[width=.49\linewidth]{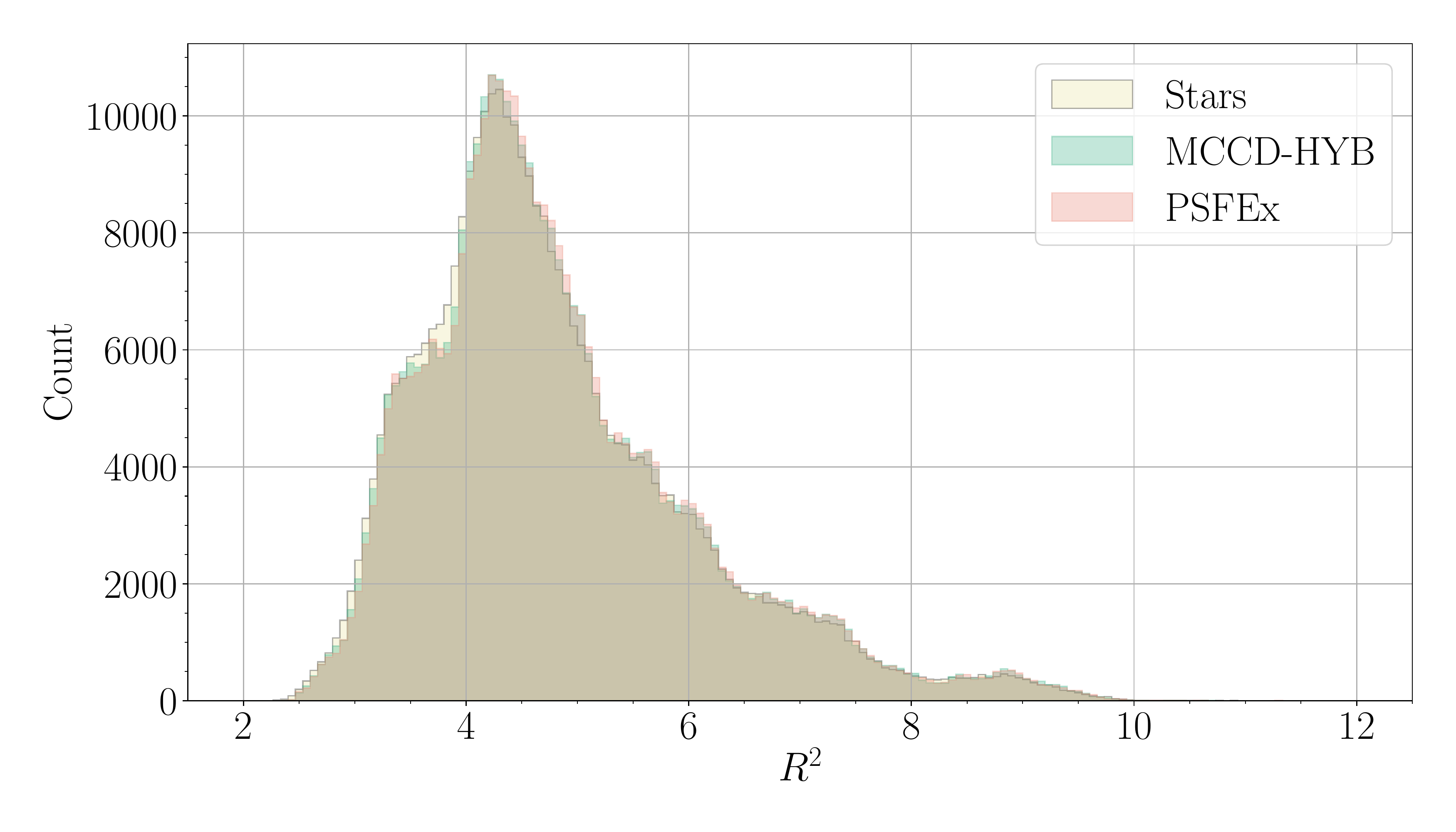}\\
    \caption{Histograms showing the distribution of the size metric over the train stars and their residuals for both PSF models, \texttt{PSFEx} and MCCD-HYB.}
    \label{fig:CFIS_moment_histograms}
\end{figure*}

\autoref{fig:CFIS_moment_histograms} presents in the left column the histogram of the residuals and in the right column the histograms of the size metrics.
We can notice that the MCCD-HYB algorithm has a sharper residual size around zero. The figure also includes the mean of the residuals for each PSF model. This shows that both models tend to overestimate the size of the PSF. However, the MCCD-HYB model presents a $30\%$ gain in the mean residual size with respect to \texttt{PSFEx}, indicating a smaller bias in the shape.

%-------------------------------------------------------------------

% Conclusions
\section{Reproducible research}

In the spirit of reproducible research, the MCCD-RCA algorithm will be publicly available on the CosmoStat's Github\footnote{\url{https://github.com/CosmoStat/mccd}}, including the material needed to reproduce the simulated experiences. The MCCD PSF modelling software will be included in the CFIS shape measurement pipeline \citep{guinot2020}.

\section{Conclusion}\label{sec:conclusion}

We have presented a family of non-parametric PSF (Point Spread Function) modelling methods coined MCCD, including its best-performing extension MCCD-HYB, which are built upon the existing RCA (Resolved Component Analysis) method and are capable of constructing PSF models that span all the CCDs (Charge-Coupled Device) from an instrument's focal plane at once. Naturally, the use of more stars for the training allows us to build more complex models that can capture evasive features. Our model is composed of global components, spanning all the CCDs, and local components that are CCD-specific. By using this structure we can better capture global patterns and features that might be lost when using only a local model like in RCA or, the widely used algorithm, \texttt{PSFEx}.

The method was first tested with a set of simulated PSFs following a real star spatial distribution  over MegaCam's focal plane, an instrument from the CFHT (Canada-France-Hawaii Telescope). Its use leads to better performance in all the evaluated metrics when compared to \texttt{PSFEx}. We then tested the method on a set of real CFIS images, an imaging survey based on CFHT, in order to confirm that it can handle real data. Our method achieves a smaller pixel RMS (Root Mean Square) residual than \texttt{PSFEx} and the estimated model is considerably less noisy. 

The performance gain of the MCCD methods over \texttt{PSFEx} is higher when using our simulated dataset than when using the real dataset. This can be explained by the fact that our simulated dataset shows more intricate variations in the PSF than the real data does, and MCCD is better at capturing such strong variations. 

The proposed method can naturally handle more complex PSF profiles, such as those expected from space-based instruments. The RCA method 
was tested with Euclid-like simulated PSFs and has shown a better performance than \texttt{PSFEx} \citep{ngole2016, schmitz2020}. Therefore, we expect to have an even superior performance in this scenario with MCCD. 
Thanks to its formulation, it can also handle super-resolution, making it suitable for under-sampled data.

Despite the good performance of the method, there is still room for improvement. 
A natural straight-forward extension for the MCCD algorithms would be to replace the denoising strategy by one more suited for the specificities of the PSFs we work with.  This could be accomplished by using a deep neural network as the denoiser \citep{ronneberger2015u, ye2018deep}.

% Acknowledgements
\begin{acknowledgements}
      The authors would like to thank the anonymous referee for the fruitful comments on the paper.
      This work is based on data obtained as part of the Canada-France Imaging Survey, a CFHT large program of the National Research Council of Canada and the French Centre National de la Recherche Scientifique. Based on observations obtained with MegaPrime/MegaCam, a joint project of CFHT and CEA Saclay, at the Canada-France-Hawaii Telescope (CFHT) which is operated by the National Research Council (NRC) of Canada, the Institut National des Science de l'Univers (INSU) of the Centre National de la Recherche Scientifique (CNRS) of France, and the University of Hawaii. This research used the facilities of the Canadian Astronomy Data Centre operated by the National Research Council of Canada with the support of the Canadian Space Agency. \\
      This work has made use of the CANDIDE Cluster at the Institut d'Astrophysique de Paris and made possible by grants from the PNCG and the DIM-ACAV.\\
      Software: Numpy \citep{numpy2011}, Scipy \citep{scipy2020},  Astropy \citep{Robitaille2013,PriceWhelan2018}, GalSim \citep{rowe2015}, IPython \citep{Perez2007}, Jupyter \citep{Kluyver2016}, Matplotlib \citep{Hunter2007}, PySAP \citep{farrens2020}.
%      \AckECon
\end{acknowledgements}

% Bibliography
\bibliographystyle{aa}
\bibliography{PSF_prop}

%-------------------------------------------------------------------

% Appendix
\begin{appendix}
%--------------------------------
\section{RCA regularisations}\label{appdx:rca_reg}

In this section we give a more detailed description of each regularisation we use in our local RCA model:

\begin{itemize}

\item[1.] \textit{Low rank:} PSF variations can be explained by a small number of eigenPSFs. This constraint can be enforced by the proper choice of two parameters, the number of local, $r_k$, and global, $\tilde{r}$, eigenPSFs. These parameters are directly linked with the complexity of the model we will be addressing and its selection will naturally depend on the PSF field we will be facing. It is important to allow the model a certain complexity so that it can correctly capture the PSF field's variations but it should not be much more complex as the model will tend to overfit the noisy observations and therefore lose its generalising power to estimate the PSF in galaxy positions.

\item[2.] \textit{Positivity:} the reconstructed PSFs $\hat{H}$ should only contain non-negative pixel values.

\item[3.] \textit{Sparsity:} the observed PSFs are structured images; a way to promote our model to follow this structured behaviour is to enforce the sparsity of the eigenPSFs in an appropriate basis.

\item[4.] \textit{Spatial constraints:} the regularity of the PSF field $\mathcal{H}$ means that the smaller the distance between two PSFs positions $u_i , u_j$ the smaller the difference between their representations should be $\mathcal{H}(u_i), \mathcal{H}(u_j)$. This regularity can be achieved by enforcing constraint in the coefficient matrices $A_k, \Tilde{A}_k$; for example, the line $l$ of $A_k$ corresponds to the contribution of eigenPSF $l$ to the $n_{\rm star}^{k}$ stars in CCD $k$ located in positions $\left( u_i \right)_{i=1}^{n_{\rm star}^{k}}$. The closer the positions, the closer the coefficient values should be. 

\end{itemize}

%--------------------------------
\section{Shape and size definitions}\label{appdx:shape_defs}

The ellipticity parameters and the size are defined in terms of the moments of the surface brightness profile $I(x,y)$ following \citep{hirata2003}:

\begin{align}
    \bar{\mu} & = \frac{\int \mu \; I(x,y) \; w(x,y) \; \text{d}x \text{d}y }{\int I(x,y) \; w(x,y) \; \text{d}x \text{d}y},  \\
    M_{\mu \nu} &= \frac{\int I(x,y) \; (\mu - \bar{\mu}) \; (\nu - \bar{\nu}) \; w(x,y) \; \text{d}x \text{d}y}{\int I(x,y) w(x,y) \; \text{d}x \text{d}y},
\end{align}
where $\mu , \nu \in \{x,y \}$ and $w(x,y)$ is weight window to avoid noise related issues. The size is defined as:

\begin{equation}
    T = R^{2} = M_{x x} + M_{y y},
\end{equation}
and the ellipticities are defined as:

\begin{equation}
    e = e_1 + {\rm i} e_2 = \frac{(M_{x x} - M_{y y}) + {\rm i} \, 2 M_{x y}}{T} .
\end{equation}
The adaptive moment measurement from HSM gives $\sigma$ as output which relates to our size metric as $R^2 = 2\sigma^2$.

%--------------------------------
\section{Additional figures}\label{appdx:plots}

\begin{figure}[ht]
    \centering
    \includegraphics[width=.99\linewidth]{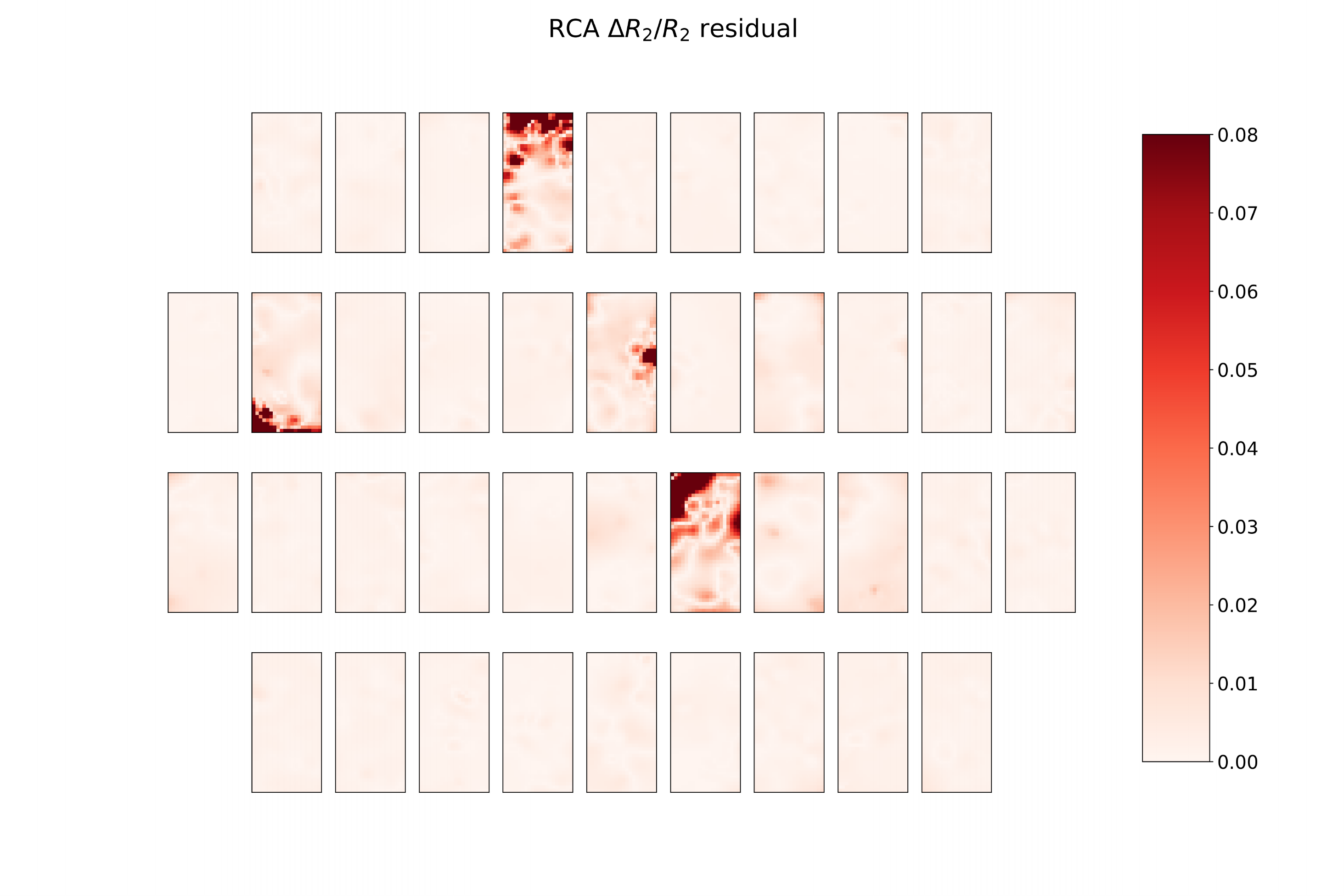}
    \caption{Residual $\Delta R_2 / R_2$ map of the RCA algorithm of stars with a SNR of 50. The CCDs where the RCA model is having degeneracies that can be clearly spotted on the map.}    
    \label{fig:rca_degen}
\end{figure}

\begin{figure}[ht]
    \centering
        \begin{subfigure}[t]{.99\linewidth}
            \includegraphics[width=.99\linewidth]{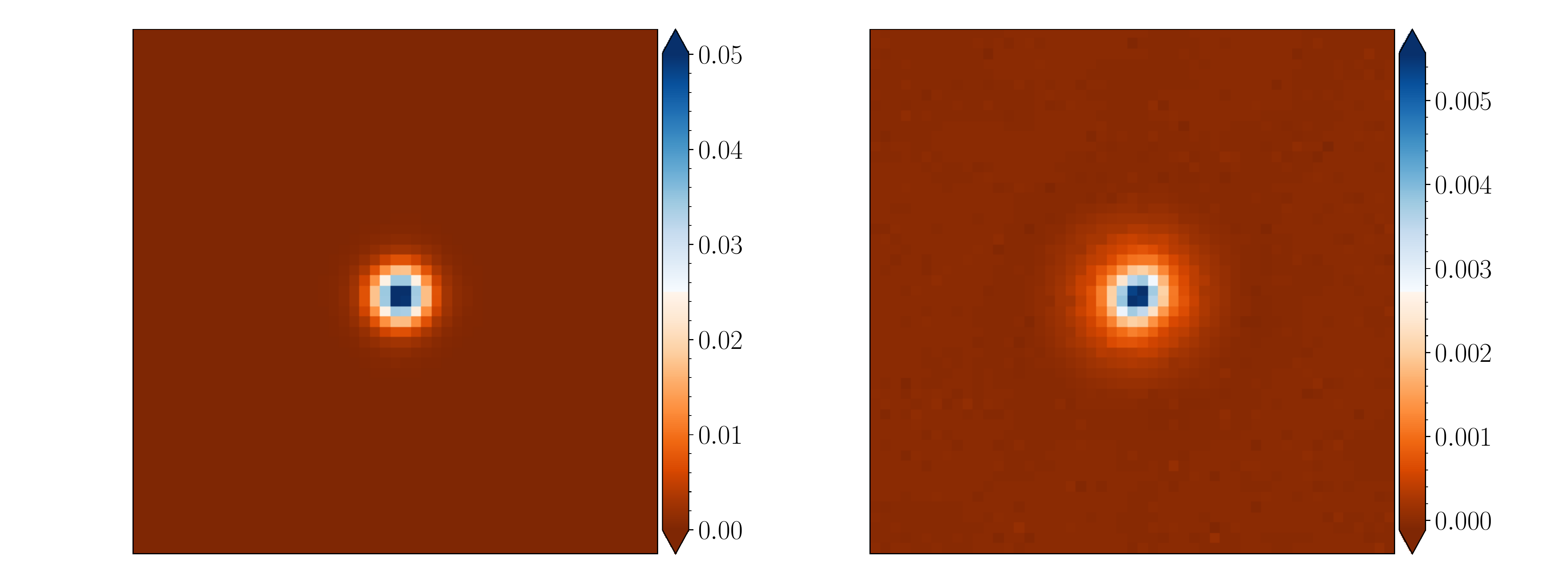}\\
            \includegraphics[width=.99\linewidth]{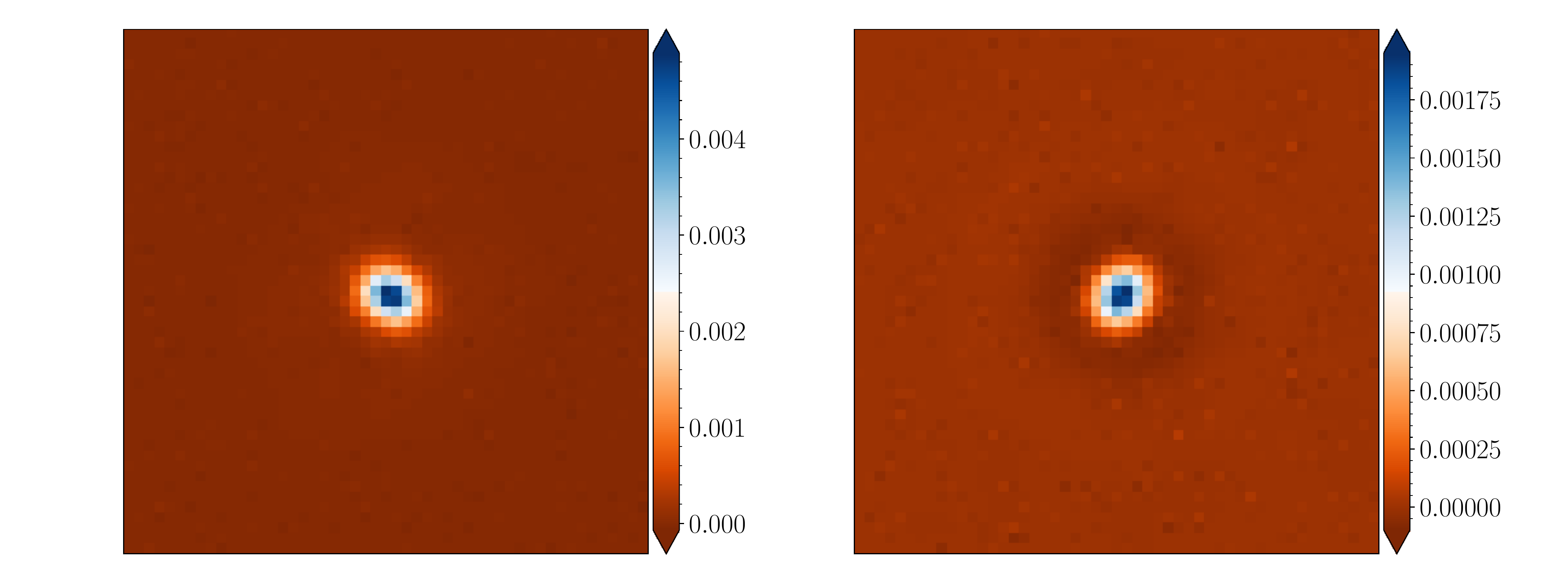}\\
            \includegraphics[width=.99\linewidth]{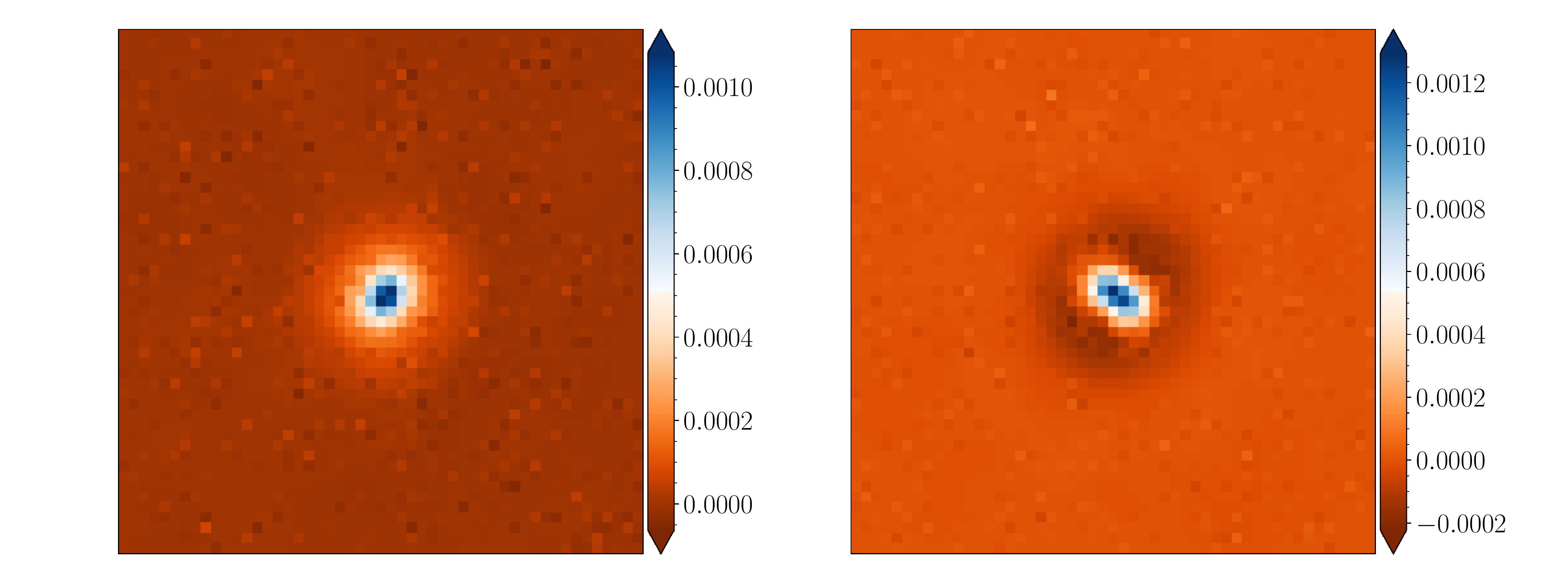}
            \caption{Global eigenPSFs.}
        \end{subfigure}
        \begin{subfigure}[t]{.99\linewidth}
            \includegraphics[width=.99\linewidth]{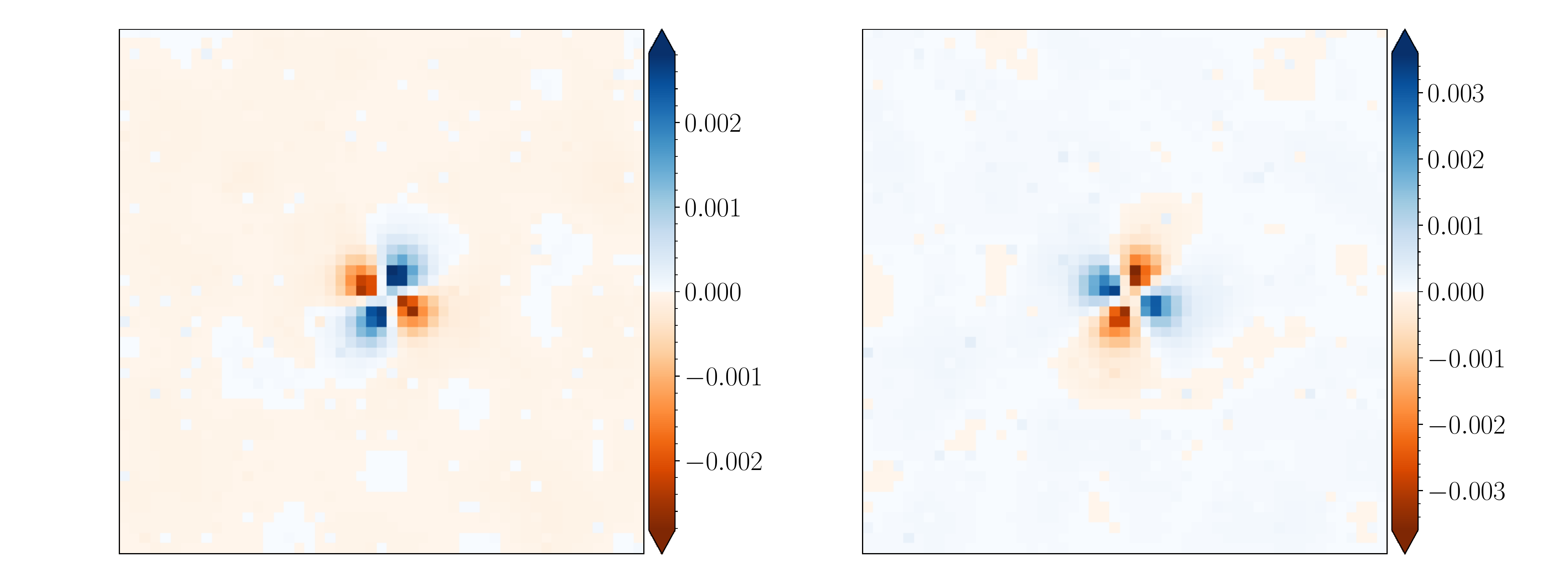}\\
            \includegraphics[width=.99\linewidth]{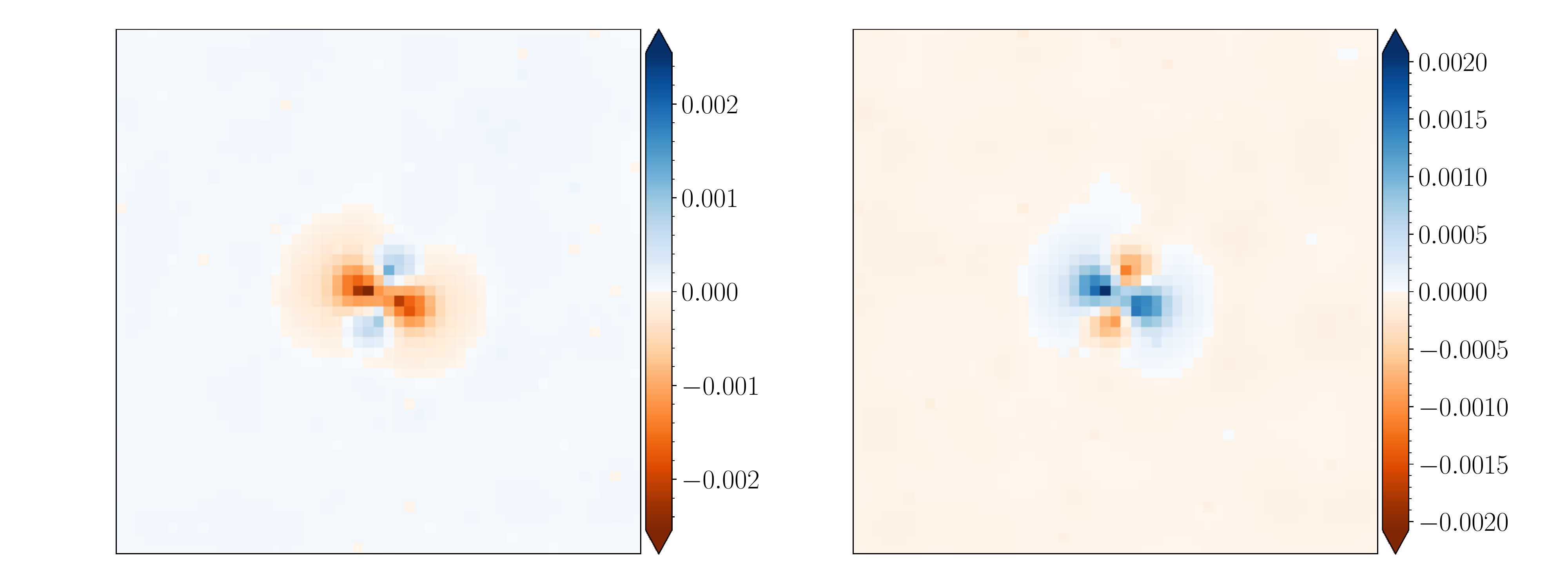}
            \caption{Local eigenPSFs.} % CCD 12
        \end{subfigure}
    \caption{Example eigenPSFs extracted from the MCCD-HYB PSF model trained on the simulated dataset with a SNR of $70$. The local eigenPSFs were extracted from the graph's spatial constraint of a central CCD. It can be seen from the eigenPSFs that the global model is specialising on the shape of the PSF while the local model specialises on capturing its ellipticity. It is also worth to mention that the first global eigenPSF found on the first row provides the baseline isotropic PSF the model uses.}
    \label{fig:example_eigenPSFs}
\end{figure}

In this appendix we include the additional figures, \autoref{fig:rca_degen} and \autoref{fig:example_eigenPSFs}.

%--------------------------------
\section{Optimisation methods}\label{appdx:optim}

In this appendix we include details on the practical resolution of the four optimisation problems seen in Algorithm \ref{alg:mccd}. For more information about proximal operators and proximal algorithms we refer the reader to \citep{parikh2014} and \citep{beck2017}.

% ---------- %
\subsection{Problem (III)}\label{appdx:optim_prob_III}

As in most of the optimisation problems, the algorithm used depends on the objective function we work with. In this case, we use the primal-dual algorithm 3.1 in \cite{condat2013}\footnote{We use the implementation found in the python package \url{https://github.com/CEA-COSMIC/ModOpt} from \cite{farrens2020}.}. The main motivation resides in the nature of the constraints we use when optimising over $S_k$, as we face one smooth and two non-smooth terms, and a linear operator.
The optimisation algorithm aims at solving the following problem:

\begin{equation}
    \text{Find } \hat{x} \in \argmin_{x \in \mathcal{X}} \left[ F(x) + G(x) + H(L(x)) \right],
    \label{eq:condat_optim}
\end{equation}
where: i) $F$ is convex, differentiable and its gradient is L-Lipschitz continuous; ii) $G$ and $H$ are proximable functions that should have closed form proximal operators; iii) $L$ is a bounded linear operator; and iv) the set of minimisers of the aforementioned optimisation problem is nonempty. 
It is straightforward to identify the different functions in the optimisation of the local $S_k$ matrix which match the formulation of \autoref{eq:condat_optim}. 
Following the notation we used throughout the article, let $F_k(x=(S_1, \ldots , S_N , \Tilde{S}, \alpha_1 , \ldots , \alpha_N , \Tilde{\alpha})) = \frac{1}{2} \| Y_k - \mathcal{F}_k(\hat{H}_k) \|_F^2$, with $\hat{H}_k = S_k \alpha_k V_k^\top + \Tilde{S} \Tilde{\alpha} \Pi_k$, and $G(S_k) =  \sum_i \| \mathbf{\mathbf{w}}_{k,i} \odot \Phi \mathbf{s}_{k,i} \|_1 $. Let $H(S_k) = \iota_+ (S_k) $ and the linear operator $L$ be $L(S_k) = S_k \alpha_k^{(l)} V_k^\top + \Tilde{S}^{(l)} \Tilde{\alpha}^{(l)} \Pi_k$. For the moment, we will consider $\Phi$ to be the identity. 

To solve the algorithm we need the proximal operator of $H^{*}$, the adjoint function of $H$, the proximal operator of $G$ and the gradient of $F$ with its Lipschitz constant.

Starting with $H$, the proximal operator of $H^{*}$ can be calculated directly using the proximal operator of the function $H$ itself by means of the Moreau decomposition \citep[Theorem 6.44]{beck2017}. The proximal operator of an indicator function over a set $\mathcal{C}$ is the orthogonal projection over that set. Therefore, we note $[X]_{+}$ the projection of $X \in \mathbb{R}^{n \times m}$ onto the positive orthant, that is

\begin{align}
     \text{prox}_{\iota_{+}(\cdot)}(X) = [X]_{+} \rightarrow [X_{i,j}]_{+} = 
    \begin{cases}
       X_{i,j} &\mbox{if $X_{i,j} \geq 0$ },\\
       0 &\mbox{otherwise}.
    \end{cases}
\end{align}
Continuing with $G$, the proximal operator of the $\ell_1$ norm is the soft thresholding operator which can be defined component-wise, for $x,\lambda \in \mathbb{R}$, as

\begin{align}
     \text{SoftThresh}_{\lambda}(x) = (|x| - \lambda)_{+} \text{sign}(x) =  
    \begin{cases}
       x - \lambda,   & x \geq \lambda ,\\
       0,       & |x| < \lambda , \\
       x + \lambda,   & x \leq - \lambda .
    \end{cases}
\end{align}
We name $L_{\nabla_{S_k} F(\cdot)}$ the Lipschitz constant of $F$'s gradient. The next equations resume what we need to use the chosen optimisation algorithm:

\begin{align}
    & \nabla_{S_k} F(S_k) =  - \mathcal{F}_k^*(Y_k - \mathcal{F}_k(\hat{H}_k)) V_k \alpha_k^\top, \\
    & L_{\nabla_{S_k} F(\cdot)} = \rho(\mathcal{F}_k^* \circ \mathcal{F}_k) \rho(\alpha_k V_k^\top (\alpha_k V_k^\top)^\top), \\
    & \text{prox}_{\tau G(\cdot)}([\mathbf{s}_{k,i}]_{j}) = \text{SoftThresh}_{\tau [\mathbf{w}_{k,i}]_{j}}([\mathbf{s}_{k,i}]_{j} ), \\
    & \text{prox}_{\sigma H^{*}(\cdot)}(X) = X - (X)_{+},
\end{align}
where the proximal operator of $G$ is defined component-wise, the notation $[\mathbf{s}_{k,i}]_{j}$ represents the element $j$ of the $i$ column vector of matrix $S_k$, $\mathcal{F}_k^*$ is the adjoint operator of $\mathcal{F}_k$, and $\rho(\cdot)$ is the spectral radius\footnote{The spectral radius can be defined as $\rho(B) = \max \{ |\lambda_1(B)|, \dots, |\lambda_n(B)| \}$ where $\lambda_i(B)$ are the eigenvalues of the matrix $B$.} that we calculate using the power method \citep{golub1996}.
For the algorithm's parameters $\tau$ and $\sigma$, based on Theorem 3.1 from \cite{condat2013}, we use:

\begin{equation}
    \tau = \frac{1}{\alpha L_{\nabla_{S_k} F(\cdot)}} , \qquad \sigma = \frac{\alpha L_{\nabla_{S_k} F(\cdot)}}{2 \| L \|_{\rm op}^{2}},
\end{equation}
where $\| \cdot \|_{\rm op}$ is the operator norm \citep{aliprantis2007} and $\alpha$ is a parameter we set to $3/2$. Being $L$ a bounded linear operator we can calculate $\| L \|_{\rm op}$ as $\sqrt{\rho(L^*L)}$ being $L^*$ its adjoint operator. 

We now consider the case where $\Phi$ is not the identity, but it is orthonormal, $\Phi^{T}\Phi = I$. We can adapt the soft thresholding operator in order to cope with the $G$ term. This would be $\mathbf{s}_k \rightarrow \Phi^{T} \text{SoftThresh}_{\tau \mathbf{w}_k} (\Phi \mathbf{s}_k)$. When using undecimated wavelets as the starlets, the orthonormal condition is not met. Nevertheless, they are tight frames whose Gram matrix is close to the identity which means that the presented formulation will be a good approximation. We redirect the reader to \cite{starck2015} for more information on wavelets. 

% ---------- %
\subsection{The remaining optimisation problems}\label{appdx:other_optim_problems}

We deal in a similar way with the problems (II) and (IV) from Algorithm \ref{alg:mccd} using the same optimisation method proposed in \cite{condat2013}. On the other hand, for problem (I), we use the optimisation algorithm in \cite{liang2018}. This is due to the fact that we are neglecting the positivity constraint as we account for it when optimising over the other variables. 
In order to use these algorithms we need to compute the gradients of the differentiable term of each problem as follows:  

\begin{align}
    \nabla_{S_k} F_k(x) =  - & \mathcal{F}_k^*(Y_k - \mathcal{F}_k(\hat{H}_k)) (\alpha_k V_k^\top)^\top , \\
    \nabla_{\alpha_k} F_k(x) = - & S_k^\top \mathcal{F}_k^*(Y_k - \mathcal{F}_k(\hat{H}_k)) V_k  , \\
    \nabla_{\Tilde{S}} F(x) =  \sum_{k = 1}^N & \nabla_{\Tilde{S}} F_k(x) = \sum_{k = 1}^N  - \mathcal{F}_k^*(Y_k - \mathcal{F}_k(\hat{H}_k)) (\Tilde{\alpha} \Pi_k)^\top, \\
    \nabla_{\Tilde{\alpha}} F(x) =  \sum_{k = 1}^N & \nabla_{\Tilde{\alpha}} F_k(x) =  \sum_{k = 1}^N - \Tilde{S}^\top \mathcal{F}_k^*(Y_k - \mathcal{F}_k(\hat{H}_k)) \Pi_k^\top ,
\end{align}
where $F = \sum_{k = 1}^N F_k = \frac{1}{2} \| Y - \mathcal{F}(H + \Tilde{S} \Tilde{\alpha} \Pi) \|_F^2$. Concerning the global optimisation over $\Tilde{S}$ and $\Tilde{A}$ we need to consider all the CCDs when computing the gradient. So we can reformulate the global formulas as:

\begin{align}
    \nabla_{\Tilde{S}} F(x) = & - \mathcal{F}^*(Y - \mathcal{F}(\hat{H})) (\Tilde{\alpha} \Pi)^\top , \\
    \nabla_{\Tilde{\alpha}} F(x) = & - \Tilde{S}^\top \mathcal{F}^*(Y - \mathcal{F}(\hat{H})) \Pi^\top.
\end{align}
An approximation for the Lipschitz constants of the different gradients can be calculated as:

\begin{align}
    L_{S_k} = &\ \rho(\mathcal{F}_k^* \circ \mathcal{F}_k) \rho(\alpha_k V_k^\top (\alpha_k V_k^\top)^\top)  \\
    L_{\alpha_k} = &\ \rho(\mathcal{F}_k^* \circ \mathcal{F}_k)  \rho(S_k^\top S_k) \rho(V_k^\top V_k) \\
    L_{\Tilde{S}} = &\ \rho(\mathcal{F}^* \circ \mathcal{F}) \rho(\Tilde{\alpha} \Pi (\Tilde{\alpha} \Pi)^\top )  \\
    L_{\Tilde{\alpha}} = &\ \rho(\mathcal{F}^* \circ \mathcal{F}) \rho(\Tilde{S}^\top \Tilde{S}) \rho(\Pi \, \Pi^\top ) 
\end{align}
where $\rho(\cdot)$ is the spectral radius. 

Finally, we also need the proximal operator of the indicator function over the unit-ball $\iota_{\mathcal{B}}(\cdot)$, where $\mathcal{B} = \left\{ x \in \mathbb{R}^{n} \; | \; \| x \|_2 = 1\right\}$. It can be computed as:

\begin{equation}
    \text{prox}_{\iota_{\mathcal{B}}(\cdot)}(x) = \frac{x}{\| x \|_{2}}.
\end{equation}

% ---------- %
\subsection{Sparsity enforcement parameters}\label{appdx:sparsity_strategies}

There are two moments when we enforce sparsity during the optimisation. 
First, when we denoise the eigenPSFs by the use of the $\ell_1$ norm as in \autoref{eq:min_problem}. The $\mathbf{w}$ weights are set depending on a noise estimation of the observed images, and the parameters $K_{\sigma}^{\rm Loc}$ and $K_{\sigma}^{\rm Glob}$. The noise standard deviation is estimated using the median absolute deviation. The higher the $K_{\sigma}$ parameters are set, the higher the  thresholding and the denoising will be.
Second, when we enforce the spatial constraints through $\alpha$ sparsity. In this case, we follow the sparsity enforcement proposed in \cite{ngole2016}.

\end{appendix}

\end{document}